# Gaussian Interference Channel Capacity to Within One Bit*

Raul H. Etkin †, David N. C. Tse ‡, Hua Wang §
raul.etkin@hp.com, dtse@eecs.berkeley.edu, huawang@uiuc.edu

February 1, 2008

**Abstract**

The capacity of the two-user Gaussian interference channel has been open for thirty years. The understanding on this problem has been limited. The best known achievable region is due to Han-Kobayashi but its characterization is very complicated. It is also not known how tight the existing outer bounds are. In this work, we show that the existing outer bounds can in fact be arbitrarily loose in some parameter ranges, and by deriving new outer bounds, we show that a simplified Han-Kobayashi type scheme can achieve to within a single bit the capacity for all values of the channel parameters. We also show that the scheme is asymptotically optimal at certain high SNR regimes. Using our results, we provide a natural generalization of the point-to-point classical notion of degrees of freedom to interference-limited scenarios.

## 1 Introduction

Interference is a central phenomenon in wireless communication when multiple uncoordinated links share a common communication medium. Most state-of-the-art wireless systems deal with interference in one of two ways:

- orthogonalize the communication links in time or frequency, so that they do not interfere with each other at all;

*The ordering of the authors in this paper is alphabetical.

†R. Etkin is with Hewlett-Packard Laboratories, Palo Alto, CA 94304, USA. This work was done when R. Etkin was a graduate student at U.C. Berkeley. R. Etkin was supported by the National Science Foundation under an ITR grant: “the 3R’s of Spectrum Management: Reuse, Reduce and Recycle”.

‡D. Tse is with Wireless Foundations, University of California, Berkeley, Berkeley, CA 94720, USA. D. Tse is supported by the National Science Foundation under an ITR grant: “the 3R’s of Spectrum Management: Reuse, Reduce and Recycle”.

§H. Wang is with the Dept. of Electrical and Computer Engineering, University of Illinois, Urbana-Champaign, Urbana, IL 61801, USA. This work was done when H. Wang was visiting U.C. Berkeley. H. Wang is supported by a Vodafone US Foundation Graduate Fellowship and NSF CCR CAREER 0237549.

- allow the communication links to share the same degrees of freedom, but treat each other's interference as adding to the noise floor.

It is clear that both approaches can be sub-optimal. The first approach entails an *a priori* loss of degrees of freedom in both links, no matter how weak the potential interference is. The second approach treats interference as pure noise while it actually carries information and has structure that can potentially be exploited in mitigating its effect.

These considerations lead to the natural question of what is the best performance one can achieve without making any *a priori* assumptions on how the common resource is shared. A basic information theory model to study this question is the two-user Gaussian interference channel, where two point-to-point links with additive white Gaussian noise interfere with each other (Figure 1). The capacity region of this channel is the set of all simultaneously achievable rate pairs $(R_1, R_2)$ in the two interfering links, and characterizes the fundamental tradeoff between the performance achievable in the two links in face of interference. Unfortunately, the problem of characterizing this region has been open for over thirty years. The only case in which the capacity is known is in the *strong* interference case, where each receiver has a better reception of the other user's signal than the intended receiver [1, 2]. The best known strategy for the other cases is due to Han and Kobayashi [1]. This strategy is a natural one and involves splitting the transmitted information of both users into two parts: private information to be decoded only at own receiver and common information that can be decoded at both receivers. By decoding the common information, part of the interference can be cancelled off, while the remaining private information from the other user is treated as noise. The Han-Kobayashi strategy allows arbitrary splits of each user's transmit power into the private and common information portions as well as time sharing between multiple such splits. Unfortunately, the optimization among such myriads of possibilities is not well-understood, so while it is clear that it will be no worse than the above-mentioned strategies as it includes them as special cases, it is not very clear how much improvement can be obtained and in which parameter regime would one get significant improvement. More importantly, it is also not clear how close to capacity can such a scheme get and whether there will be other strategies that can do significantly better.

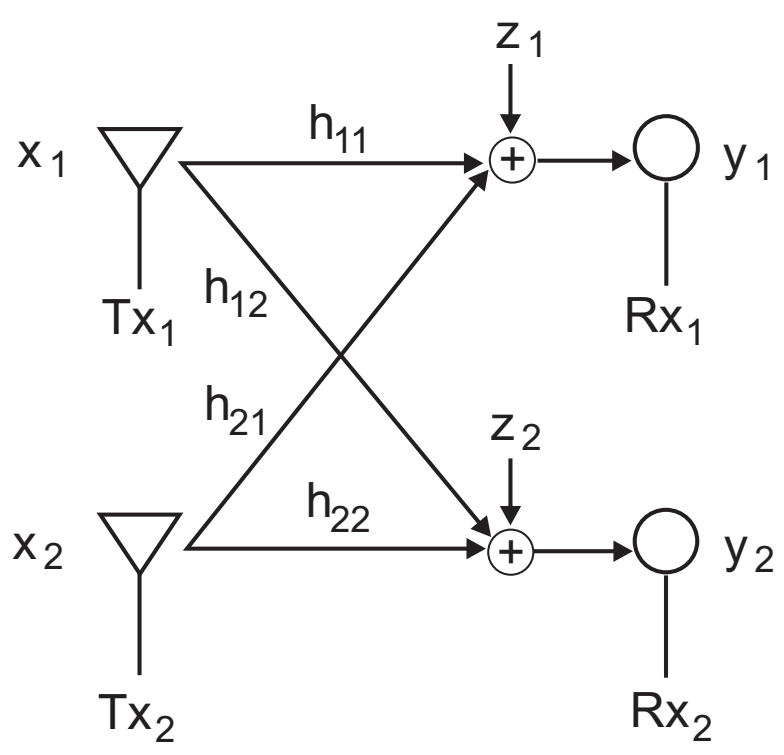

Figure 1: Two-user Gaussian interference channel.

In this paper, we make progress on this state of affairs by showing that a very simple Han-Kobayashi type scheme can in fact achieve rates within 1 bits/s/Hz of the capacity of the channel for *all* values of the channel parameters. That is, this scheme can achieve the rate pair $(R_1 - 1, R_2 - 1)$ for any $(R_1, R_2)$ in the interference channel capacity region. This result is particularly relevant in the high signal-to-noise-ratio (SNR) regime, where the achievable rates are high and grow unbounded as the noise level goes to zero. In fact, in some high SNR regimes, we can strengthen our results to show that our scheme is asymptotically optimal. The high SNR regime is the interference-limited scenario: when the noise is small, interference from one link will have a significant impact on the performance of the other. The low SNR regime is less interesting since here the performance of each link is primarily noise-limited and interference is not having a significant effect.

The key feature of the scheme is that the power of the private information of each user should be set such that it is received at the level of the Gaussian noise at the other receiver. In this way, the interference caused by the private information has a small effect on the other link as compared to the impairments already caused by the noise. At the same time, quite a lot of private information can be conveyed in the own link if the direct gain is appreciably larger than the cross gain.

To prove our result, we need good outer bounds on the capacity region of the interference channel. The best known outer bound is based on giving extra side information to one of the receivers so that it can decode all of the information from the other user (the one-sided interference channel and related bounds). It turns out that while this bound is sufficiently tight in some parameter regimes, it can get arbitrarily loose in others. We derive new outer bounds to cover for the other parameter ranges.

At high SNR, it is well known that the capacity of a point-to-point additive white Gaussian noise (AWGN) link, in bits/s/Hz, is approximately:

$$C_{\text{awgn}} \approx \log \mathsf{SNR} \tag{1}$$

Using our results, we can derive analogous approximations of the Gaussian interference channel capacity, accurate to within one bit/s/Hz. Just to give a flavor of the results, let us consider the symmetric case where the signal-to-noise ratios at the two receivers are the same (denoted by $\mathsf{SNR}$) and the interference-to-noise ratios at the receivers are also the same (denoted by $\mathsf{INR}$). The symmetric capacity, i.e. the best rate that both users can simultaneously achieve, is approximately:

$$C_{\text{sym}} \approx \begin{cases} \log(\frac{\mathsf{SNR}}{\mathsf{INR}}) & \log \mathsf{INR} < \frac{1}{2}\log \mathsf{SNR} & \text{(regime 1)} \\ \log \mathsf{INR} & \frac{1}{2}\log \mathsf{SNR} < \log \mathsf{INR} < \frac{2}{3}\log \mathsf{SNR} & \text{(regime 2)} \\ \log \frac{\mathsf{SNR}}{\sqrt{\mathsf{INR}}} & \frac{2}{3}\log \mathsf{SNR} < \log \mathsf{INR} < \log \mathsf{SNR} & \text{(regime 3)} \\ \log \sqrt{\mathsf{INR}} & \log \mathsf{SNR} < \log \mathsf{INR} < 2\log \mathsf{SNR} & \text{(regime 4)} \\ \log \mathsf{SNR} & \log \mathsf{INR} > 2\log \mathsf{SNR} & \text{(regime 5)} \end{cases} \tag{2}$$

Note that there are five regimes in which the qualitative behaviors of the capacity are different.

The fifth regime is the *very strong interference* regime [2]. Here the interference is so strong that each receiver can decode the other transmitter's information, treating its own signal as noise, before decoding its own information. Thus, interference has no impact on the performance of the other link. The fourth regime is the *strong interference* regime, where the optimal strategy is for both receivers to decode entirely each other's signal, i.e. all the transmitted information is common information. Here, the capacity increases monotonically with $\mathsf{INR}$ because increasing $\mathsf{INR}$ increases the common information rate.

The capacity in the fourth and fifth regimes follow from previous results. The first three regimes fall into the *weak interference* regime, and the capacity in these regimes is a consequence of the new results that we obtain. In these regimes, the interference is not strong enough to be decoded in its entirety. In fact, the capacity approximation (2) in regime 1 implies that if the interference is very weak, then treating interference as noise is optimal. The capacity expression for regimes 2 and 3 however imply that if the interference is not very weak, decoding it partially can significantly improve performance. Interestingly, the capacity is *not* monotonically decreasing with $\mathsf{INR}$ in the weak interference regime.

In point-to-point links, the notion of *degrees of freedom* is a fundamental measure of channel resources. It tells us how many signal dimensions are available for communication. In the (scalar) AWGN channel, there is one degree of freedom per second per Hz. When multiple links share the communication medium, one can think of the mutual interference as reducing the available degrees of freedom for useful communication. Our results quantify this reduction. Define

$$\alpha := \frac{\log \mathsf{INR}}{\log \mathsf{SNR}}$$

as the ratio of the interference-to-noise ratio and the signal-to-noise ratio in dB scale, and

$$d_{\text{sym}} := \frac{C_{\text{sym}}}{C_{\text{awgn}}}$$

as the *generalized degrees of freedom* per user. Then (2) yields the following characterization:

$$d_{\text{sym}} = \begin{cases} 1-\alpha & 0 \le \alpha \le \frac{1}{2} \\ \alpha & \frac{1}{2} \le \alpha \le \frac{2}{3} \\ 1-\frac{\alpha}{2} & \frac{2}{3} \le \alpha \le 1 \\ \frac{\alpha}{2} & 1 \le \alpha \le 2 \\ 1 & \alpha \ge 2. \end{cases} \tag{3}$$

This is plotted in Figure 2, together with the performance of our baseline strategies of orthogonalizing and treating interference as noise. Note that orthogonalizing between the links, in which each link achieves half the degrees of freedom, is strictly sub-optimal except when $\alpha = \frac{1}{2}$ and $\alpha = 1$. Treating interference as noise, on the other hand, is strictly sub-optimal except for $\alpha \le \frac{1}{2}$. Note also the fundamental importance of comparing the signal-to-noise and the interference-to-noise ratios in dB scale.

The rest of the paper is structured as follows. In Section 2, we describe the model. Section 3 focuses on the symmetric rate point in the symmetric interference channel, where the results can be described in the simplest form. Results on the entire capacity region for the general two-user channel are explained in Section 4. Section 5 investigates how the generalized degrees of freedom in the general case depend on the various channel parameters. Section 6 provides intuition about the simple Han-Kobayashi scheme used in this paper. Section 7 explores some analogies between our results and those of El Gamal and Costa on a deterministic interference channel [3].

Regarding notation, we will use lowercase or uppercase letters for scalars, lowercase boldface letters for vectors, and calligraphic letters for sets. For example we write $h$ or $P$ for scalars, $\mathbf{x}$ for a vector, and $\mathcal{R}$ for a set. We use $H(\cdot)$ to denote binary entropy of a discrete random variable or vector, $h(\cdot)$ to denote binary differential entropy of a continuous random variable or vector, and $I(\cdot;\cdot)$ to denote mutual information. In addition, unless otherwise stated, all logarithms are to the base 2.

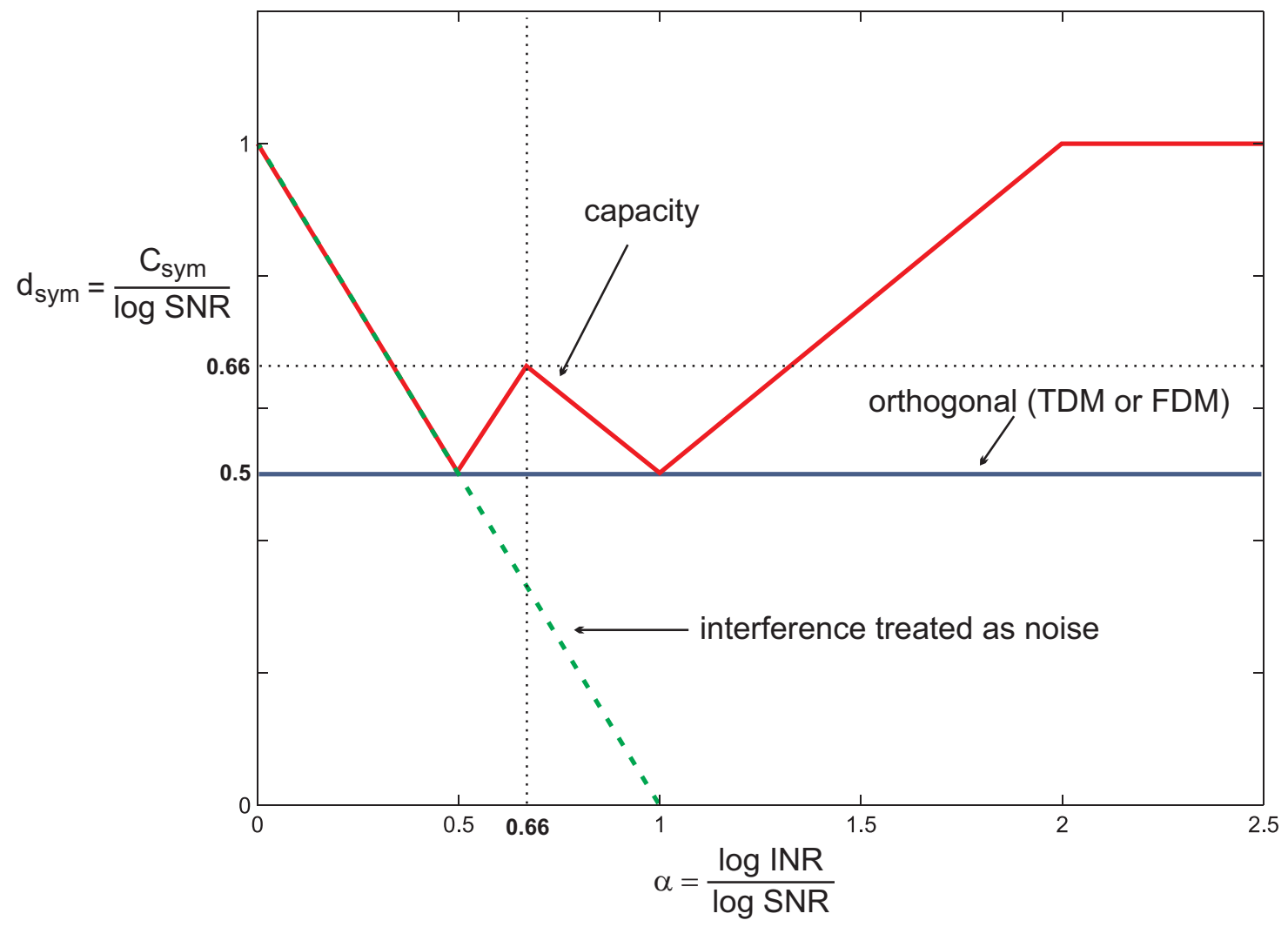

Figure 2: Generalized degrees of freedom for two suboptimal schemes vs. capacity. These suboptimal schemes are treating interference as noise and orthogonalizing the users over time or frequency.

## 2 Model

In this section we describe the model to be used in the rest of this part. We consider a two-user Gaussian interference channel. In this model there are two transmitter-receiver pairs, where each transmitter wants to communicate with its corresponding receiver (cf. Figure 1).

This channel is represented by the equations:

$$\begin{aligned} y_1 &= h_{11}x_1 + h_{21}x_2 + z_1 \\ y_2 &= h_{12}x_1 + h_{22}x_2 + z_2 \end{aligned} \tag{4}$$

where for $i = 1, 2$, $x_i \in \mathbb{C}$ is subject to a power constraint $P_i$, i.e. $E[|x_i|^2] \leq P_i$, and the noise processes $z_i \sim \mathcal{CN}(0, N_0)$ are i.i.d. over time. For convenience we will denote the power gains of the channels by $|h_{ij}|^2 = |h_{ij}|^2$, $i, j = 1, 2$.

It is easy to see that the capacity region of the interference channel depends only on four parameters: the signal to noise and interference to noise ratios. For $i = 1, 2$, let $\mathsf{SNR}_i = |h_{ii}|^2 P_i / N_0$ be the signal to noise ratio of user $i$, and $\mathsf{INR}_1 = |h_{21}|^2 P_2 / N_0$ ($\mathsf{INR}_2 = |h_{12}|^2 P_1 / N_0$) be the interference to noise ratio of user 1 (2). As will become apparent from our analysis, this parameterization in terms of SNR and INR is more natural for the interference channel, because it puts in evidence the main factors that determine the channel capacity.

For a given block length $n$, user $i$ communicates a message $m_i \in \{1, \ldots, 2^{nR_i}\}$ by choosing a codeword from a codebook $\mathcal{C}_{i,n}$, with $|\mathcal{C}_{i,n}| = 2^{nR_i}$. The codewords $\{\mathbf{c}_i(m_i)\}$ of this codebook must satisfy the average power constraint:

$$\frac{1}{n}\sum_{t=1}^{n} |c_i(m_i)[t]|^2 \leq P_i$$

Receiver $i$ observes the channel outputs $\{y_i[t] : t = 1, \ldots, n\}$ and uses a decoding function $f_{i,n} : \mathbb{C}^n \to \mathbb{N}$ to get the estimate $\hat{m}_i$ of the transmitted message $m_i$. The receiver is in error whenever $\hat{m}_i \neq m_i$. The average probability of error for user $i$ is given by

$$\epsilon_{i,n} = E[P(\hat{m}_i \neq m_i)]$$

where the expectation is taken with respect to the random choice of the transmitted messages $m_1$ and $m_2$.

A rate pair $(R_1, R_2)$ is achievable if there exists a family of codebook pairs $\{(\mathcal{C}_{1,n}, \mathcal{C}_{2,n})\}_n$ with codewords satisfying the power constraints $P_1$ and $P_2$ respectively, and decoding functions $\{(f_{1,n}(\cdot), f_{2,n}(\cdot)\}_n$, such that the average decoding error probabilities $\epsilon_{1,n}, \epsilon_{2,n}$ go to zero as the block length $n$ goes to infinity.

The capacity region $\mathcal{R}$ of the interference channel is the closure of the set of achievable rate pairs.

# 3 Symmetric Gaussian Interference Channel

## 3.1 Symmetric channel and symmetric rate point

In order to introduce the main ideas and results in the simplest possible setting, we start our analysis of the interference channel capacity region by considering a symmetric interference channel and the symmetric rate point.

In the symmetric interference channel we have $|h_{11}|^2 = |h_{22}|^2 = |h_d|^2$, $|h_{12}|^2 = |h_{21}|^2 = |h_c|^2$ and $P_1 = P_2 = P$, or equivalently, $\mathsf{SNR}_1 = \mathsf{SNR}_2$ and $\mathsf{INR}_1 = \mathsf{INR}_2$. In addition, the symmetric capacity is the solution to the following optimization problem:

$$C_{\text{sym}} = \begin{cases} \text{Maximize:} & \min\{R_1, R_2\} \\ \text{Subject to:} & (R_1, R_2) \in \mathcal{R} \end{cases}$$

where $\mathcal{R}$ is the capacity region of the interference channel.

Due to the convexity and symmetry of the capacity region of the symmetric channel, the symmetric capacity is attained at the point of the capacity region that maximizes the sum rate $R_1 + R_2$. Since the capacity region is known in the strong interference case when $\mathsf{INR}/\mathsf{SNR} \geq 1$, we will focus on the case where $0 < \mathsf{INR}/\mathsf{SNR} < 1$. In addition, we will concentrate on the situation where $\mathsf{INR} \geq 1$, i.e. the interfering signal power is at least as large as the noise power. The case $\mathsf{INR} < 1$ is not so interesting because the communication is essentially limited by noise. We will address this case briefly later.

### 3.2 A simple communication scheme

We will use a simple communication scheme that is a special case of the general type of schemes introduced by Han and Kobayashi in [1]. For a given block length $n$ user $i$ chooses a private message from codebook $\mathcal{C}_{i,n}^u$ and a common message from codebook $\mathcal{C}_{i,n}^w$. These codebooks satisfy the power constraints $P_u$ and $P_w$ with $P_u + P_w = P$. The sizes of these codebooks are such that $|\mathcal{C}_{i,n}^u| \cdot |\mathcal{C}_{i,n}^w| = 2^{nR_i}$. After selecting the corresponding codewords user $i$ transmits the signal $\mathbf{x}_i = \mathbf{c}_i^u + \mathbf{c}_i^w$ by adding the private and common codewords. The private codewords are meant to be decoded by receiver $i$, while the common codewords must be decoded by both receivers.

The general Han and Kobayashi scheme allows to generate the codebooks using arbitrary input distributions, and allows to do time sharing between multiple strategies. We will consider a simple scheme where the codebooks are generated by using i.i.d. random Gaussian random variables with the appropriate variances. Let $\mathsf{INR}_p = |h_c|^2 P_u / N_0$, that is, $\mathsf{INR}_p$ is the interference to noise ratio created onto the non-intending receiver by the private message. We choose $\mathsf{INR}_p = 1$, i.e. the interference created by the private message has the same power as the Gaussian noise[1]. In addition, we use a fixed strategy, i.e. we don't do time sharing.

Why do we choose $\mathsf{INR}_p = 1$ ? From the point of view of a single user, that is, if we don't take interference into account, one should make the private message power as large as possible (i.e. set $\mathsf{INR}_p = \mathsf{INR}$). However, due to interference, it may be convenient to reduce the private message power, so that part of the interfering signal (the common message) can be decoded and subtracted at the other receiver. We see that there is a tradeoff between achieving a large rate at one's link and minimizing the interference caused at the other user's link. In Figure 3 we plot the single user rate as a function of the interference power created by the private message of the other user. We can see that if we choose $\mathsf{INR}_p = 1$ the effect of the interference caused by the private message is small. At the same time, $\mathsf{INR}_p = 1$ allows to obtain a relatively large private message rate in the direct link. We will give a deeper explanation later on in Section 6.

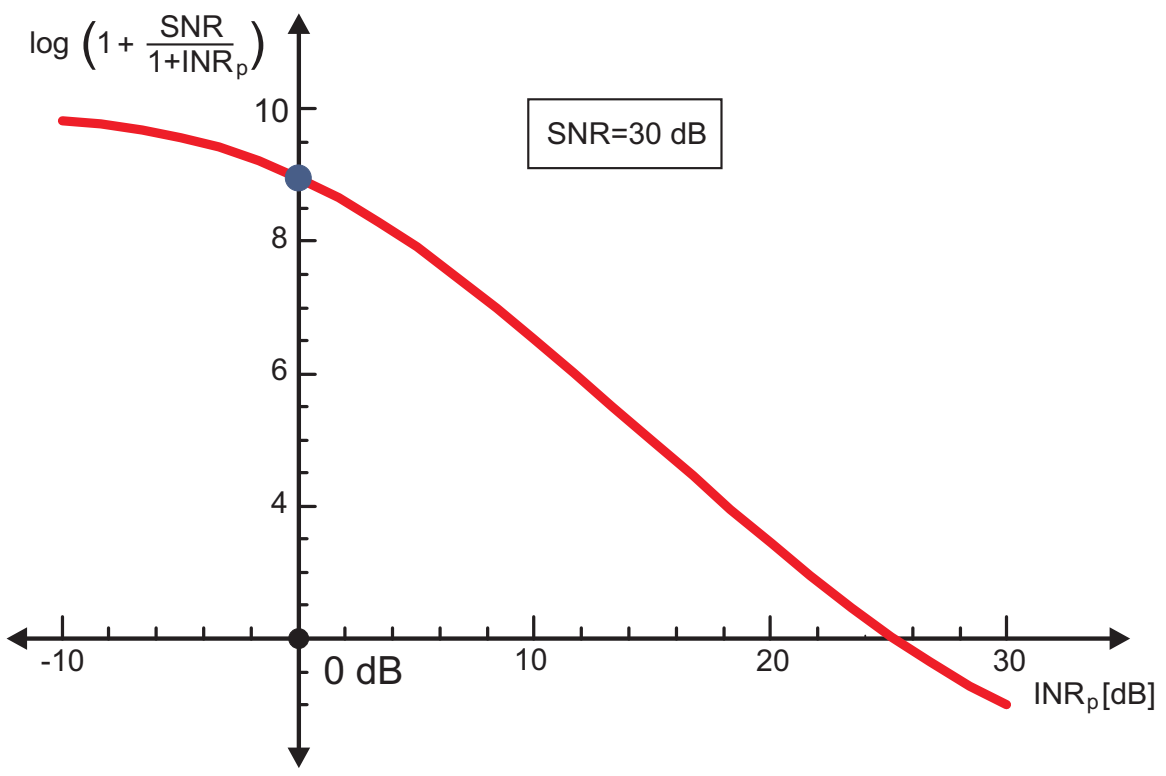

Figure 3: Rate vs. interference power level. The choice $\mathsf{INR}_p = 1$ (0 dB) does not create too much interference and it achieves a large private message rate.

[1]Note that this is possible with the available power under the assumption $\mathsf{INR} \geq 1$. If $\mathsf{INR} < 1$ one can choose $\mathsf{INR}_p = \mathsf{INR}$, but will not consider this case in this section.

We will show that this simple scheme allows us to achieve a symmetric rate close to the symmetric rate capacity of the channel. In order to determine the symmetric rate that we can achieve with this scheme, it is useful to think of each user as being split into two virtual users: private user $U_i$ and common user $W_i$. Let $MAC_1$ be the multiple access channel formed by virtual users $U_1$, $W_1$, and $W_2$, and receiver 1, with the signal from virtual user $U_2$ being treated as noise. In a similar way, let $MAC_2$ be the multiple access channel formed by virtual users $U_2$, $W_1$, and $W_2$, and receiver 2, with the signal from virtual user $U_1$ being treated as noise. Since the common messages must be decoded by both receivers, while the private messages must be decoded only by the intending receiver, we see that the rates achievable by a the Han and Kobayashi scheme correspond to the intersection of the capacity regions of $MAC_1$ and $MAC_2$.

Among all the possible rate assignments for the private and common messages of this scheme, we choose the private rates of both users, as well as the common rate of both users to be equal, i.e. $R_{u,1} = R_{u,2}$ and $R_{w,1} = R_{w,2}$. We also fix a decoding order at each receiver, so that the common messages are decoded first, while the private message is decoded last. This choice of rates and decoding order allows for an easy analysis of the scheme and, as will be shown later, also achieves a symmetric rate close to capacity.

Since the private message is decoded last, while the private message of the other user is treated as noise, the private rate of each user is given by:

$$R_u = \log\left(1 + \frac{\mathsf{SNR}\cdot\mathsf{INR}_p}{\mathsf{INR}(1+\mathsf{INR}_p)}\right) = \log\left(1+\frac{\mathsf{SNR}}{2\mathsf{INR}}\right).$$

Since each receiver decodes the common messages first, both private messages are treated as noise when decoding the common messages. With this decoding order, the sum rate of the common messages must satisfy two constraints:

$$R_{w,1} + R_{w,2} \leq \log\left(1 + \frac{(\mathsf{INR}-1)(\mathsf{SNR}+\mathsf{INR})}{\mathsf{SNR}+2\mathsf{INR}}\right) \tag{5}$$

and

$$R_{w,1} + R_{w,2} \leq 2\log\left(1 + \frac{\mathsf{INR}(\mathsf{INR}-1)}{\mathsf{SNR}+2\mathsf{INR}}\right) \tag{6}$$

where (5) arises from the sum rate constraint of the MAC formed by virtual users $W_1$ and $W_2$ at receiver 1 (or receiver 2) when the messages from virtual users $U_1$ and $U_2$ are treated as noise, and (6) arises from the individual rate constraint of decoding the message of virtual user $W_1$ at receiver 2 and virtual user $W_2$ at receiver 1, treating the messages from virtual users $U_1$ and $U_2$ as noise (see Figure 4).

Therefore, with the simple Han and Kobayashi scheme we obtain a symmetric rate:

$$\begin{aligned} R_{HK} &= \log\left(1+\frac{\mathsf{SNR}}{2\mathsf{INR}}\right) + \min\left\{\frac{1}{2}\log\left(1+\frac{(\mathsf{INR}-1)(\mathsf{SNR}+\mathsf{INR})}{\mathsf{SNR}+2\mathsf{INR}}\right), \log\left(1+\frac{\mathsf{INR}(\mathsf{INR}-1)}{\mathsf{SNR}+2\mathsf{INR}}\right)\right\} \\ &= \min\left\{\frac{1}{2}\log\left(1+\mathsf{SNR}+\mathsf{INR}\right) + \frac{1}{2}\log\left(2+\frac{\mathsf{SNR}}{\mathsf{INR}}\right) - 1, \log\left(1+\mathsf{INR}+\frac{\mathsf{SNR}}{\mathsf{INR}}\right) - 1\right\}. \end{aligned} \tag{7}$$

By comparing (5) and (6) we can determine the parameter ranges in which each of the terms of the $\min\{\cdot,\cdot\}$ in (7) is active. Define:

$$\begin{aligned} \mathcal{B}_1 &= \left\{(\mathsf{SNR},\mathsf{INR}) : \mathsf{INR}\geq 1 \text{ and } \mathsf{SNR}(\mathsf{SNR}+\mathsf{INR}) < \mathsf{INR}^2(\mathsf{INR}+1)\right\} \\ \mathcal{B}_2 &= \left\{(\mathsf{SNR},\mathsf{INR}) : \mathsf{INR}\geq 1 \text{ and } \mathsf{SNR}(\mathsf{SNR}+\mathsf{INR}) \geq \mathsf{INR}^2(\mathsf{INR}+1)\right\}. \end{aligned} \tag{8}$$

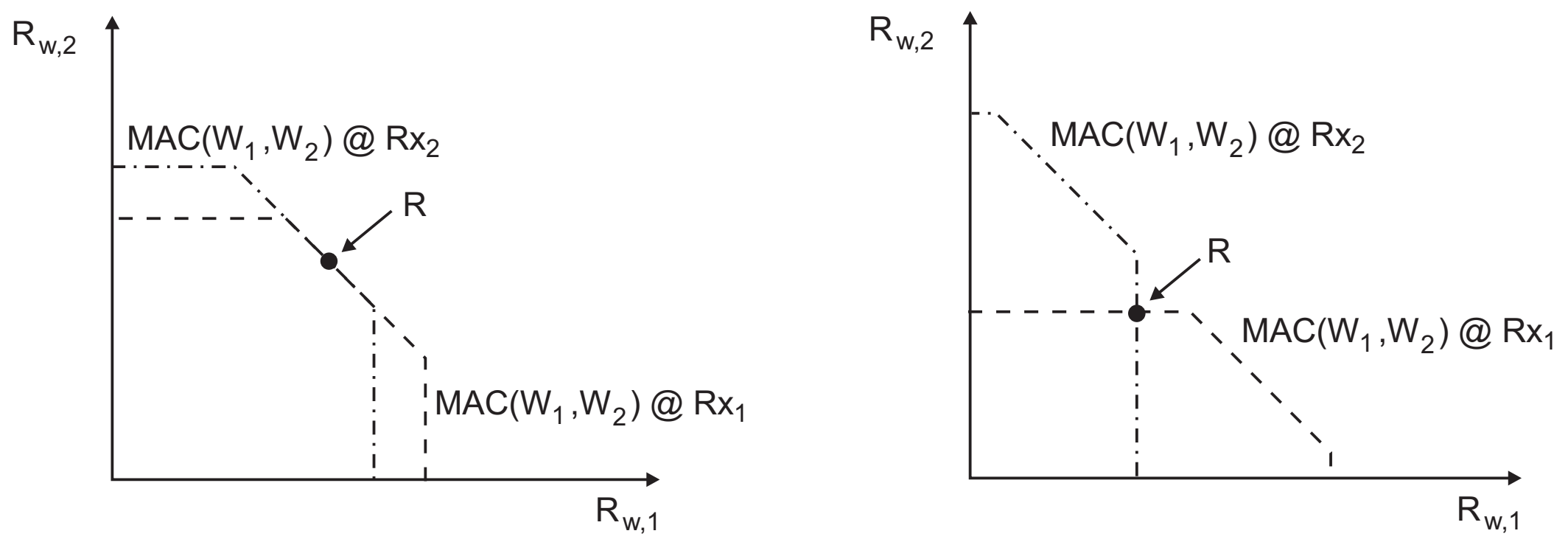

Figure 4: Intersection of the multiple access channel regions corresponding to virtual users $W_1$ and $W_2$ at receivers 1 and 2, when the signals from virtual users $U_1$ and $U_2$ are treated as noise. The left figure corresponds to the case in which the sum rate constraint (5) is active, while the right figure corresponds to the case in which the sum rate constraint(6) is active. In both cases, the symmetric rate point is indicated. Note that due to the symmetry in the channel and power allocations, the regions MAC$(W_1, W_2)$ at receivers 1 and 2 are mirror images of each other with respect to the line $R_{w,1} = R_{w,2}$.

Then, the first (second) term of the $\min\{\cdot,\cdot\}$ is active in $\mathcal{B}_1$ ($\mathcal{B}_2$). We denote by $R_{HK_1}$ ($R_{HK_2}$) the symmetric rate expression that results in $\mathcal{B}_1$ ($\mathcal{B}_2$).

We can gain further insight into the achievable rate (7) and the different parameter regimes $\mathcal{B}_1, \mathcal{B}_2$, by considering how the ratio $R_{HK}/C_{\text{awgn}}$ varies for different interference levels. Dividing (7) by $\log(1+\mathsf{SNR})$, we obtain for large $\mathsf{SNR}, \mathsf{INR}$:

$$\begin{aligned}
\frac{R_{HK}}{C_{\text{awgn}}} &= \frac{\min\left\{\frac{1}{2}\log\left(1+\mathsf{SNR}+\mathsf{INR}\right)+\frac{1}{2}\log\left(2+\frac{\mathsf{SNR}}{\mathsf{INR}}\right)-1, \log\left(1+\mathsf{INR}+\frac{\mathsf{SNR}}{\mathsf{INR}}\right)-1\right\}}{\log(1+\mathsf{SNR})} \\
&\approx \frac{\min\left\{\frac{1}{2}\log\left(\mathsf{SNR}\right)+\frac{1}{2}\left[\log\left(\mathsf{SNR}\right)-\log\left(\mathsf{INR}\right)\right], \max\left\{\log\left(\mathsf{INR}\right), \log\left(\mathsf{SNR}\right)-\log\left(\mathsf{INR}\right)\right\}\right\}}{\log(\mathsf{SNR})} \\
&\approx \min\left\{1-\frac{1}{2}\frac{\log\mathsf{INR}}{\log\mathsf{SNR}}, \max\left\{\frac{\log\mathsf{INR}}{\log\mathsf{SNR}}, 1-\frac{\log\mathsf{INR}}{\log\mathsf{SNR}}\right\}\right\}. \qquad (9)
\end{aligned}$$

We define the interference level $\alpha$ as the ratio of $\mathsf{INR}$ and $\mathsf{SNR}$ in dB, that is:

$$\alpha := \frac{\log\mathsf{INR}}{\log\mathsf{SNR}}$$

and rewrite (9) as a function of $\alpha$:

$$\frac{R_{HK}(\alpha)}{C_{\text{awgn}}} \approx \min\left\{1-\frac{\alpha}{2}, \max\left\{\alpha, 1-\alpha\right\}\right\} \qquad (10)$$

By inspecting (10) we can readily identify three different regimes. The first term of the $\min\{\cdot,\cdot\}$ is active in (10) when $2/3 < \alpha < 1$. This corresponds to the parameter range $\mathcal{B}_1$, in which the MAC constraint (5) is active. For $0 < \alpha < 2/3$ the second term of the $\min\{\cdot,\cdot\}$ is active in (10), which corresponds to the parameter range $\mathcal{B}_2$. In this range the MAC constraint (6) is active. In addition, we can further identify two subregimes, depending on whether $1/2 < \alpha < 2/3$ (the first term of the $\max\{\cdot,\cdot\}$ is active) or $0 < \alpha < 1/2$ (the second term of the $\max\{\cdot,\cdot\}$ is active). Figure 5 shows how $R_{HK}/C_{\text{awgn}}$ varies with $\alpha$ in the different parameter regimes.

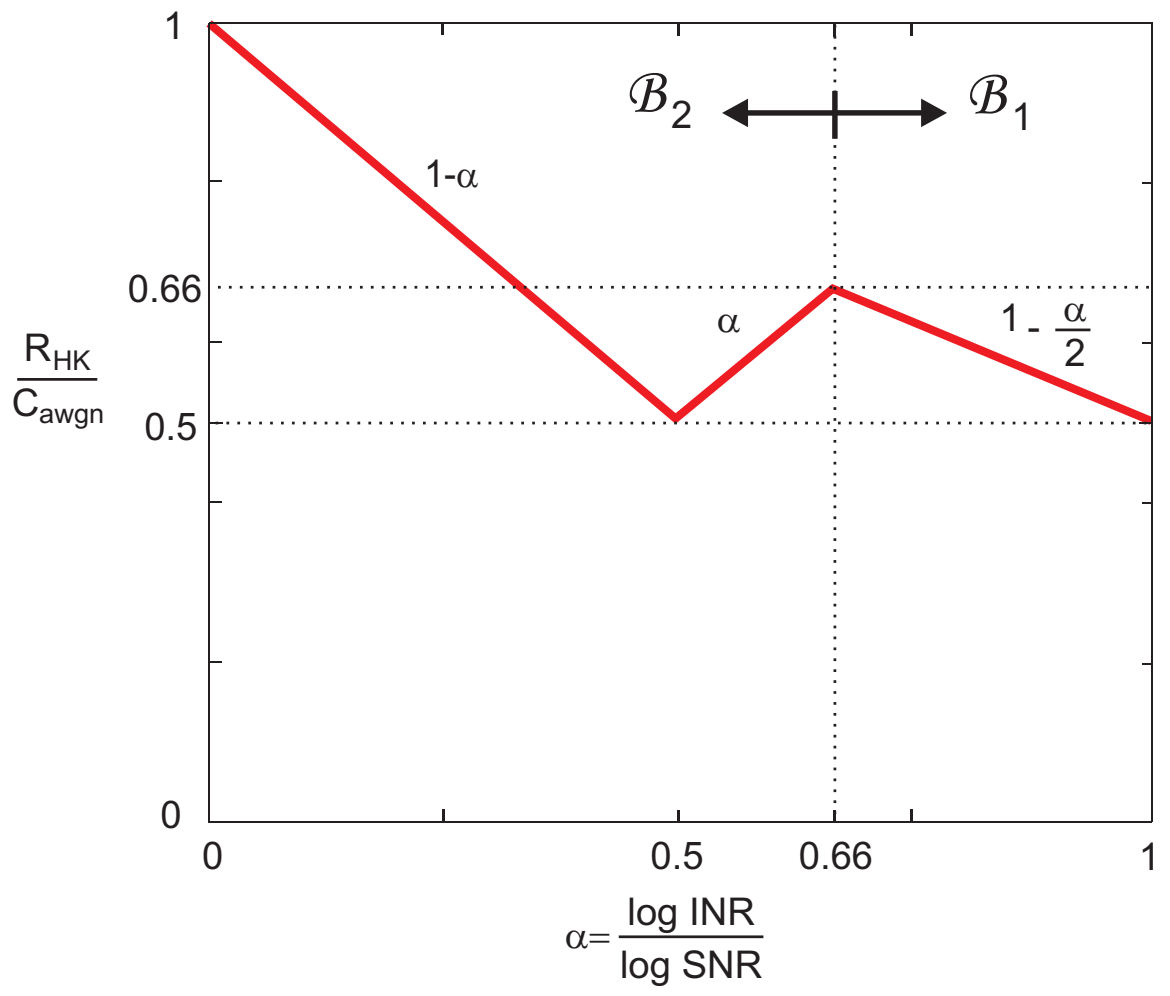

Figure 5: Achievable rate as a fraction of single user capacity vs. interference level.

### 3.3 Known upper bounds

In order to evaluate the performance of our communication scheme, we can compare the symmetric rate achieved with an upper bound. We can obtain this upper bound by considering any outer bound to the interference channel capacity region evaluated at $R_1 = R_2$. The best known outer bound to the interference channel capacity region is that given in [6] Theorem 2. We analyze this bound in Appendix A, and provide in this section an alternative bound that has similar performance and is easier to obtain and analyze.

We will consider a general interference channel so that the upper bounds that we derive are not restricted to the symmetric interference channel. Consider a modified interference channel, where a genie provides the side information $x_2$ to receiver 1 (see Figure 6).

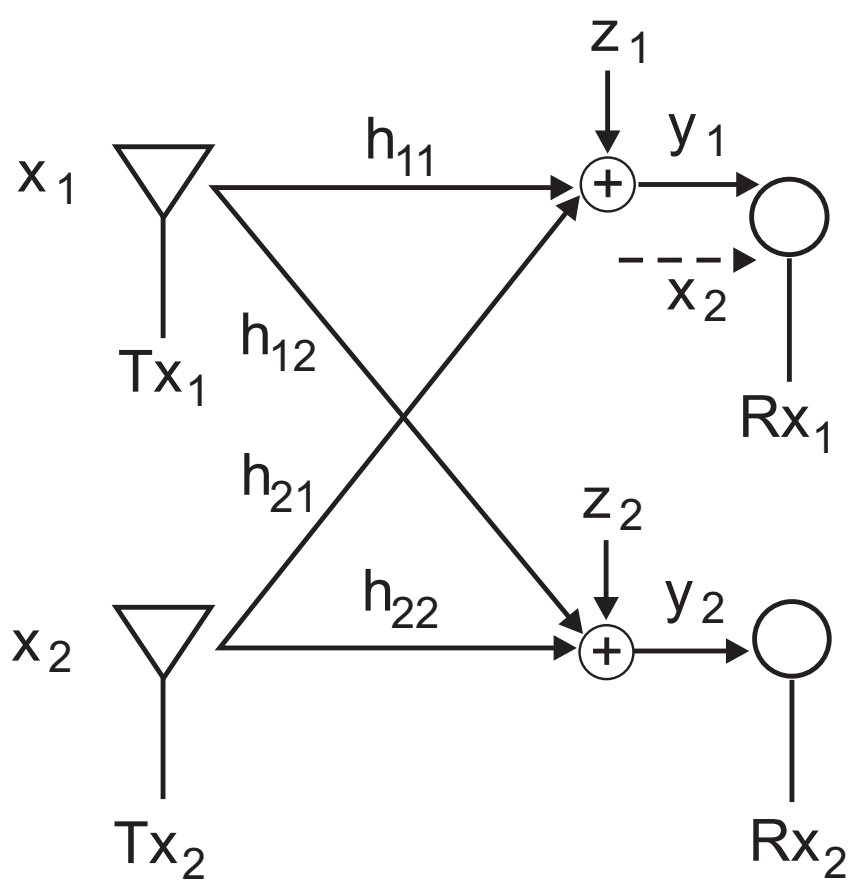

Figure 6: Genie-aided two-user interference channel. A genie provides signal $x_2$ to receiver 1.

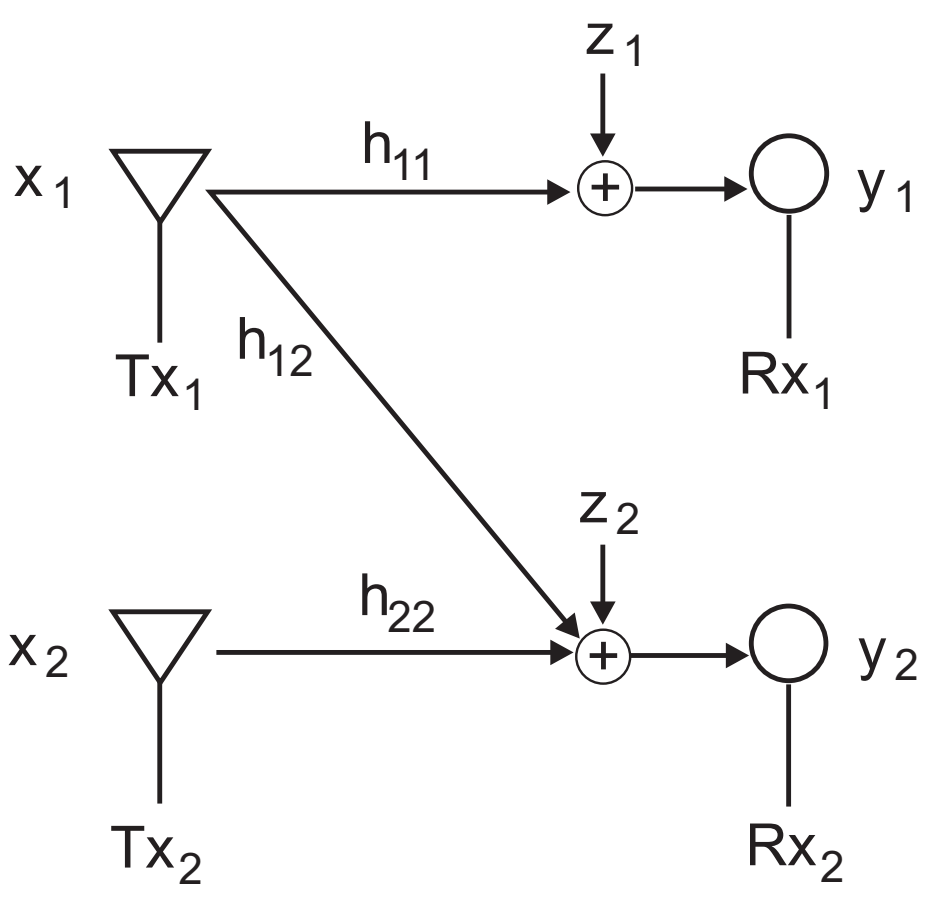

Figure 7: One-sided interference channel.

Since $\mathbf{x}_1^n$ is independent of $\mathbf{x}_2^n$ we can write for any block of length $n$:

$$I(\mathbf{x}_1^n; \mathbf{y}_1^n, \mathbf{x}_2^n) = I(\mathbf{x}_1^n; \mathbf{x}_2^n) + I(\mathbf{x}_1^n; \mathbf{y}_1^n | \mathbf{x}_2^n) = I(\mathbf{x}_1^n; h_{11}\mathbf{x}_1^n + \mathbf{z}_1^n)$$

and it follows that receiver 1 can get an interference-free signal by subtracting the interference $h_{21}x_2$ using the side information provided by the genie. Therefore we obtain that the genie-aided channel is equivalent to the one-sided interference channel depicted in Figure 7.

The sum rate capacity of a one-sided interference channel for the case of $\mathsf{INR}_2 < \mathsf{SNR}_1$ is known from previous results [8] and will be explicitly derived in Section 4.1. It is given by:

$$R_{sum}(\text{one-sided IC}) = \log(1 + \mathsf{SNR}_1) + \log\left(1 + \frac{\mathsf{SNR}_2}{1 + \mathsf{INR}_2}\right) \tag{11}$$

and since the aid of the genie can only increase the capacity region of the interference channel, we obtain the upper bound for the symmetric rate:

$$R_{UB} = \frac{1}{2}\log(1 + \mathsf{SNR}) + \frac{1}{2}\log\left(1 + \frac{\mathsf{SNR}}{1 + \mathsf{INR}}\right). \tag{12}$$

In order to compare this bound with the rate obtained with our simple Han-Kobayashi scheme we approximately compute the ratio $R_{UB_2}/C_{\text{awgn}}$ for large $\mathsf{SNR}, \mathsf{INR}$:

$$\frac{R_{UB}}{C_{\text{awgn}}} \approx 1 - \frac{1}{2}\frac{\log \mathsf{INR}}{\log \mathsf{SNR}} = 1 - \frac{\alpha}{2}. \tag{13}$$

We see that (13) coincides with (10) when $2/3 < \alpha < 1$, but (13) and (10) differ when $0 < \alpha < 2/3$ (see Figure 8).

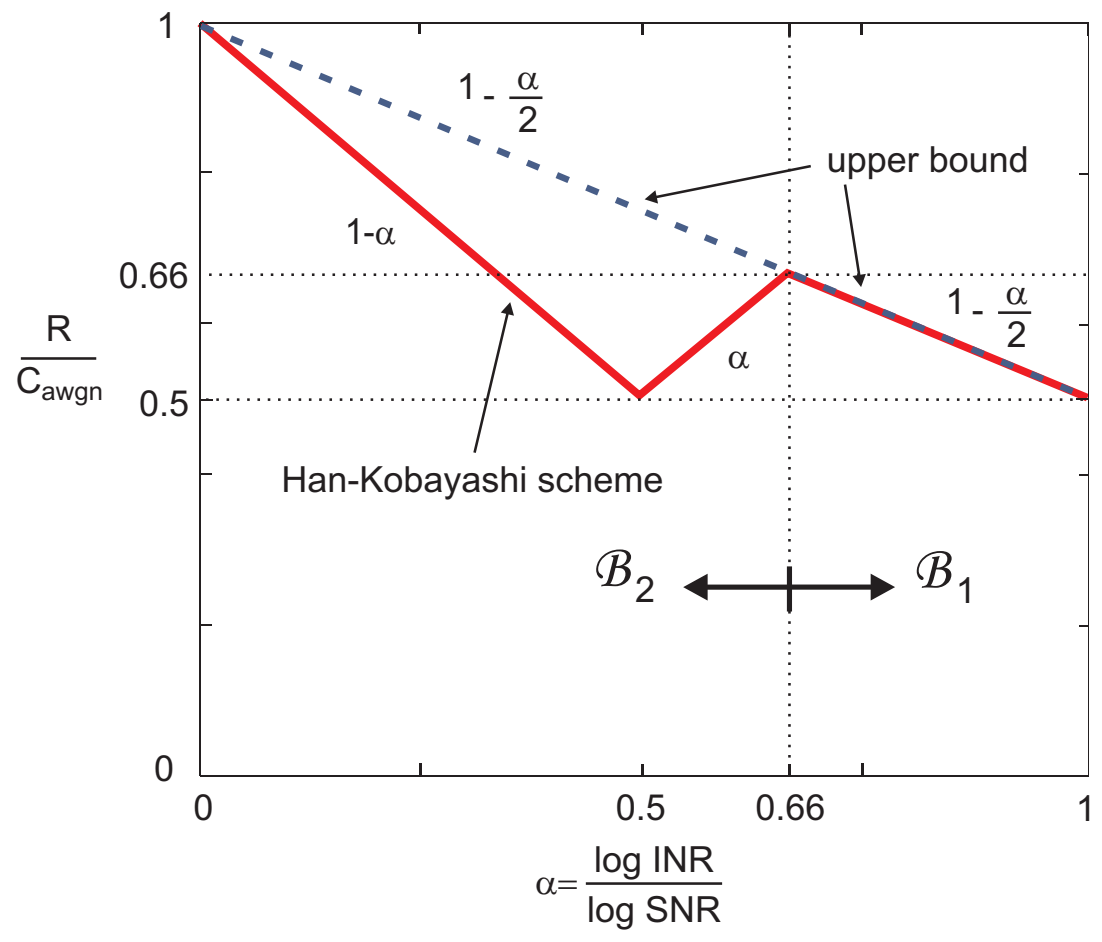

Figure 8: Upper bound and achievable Han-Kobayashi rate (relative to single user capacity) as a function of the interference level $\alpha$.

Figure 8 suggests that the bound (12) is reasonably tight in the parameter range $\mathcal{B}_1$. It turns out that the upper bound (12) and the lower bound (7) differ by at most 1 bit/s/Hz in the parameter range $\mathcal{B}_1$. This can be checked by writing for the parameter range $\mathcal{B}_1$:

$$\begin{aligned}
R_{UB} - R_{HK_1} &= \frac{1}{2}\log\left(1+\mathsf{SNR}\right) + \frac{1}{2}\log\left(1+\frac{\mathsf{SNR}}{1+\mathsf{INR}}\right) - \frac{1}{2}\log\left(1+\mathsf{SNR}+\mathsf{INR}\right) \\
&\quad -\frac{1}{2}\log\left(2+\frac{\mathsf{SNR}}{\mathsf{INR}}\right)+1 \\
&= \frac{1}{2}\log\left(\frac{1+\mathsf{SNR}}{1+\mathsf{INR}}\right) - \frac{1}{2}\log\left(\frac{2\mathsf{INR}+\mathsf{SNR}}{\mathsf{INR}}\right)+1 \\
&< \frac{1}{2}\log\left(\frac{1+\mathsf{SNR}}{2\mathsf{INR}+\mathsf{SNR}}\right)+1 \\
&\leq 1 \qquad (14)
\end{aligned}$$

where we used the assumption $\mathsf{INR} \geq 1$ in the last inequality.

We also observe in Figure 8 that the gap between the upper bound and the achievable rate can be arbitrarily large in the parameter range $\mathcal{B}_2$[2]. This large gap could be due to a very suboptimal scheme, a loose upper bound, or both. It turns out that the large gap is due to the looseness of the upper bound.

Even though the bound (11) is not as good as the bounds presented in [6], all these bounds have the same worst case 1-bit/s/Hz gap with respect to our simple communication strategy in the parameter range $\mathcal{B}_1$. Also, in the parameter range $\mathcal{B}_2$ all these bounds are arbitrarily loose.

Why are all these bounds loose in $\mathcal{B}_2$ ? The problem is that they rely, in one way or another, in giving side information to receiver 1 so that he can eventually cancel the interfering signal from user 2. We can gain some intuition about why these bounds are loose in $\mathcal{B}_2$ by considering our simple communication scheme in the genie-aided channel of Figure 6. The side information provided by the genie allows receiver 1 to subtract the interference generated by transmitter 2. The rates of the virtual private users $U_1$ and $U_2$ are in this case:

$$\begin{aligned}
R_{u_1} &= \log\left(1+\frac{\mathsf{SNR}\cdot\mathsf{INR}_p}{\mathsf{INR}}\right) = \log\left(1+\frac{\mathsf{SNR}}{\mathsf{INR}}\right) \\
R_{u_2} &= \log\left(1+\frac{\mathsf{SNR}\cdot\mathsf{INR}_p}{\mathsf{INR}(1+\mathsf{INR}_p)}\right) = \log\left(1+\frac{\mathsf{SNR}}{2\mathsf{INR}}\right)
\end{aligned}$$

and we see that virtual user $U_1$ gains at most 1 bit/s/Hz due to the help of the genie. The sum rate of the MAC formed by virtual users $W_1$ and $W_2$ at receiver 2 does not change due to the aid of the genie. Therefore, the sum rate constraint (5) still holds. However, due to the aid of the genie receiver 1 can decode the message of virtual user $W_2$ and the sum rate constraint (6) does not appear in this case.

In $\mathcal{B}_1$ the sum rate constraint (5) is active in the original channel, and the aid of the genie does not allow to increase the sum rate by a large amount. In this regime, the bound obtained from the genie-aided channel is good. In contrast, in $\mathcal{B}_2$ the sum rate constraint (6) is active in the original channel, and the genie effectively releases this constraint by providing enough information to receiver 1 to decode the message of virtual user $W_2$. Since in $\mathcal{B}_2$ the constraint (5) is larger than (6) (and the gap between the two constraints can be made arbitrarily large), the bound obtained from the genie-aided channel is loose.

[2] Note that in Figure 8 the rates are plotted relative to $C_{\text{awgn}}$ and any non-zero gap in the figure translates into an unbounded gap in the rates as $\mathsf{SNR} \to \infty$.

### 3.4 A new upper bound

In order to derive a tighter sum rate bound for the parameter range $\mathcal{B}_2$ we will make use of the help of genies, but will avoid giving too much information to either receiver. The information that we will provide will not allow either receiver to completely decode the message of the interfering transmitter. The new sum rate upper bound is given in the following theorem, which we state for a general (not necessarily symmetric) Gaussian interference channel.

***Theorem* 1.** *For a Gaussian interference channel as defined in Section 2, equation (4), the sum rate is upper bounded by*

$$R_1 + R_2 \leq \log\left(1 + \mathsf{INR}_1 + \frac{\mathsf{SNR}_1}{1 + \mathsf{INR}_2}\right) + \log\left(1 + \mathsf{INR}_2 + \frac{\mathsf{SNR}_2}{1 + \mathsf{INR}_1}\right). \tag{15}$$

*Proof.* Define

$$\begin{aligned} s_1 &= h_{12}x_1 + z_2 \\ s_2 &= h_{21}x_2 + z_1 \end{aligned}$$

and consider the genie-aided channel where a genie provides $s_1$ to receiver 1 and $s_2$ to receiver 2 (see Figure 9). Clearly, the capacity region of this genie-aided channel is an outer bound to the capacity region of the original interference channel. Therefore, we can obtain an upper bound for the sum rate of the original channel by computing an upper bound on the sum rate of the genie aided channel. For a block of length $n$ we can bound the sum rate of the genie-aided channel in the following way:

$$
\begin{aligned}
n(R_1+R_2) &\leq I(\mathbf{x}_1^n;\mathbf{y}_1^n,\mathbf{s}_1^n)+I(\mathbf{x}_2^n;\mathbf{y}_2^n,\mathbf{s}_2^n)+n\epsilon_n \\
&= I(\mathbf{x}_1^n;\mathbf{s}_1^n)+I(\mathbf{x}_1^n;\mathbf{y}_1^n|\mathbf{s}_1^n)+I(\mathbf{x}_2^n;\mathbf{s}_2^n)+I(\mathbf{x}_2^n;\mathbf{y}_2^n|\mathbf{s}_2^n)+n\epsilon_n \\
&= h(\mathbf{s}_1^n)-h(\mathbf{s}_1^n|\mathbf{x}_1^n)+h(\mathbf{y}_1^n|\mathbf{s}_1^n)-h(\mathbf{y}_1^n|\mathbf{x}_1^n,\mathbf{s}_1^n) \\
&\quad +h(\mathbf{s}_2^n)-h(\mathbf{s}_2^n|\mathbf{x}_2^n)+h(\mathbf{y}_2^n|\mathbf{s}_2^n)-h(\mathbf{y}_2^n|\mathbf{x}_2^n,\mathbf{s}_2^n)+n\epsilon_n \\
&= h(\mathbf{s}_1^n)-h(\mathbf{z}_2^n)+h(\mathbf{y}_1^n|\mathbf{s}_1^n)-h(\mathbf{s}_2^n) \\
&\quad +h(\mathbf{s}_2^n)-h(\mathbf{z}_1^n)+h(\mathbf{y}_2^n|\mathbf{s}_2^n)-h(\mathbf{s}_1^n)+n\epsilon_n \\
&= h(\mathbf{y}_1^n|\mathbf{s}_1^n)+h(\mathbf{y}_2^n|\mathbf{s}_2^n)-h(\mathbf{z}_1^n)-h(\mathbf{z}_2^n)+n\epsilon_n \\
&\leq \sum_{i=1}^{n}[h(y_{1i}|s_{1i})+h(y_{2i}|s_{2i})-h(z_{1i})-h(z_{2i})]+n\epsilon_n \qquad (16)
\end{aligned}
$$

where the last inequality follows by the fact that removing conditioning cannot reduce differential entropy, and $\epsilon_n \to 0$ as $n \to \infty$.

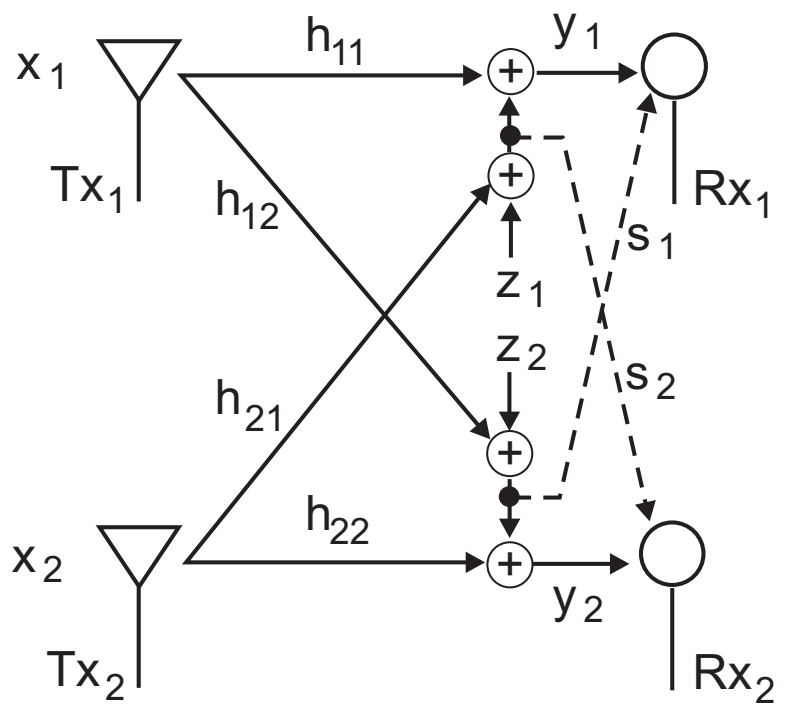

Figure 9: Genie-aided two-user Gaussian interference channel. A genie provides signals $s_1$ to receiver 1 and $s_2$ to receiver 2.

Let $\mathbb{E}[x_{1i}^2]=P_{1i}$ and $\mathbb{E}[x_{2i}^2]=P_{2i}$, we have

$$
\begin{aligned}
\frac{1}{n}\sum_{i=1}^{n}h(y_{1i}|s_{1i}) &\overset{(a)}{\leq} \frac{1}{n}\log\left[\pi e\left(N_0+|h_{21}|^2P_{2i}+\frac{|h_{11}|^2P_{1i}N_0}{N_0+|h_{12}|^2P_{1i}}\right)\right] \\
&\overset{(b)}{\leq} \log\left[\pi e\left(N_0+|h_{21}|^2\left(\frac{1}{n}\sum_{i=1}^{n}P_{2i}\right)+\frac{|h_{11}|^2\left(\frac{1}{n}\sum_{i=1}^{n}P_{1i}\right)N_0}{N_0+|h_{12}|^2\left(\frac{1}{n}\sum_{i=1}^{n}P_{1i}\right)}\right)\right] \\
&\overset{(c)}{\leq} \log\left[\pi e\left(N_0+|h_{21}|^2P_2+\frac{|h_{11}|^2P_1N_0}{N_0+|h_{12}|^2P_1}\right)\right] \qquad (17)
\end{aligned}
$$

where in step (a) we use the fact that the circularly symmetric complex Gaussian distribution maximizes conditional differential entropy for a given covariance constraint, in step (b) we use Jensen's inequality applied to a function that, as can be easily checked, is concave, and in step (c) we used the fact that the function is increasing on $P_1$ and $P_2$. Similarly, we have

$$
\frac{1}{n}\sum_{i=1}^{n}h(y_{2i}|s_{2i}) \leq \log\left[\pi e\left(N_0+|h_{12}|^2P_1+\frac{|h_{22}|^2P_2N_0}{N_0+|h_{21}|^2P_2}\right)\right]. \qquad (18)
$$

Thus we have

$$
\begin{aligned}
R_1+R_2 &\leq \frac{1}{n}\sum_{i=1}^{n}[h(y_{1i}|s_{1i})+h(y_{2i}|s_{2i})-h(z_{1i})-h(z_{2i})]+\epsilon_n \\
&\leq \log\left[\pi e\left(N_0+|h_{21}|^2P_2+\frac{|h_{11}|^2P_1N_0}{N_0+|h_{12}|^2P_1}\right)\right]-\log(\pi e N_0) \\
&\quad +\log\left[\pi e\left(N_0+|h_{12}|^2P_1+\frac{|h_{22}|^2P_2N_0}{N_0+|h_{21}|^2P_2}\right)\right]-\log(\pi e N_0)+\epsilon_n \\
&= \log\left(1+\frac{|h_{21}|^2P_2}{N_0}+\frac{|h_{11}|^2P_1}{N_0+|h_{12}|^2P_1}\right)+\log\left(1+\frac{|h_{12}|^2P_1}{N_0}+\frac{|h_{22}|^2P_2}{N_0+|h_{21}|^2P_2}\right)+\epsilon_n.
\end{aligned}
$$

Letting $n \to \infty$, $\epsilon_n \to 0$ and we get the desired upper bound. □

It is interesting to note that the upper bound of Theorem 1 can be achieved with a communication scheme where each receiver treats interference as noise. In the genie-aided channel used to derive the upper bound, the side information provided by the genie compensates for the harm that interference produces on the other link by giving a boost in the own rate in the direct link. Thus, making the signal more random by not sending any common information results in an overall improvement in the sum rate.

We now specialize the bound of Theorem 1 to the symmetric interference channel to obtain the following upper bound on the symmetric rate:

$$R_{UB_{new}} = \log\left(1 + \mathsf{INR} + \frac{\mathsf{SNR}}{1 + \mathsf{INR}}\right). \tag{19}$$

To see how this bound performs in the different regimes we compute the ratio of $R_{UB_{new}}$ and $C_{\text{awgn}}$ for large $\mathsf{SNR}, \mathsf{INR}$:

$$\frac{R_{UB_{new}}}{C_{\text{awgn}}} \approx \max\left\{\frac{\log \mathsf{INR}}{\log \mathsf{SNR}}, 1 - \frac{\log \mathsf{INR}}{\log \mathsf{SNR}}\right\} = \max\{\alpha, 1-\alpha\} \tag{20}$$

which we plot in Figure 10.

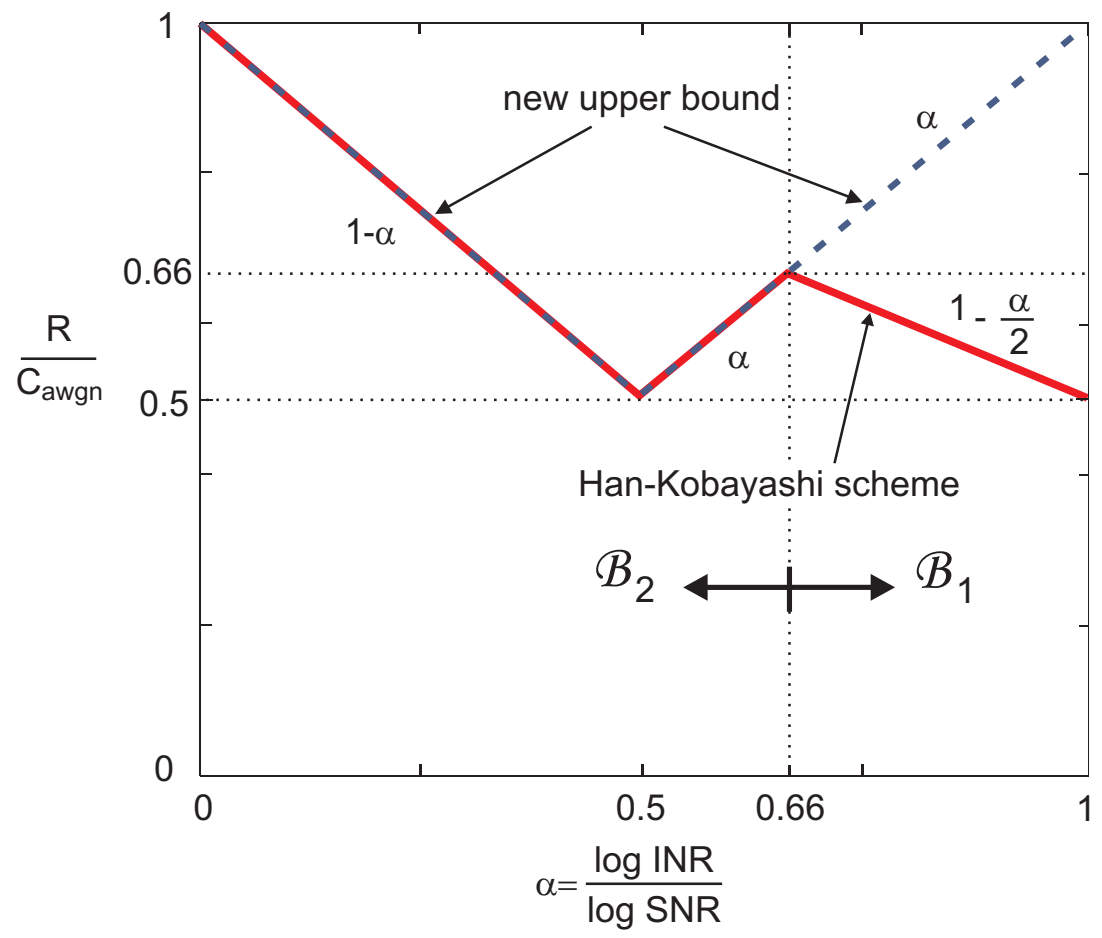

Figure 10: Upper bound of Theorem 1 and achievable Han-Kobayashi rate (relative to single user capacity) as a function of the interference level $\alpha$.

Observing Figure 10, the new upper bound seems to match the Han-Kobayashi achievable rate in the regime $\mathcal{B}_2$, where the upper bound (12) is loose. In fact, this new bound has a finite gap with respect to the achievable symmetric rate with our simple scheme in the parameter range $\mathcal{B}_2$. To verify this we compute:

$$\begin{aligned} R_{UB_{new}} - R_{HK_2} &= \log\left(1 + \mathsf{INR} + \frac{\mathsf{SNR}}{1+\mathsf{INR}}\right) - \log\left(1 + \mathsf{INR} + \frac{\mathsf{SNR}}{\mathsf{INR}}\right) + 1 \\ &< \log\left(1 + \mathsf{INR} + \frac{\mathsf{SNR}}{\mathsf{INR}}\right) - \log\left(1 + \mathsf{INR} + \frac{\mathsf{SNR}}{\mathsf{INR}}\right) + 1 \\ &= 1 \end{aligned} \tag{21}$$

and we find that the gap in the symmetric rate with respect to the new upper bound is at most 1 bit/s/Hz in $\mathcal{B}_2$.

Using (14) and (21) we see that when $\mathsf{INR} \geq 1$ our simple scheme is at most 1 bit/s/Hz away from the symmetric rate channel capacity. Proving that the simple scheme is at most 1 bit/s/Hz away from capacity when $\mathsf{INR} < 1$ is straightforward. We can set $\mathsf{INR}_p = \mathsf{INR}$ and use as a symmetric rate upper bound the single user capacity. The difference between the achievable rate and the upper bound is:

$$\log(1+\mathsf{SNR}) - \log\left(1+\frac{\mathsf{SNR}}{1+\mathsf{INR}}\right) \leq \log(1+\mathsf{SNR}) - \log\left(1+\frac{\mathsf{SNR}}{2}\right) \leq 1. \qquad (22)$$

### 3.5 Small gap between lower and upper bounds

In Sections 3.3 and 3.4 we showed that the Han-Kobayashi scheme that sets the private message power so that the interference created is at noise level achieves a symmetric rate within one bit/s/Hz of the upper bounds. Therefore, we obtained a characterization of the symmetric capacity to within one bit/s/Hz. The finite and small gap between the lower and upper bounds on the symmetric capacity was obtained by direct calculation of the difference between the bounds. In this subsection we present a more intuitive explanation for the tightness of the bounds.

We can decompose the total gap between the lower and upper bounds in two components $\Delta_1$ and $\Delta_2$, arising from the following two steps:

1. Fix Han-Kobayashi strategy (i.e. set $\mathsf{INR}_p = 1$, decode first the common messages $(w_1, w_2)$ and then the private message $u_1$ or $u_2$[3]) and see how the symmetric rate changes when varying the channel from the given interference channel to the genie aided interference channel used in the bounds. The gap $\Delta_1$ quantifies the rate change due to the side information for the fixed strategy.
2. Fix the channel to the genie aided interference channel, and change the Han-Kobayashi strategy by varying $\mathsf{INR}_p$ from $1$ to $\mathsf{INR}$. The gap $\Delta_2$ quantifies the rate change in the genie-aided channel when $\mathsf{INR}_p$ is varied.

Referring to Figure 11, $\Delta_1$ corresponds to the difference in the rates between points $A$ and $B$. $\Delta_2$ corresponds to the difference in the rates between points $B$ and $C$.

Since $\mathsf{INR}_p = \mathsf{INR}$ achieves the capacity of the genie aided channel (one can show that the sum rate upper bounds can be achieved by generating the codewords $\mathbf{x}_1^n$ and $\mathbf{x}_2^n$ with i.i.d. circularly symmetric complex Gaussian components of variance $P$, and treating interference as noise at the decoder) the sum $\Delta_1 + \Delta_2$ quantifies the rate change from the initial Han-Kobayashi strategy in the original channel (lower bound), to the capacity achieving strategy in the genie aided channel (upper bound). It follows that a small gap between the lower and upper bounds can only occur if both $\Delta_1$ and $\Delta_2$ are small.

To achieve a small value of $\Delta_1$, the help of the genie should not change the relevant rate constraints for the initial Han-Kobayashi strategy. Figure 4 and the discussion at the end of Section 3.3 describe the active constraints for the different weak interference regimes.

[3]With some abuse of notation, we use $u_i, w_i$, $i = 1, 2$ to denote the private and common messages, and also to denote the symbols of the codewords actually sent over the channel.

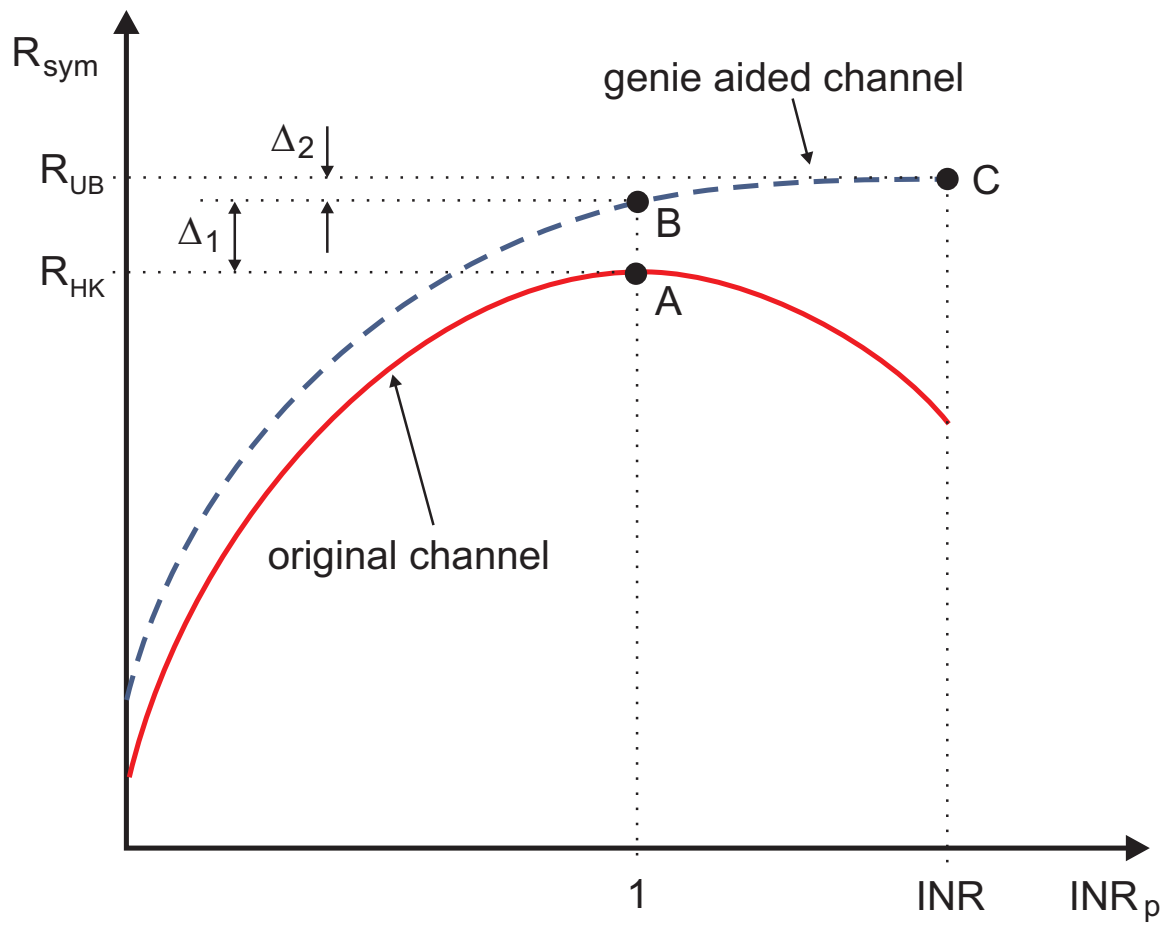

Figure 11: Gap between the achievable rate with the Han-Kobayashi scheme that sets $\mathsf{INR}_p = 1$ and the symmetric capacity upper bound decomposed into two components: $\Delta_1$ and $\Delta_2$. $\Delta_1$ results from fixing $\mathsf{INR}_p$ and changing the channel; $\Delta_2$ results from increasing $\mathsf{INR}_p$ to $\mathsf{INR}$ in the genie aided channel.

In addition to achieving a small value for $\Delta_1$, the help of the genie should result in a small value of $\Delta_2$. $\Delta_2$ arises when we vary the communication strategy from the initial Han-Kobayashi strategy to the capacity achieving strategy in the genie-aided channel. The sum rate of the genie-aided channels that we used can be explicitly computed by treating interference as noise. Unfortunately, it is hard to compute bounds for the interference channel when the interference is not treated as noise. In the original channel, setting $\mathsf{INR}_p = 1$ achieves good performance, but in general, setting $\mathsf{INR}_p = \mathsf{INR}$ (treating interference as noise) may result in very small rates. The role of the genie in the genie-aided channel is to compensate for the loss in the sum rate when $\mathsf{INR}_p$ is increased from 1 to $\mathsf{INR}$. Increasing $\mathsf{INR}_p$ beyond 1 in the original channel may produce a loss in the rate of common message due to increased interference. However, the genie provides just enough side information to compensate for this loss while making $\mathsf{INR}_p = \mathsf{INR}$ optimal.

### 3.6 Generalized degrees of freedom

In the above analysis, we see the utility of the approximations like (10), (13), (19) both in identifying the different regimes of interest as well as in developing the relevant upper bounds for the different regimes. We can formalize the approximations of this nature through the following type of definition. Define, for a fixed $\alpha \geq 0$,

$$d_{\text{sym}}(\alpha) := \lim_{\mathsf{SNR},\mathsf{INR}\to\infty;\,\frac{\log \mathsf{INR}}{\log \mathsf{SNR}}=\alpha} \frac{C_{\text{sym}}(\mathsf{INR},\mathsf{SNR})}{C_{\text{awgn}}(\mathsf{SNR})}. \tag{23}$$

If there were no interference between the two links (i.e. $\alpha = 0$), then the capacity per link is just the AWGN capacity $\log(1 + \mathsf{SNR})$. Hence $d_{\text{sym}}(0) = 1$. This can be interpreted as each link having the full degree of freedom to itself. Since interference cannot help in communicating each user's message, it follows that $C_{\text{sym}} \leq C_{\text{awgn}}$ and therefore $d_{\text{sym}}(\alpha) \leq 1$ for $\alpha > 0$. We can think of interference as effectively reducing the degrees of freedom of the channel, and thus it is natural to think of $d_{\text{sym}}(\alpha)$ as a *generalized* degree of freedom. The approximations we made can be thought of as computing analogous limits for the achievable rates and upper bounds. Since the lower and upper bounds on the symmetric capacity we derived earlier differ by at most one bit, they allow us to precisely characterize $d_{\text{sym}}(\alpha)$. For $0 \leq \alpha \leq 1$, this is plotted in Figure 10. $\alpha \geq 1$ corresponds to the strong and very strong interference regimes, and since the capacity is known in these regimes, we can compute $d_{\text{sym}}(\alpha)$ in a straightforward way.

In the very strong interference case, each user can decode the interfering message before decoding his own message [2]. After decoding the interference and subtracting it from the received signal, the user effectively gets an AWGN channel for communicating his own message. It follows that the symmetric capacity in the very strong interference case is:

$$C_{\text{sym}} = \log(1 + \mathsf{SNR}) \approx \log(\mathsf{SNR}). \tag{24}$$

The channel is in the very strong interference situation whenever $\mathsf{INR} \geq \mathsf{SNR}^2 + \mathsf{SNR}$. Taking logs and assuming $\mathsf{SNR}, \mathsf{INR} \gg 1$ the very strong interference condition becomes $\log \mathsf{INR} \geq 2 \log \mathsf{SNR}$. In this regime we obtain $d_{\text{sym}} = 1$, and therefore, interference does not reduce the available degrees of freedom of the channel.

In the strong interference regime, each receiver is able to decode both messages. The capacity region of the interference channel is given by the intersection of the capacity regions of the two multiple access channels (MAC) formed by the two transmitters and each of the receivers. In the symmetric case, the sum capacity of both MACs is the same and the corresponding symmetric capacity is given by:

$$C_{\text{sym}} = \frac{1}{2} \log \left(1 + \mathsf{SNR} + \mathsf{INR}\right). \tag{25}$$

The symmetric channel is in the strong interference situation whenever it is not in very strong interference and $\mathsf{INR} \geq \mathsf{SNR}$, which after taking logs becomes $\log \mathsf{SNR} \leq \log \mathsf{INR} < 2 \log \mathsf{SNR}$. This condition and (25) together with the assumption $\mathsf{SNR}, \mathsf{INR} \gg 1$ imply that $C_{\text{sym}} \approx \frac{1}{2} \log \mathsf{INR}$. It follows that under strong interference the generalized degrees of freedom are:

$$d_{\text{sym}} = \frac{\log \sqrt{\mathsf{INR}}}{\log \mathsf{SNR}} = \frac{\alpha}{2}. \tag{26}$$

We now have the complete picture:

$$C_{\text{sym}} \approx \begin{cases} \log(\frac{\mathsf{SNR}}{\mathsf{INR}}) & \log \mathsf{INR} < \frac{1}{2} \log \mathsf{SNR} \\ \log \mathsf{INR} & \frac{1}{2} \log \mathsf{SNR} < \log \mathsf{INR} < \frac{2}{3} \log \mathsf{SNR} \\ \log \frac{\mathsf{SNR}}{\sqrt{\mathsf{INR}}} & \frac{2}{3} \log \mathsf{SNR} < \log \mathsf{INR} < \log \mathsf{SNR} \\ \log \sqrt{\mathsf{INR}} & \log \mathsf{SNR} < \log \mathsf{INR} < 2 \log \mathsf{SNR} \\ \log \mathsf{SNR} & \log \mathsf{INR} > 2 \log \mathsf{SNR} \end{cases} \tag{27}$$

and

$$d_{\text{sym}} = \begin{cases} 1 - \alpha & 0 \leq \alpha < \frac{1}{2} \\ \alpha & \frac{1}{2} \leq \alpha < \frac{2}{3} \\ 1 - \frac{\alpha}{2} & \frac{2}{3} < \alpha \leq 1 \\ \frac{\alpha}{2} & 1 \leq \alpha < 2 \\ 1 & \alpha \geq 2. \end{cases} \tag{28}$$

The generalized degrees of freedom are plotted in Figure 12, together with the performance of the baseline strategies of orthogonalizing the users (in frequency or time) and treating interference as noise. Note that orthogonalizing between the links, in which each link achieves half the degrees of freedom, is strictly sub-optimal except when $\alpha = \frac{1}{2}$ and $\alpha = 1$. Treating interference as noise, on the other hand, is strictly sub-optimal except for $\alpha \leq \frac{1}{2}$.

Note that there are five regimes in which the qualitative behaviors of the capacity are different. The first three regimes fall into the weak interference regime, and the characterization of the symmetric capacity in these regimes is a consequence of the new results that we obtained. In these regimes, the interference is not strong enough to be decoded in its entirety. In fact, in regime 1 where the interference is very weak, treating interference as noise is optimal. In regimes 2 and 3 where the interference is not very weak, decoding it partially can significantly improve performance.

Interestingly, the capacity is *not* monotonically decreasing with $\mathsf{INR}$ in the weak interference regime. Increasing $\mathsf{INR}$ has two opposing effects: more common information can be decoded and cancelled but less private information can be sent under the constraint $\mathsf{INR}_p = 1$. Depending on which of these two effects dominates, the capacity increases or decreases with $\mathsf{INR}$.

In regime 1 where treating interference as noise is optimal, the common messages carry negligible information. In this regime, the loss in the private rate due to the increase in $\mathsf{INR}$ makes the capacity decrease with $\mathsf{INR}$. However, once interference becomes strong enough to reach regime 2, the users can start using common information to partially cancel interference. As the interference level increases, more and more of this common information can be decoded and partially cancelled, and this effects dominates the behavior of capacity with $\mathsf{INR}$. Therefore, capacity increases with $\mathsf{INR}$ in regime 2. However, as $\mathsf{INR}$ increases further to reach regime 3, the gains obtained by partially cancelling interference through the common messages are not enough to offset the loss of rate in the private information. In this regime capacity decreases with $\mathsf{INR}$ until the strong interference regime is reached. Since in the strong interference regime all the information is common information, increasing $\mathsf{INR}$ increases capacity. Finally, in the very strong interference regime all the interference can be cancelled before decoding the useful information, and interference does not have any effect on capacity.

### 3.7 Tight characterization of symmetric capacity

Our simple Han and Kobayashi type scheme, together with the symmetric capacity upper bounds (12) and (19) allowed us to characterize the symmetric capacity to within one bit/s/Hz. We will now show that in some parameter ranges, the gap between the upper bound (19) and the rates achievable with some improved communication schemes vanishes for $\mathsf{SNR}, \mathsf{INR} \to \infty$.

The communication scheme that sets the private message power so that the interference generated onto the other receiver is at noise level (i.e. $\mathsf{INR}_p = 1$) is "universal" in the sense that the same scheme can be used to achieve a symmetric rate within one bit/s/Hz of capacity in the weak interference regime, regardless of the values of the parameters.

However, we can further improve the achievable symmetric rate by modifying the communication scheme for different parameter ranges. In regime 1, when $\log \mathsf{INR} < \frac{1}{2} \log \mathsf{SNR}$ we can simply assign $\mathsf{INR}_p = \mathsf{INR}$ and not use common messages at all. As stated in the previous subsection, this scheme achieves a symmetric rate:

$$R = \log\left(1 + \frac{\mathsf{SNR}}{1 + \mathsf{INR}}\right) \tag{29}$$

and the gap between this rate and the upper bound (19) is:

$$\begin{aligned} R_{UB_{new}} - R &= \log\left(1 + \mathsf{INR} + \frac{\mathsf{SNR}}{1 + \mathsf{INR}}\right) - \log\left(1 + \frac{\mathsf{SNR}}{1 + \mathsf{INR}}\right) \\ &= \log\left(1 + \frac{\mathsf{INR}(1 + \mathsf{INR})}{1 + \mathsf{INR} + \mathsf{SNR}}\right) \approx \log\left(1 + \frac{\mathsf{INR}^2}{\mathsf{SNR}}\right). \end{aligned} \tag{30}$$

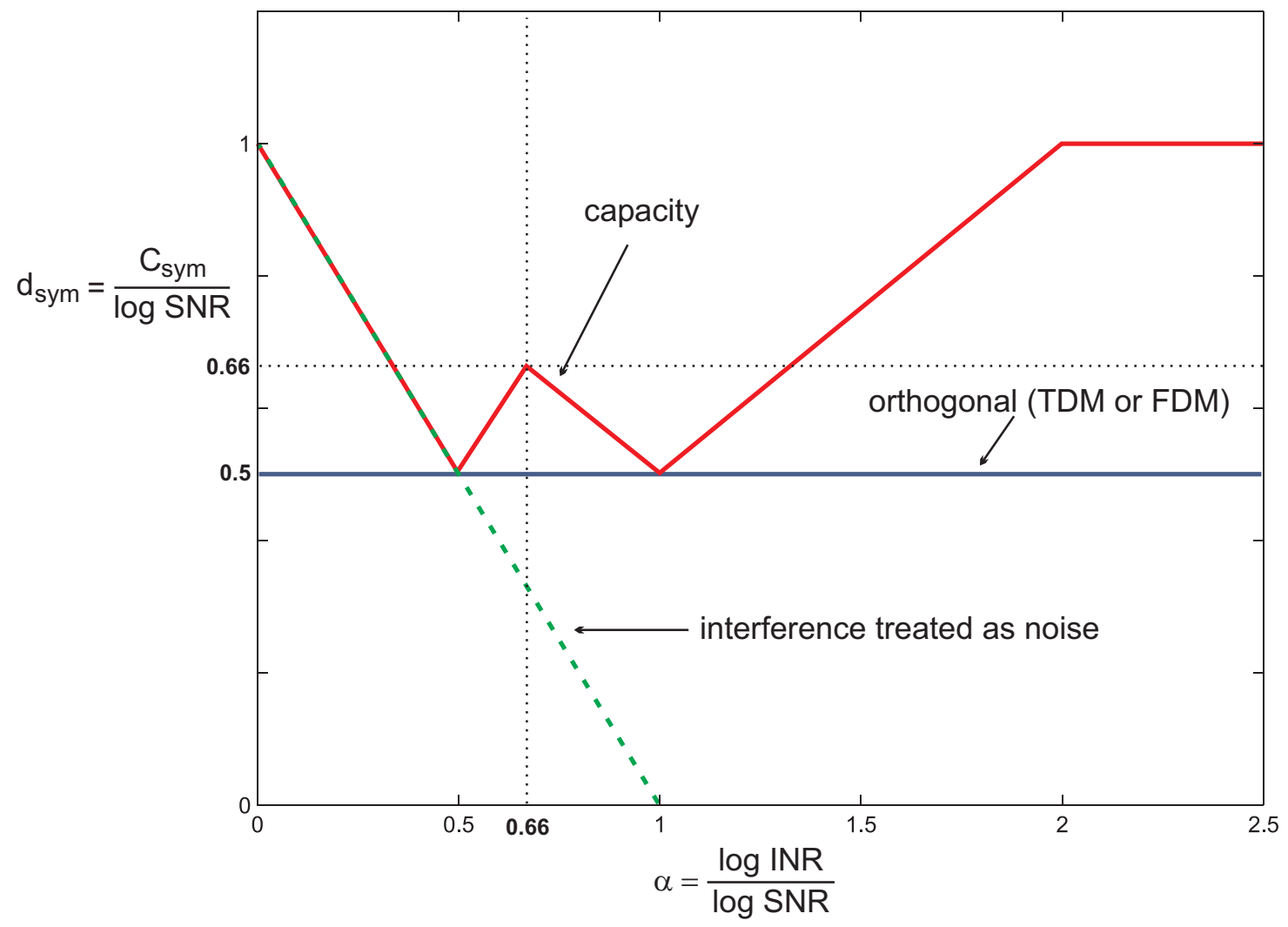

Figure 12: Generalized degrees of freedom for two suboptimal schemes vs. capacity. These suboptimal schemes are treating interference as noise and orthogonalizing the users over time or frequency.

Note that the same gap would be obtained with any scheme that uses $\mathsf{INR}_p$ such that $\mathsf{INR}_p \to \infty$ as $\mathsf{SNR}, \mathsf{INR} \to \infty$.

Recall that $\alpha = \frac{\log \mathsf{INR}}{\log \mathsf{SNR}}$. Regime 1 corresponds to $0 < \alpha < \frac{1}{2}$. In this regime, (30) implies that for fixed $\alpha$, $C_{\text{sym}} - R \to 0$ as $\mathsf{SNR}, \mathsf{INR} \to \infty$. Therefore, we have that for $0 < \alpha < 1/2$ the symmetric capacity is tightly characterized by:

$$C_{\text{sym}} = \log\left(\frac{\mathsf{SNR}}{\mathsf{INR}}\right). \tag{31}$$

In regime 2 where $\frac{1}{2} < \alpha < \frac{2}{3}$, we can choose $\mathsf{INR}_p = (\mathsf{INR}/\mathsf{SNR})^{1-\gamma}$, where

$$0 < \frac{2\alpha - 1}{1 - \alpha} < \gamma < 1 \tag{32}$$

is fixed but arbitrary. This choice of $\mathsf{INR}_p$ makes the received interference power corresponding to the private message to go to zero as $\mathsf{SNR}, \mathsf{INR} \to \infty$. Note that in regime 3 we have $\log \mathsf{SNR} < \log \mathsf{INR} < \frac{2}{3} \log \mathsf{SNR}$ and therefore $\mathsf{INR}/\mathsf{SNR} \to 0$ as $\mathsf{SNR}, \mathsf{INR} \to \infty$.

Fixing the decoding order so that the private messages are decoded last, this scheme achieves a symmetric rate:

$$\begin{aligned}
R &= \log\left[1+\frac{\mathsf{SNR}\cdot\mathsf{INR}_p}{\mathsf{INR}(1+\mathsf{INR}_p)}\right]+\min\left\{\frac{1}{2}\log\left[1+\frac{(\mathsf{SNR}+\mathsf{INR})(\mathsf{INR}-\mathsf{INR}_p)}{\mathsf{INR}+(\mathsf{SNR}+\mathsf{INR})\mathsf{INR}_p}\right],\right.\\
&\quad \left.\log\left[1+\frac{\mathsf{INR}(\mathsf{INR}-\mathsf{INR}_p)}{\mathsf{INR}+(\mathsf{SNR}+\mathsf{INR})\mathsf{INR}_p}\right]\right\}\\
&\overset{(a)}{\simeq} \log\left[1+\left(\frac{\mathsf{SNR}}{\mathsf{INR}}\right)^{\gamma}\right]+\min\left\{\frac{1}{2}\log\left[1+\frac{\mathsf{SNR}}{(\mathsf{SNR}/\mathsf{INR})^{\gamma}}\right],\log\left[1+\frac{\mathsf{INR}}{(\mathsf{SNR}/\mathsf{INR})^{\gamma}}\right]\right\}\\
&\simeq \gamma\log\left(\frac{\mathsf{SNR}}{\mathsf{INR}}\right)+\min\left\{\frac{1}{2}\log\left[\mathsf{SNR}^{1-\gamma}\mathsf{INR}^{\gamma}\right],\log\left[\frac{\mathsf{INR}^{1+\gamma}}{\mathsf{SNR}^{\gamma}}\right]\right\}\\
&= \gamma(1-\alpha)\log(\mathsf{SNR})+\min\left\{\frac{1-\gamma+\alpha\gamma}{2},\alpha(1+\gamma)-\gamma\right\}\log(\mathsf{SNR})\\
&\overset{(b)}{=} \alpha\log(\mathsf{SNR}) \qquad (33)
\end{aligned}$$

where $\simeq$ means that the the difference between the left and right hand sides goes to zero as $\mathsf{SNR},\mathsf{INR}\to\infty$, (a) follows from the fact that $\mathsf{INR}_p\to 0$ and $\mathsf{INR}/\mathsf{SNR}\to 0$ as $\mathsf{SNR},\mathsf{INR}\to\infty$, and $(b)$ follows because the second term of the $\min\{\cdot,\cdot\}$ dominates due to (32).

From the upper bound (19) we obtain:

$$\begin{aligned}
R_{UB_{new}} &= \log\left(1+\mathsf{INR}+\frac{\mathsf{SNR}}{1+\mathsf{INR}}\right)\\
&\simeq \log\left(\mathsf{SNR}^{\alpha}+\mathsf{SNR}^{1-\alpha}\right)\\
&\simeq \max\{\alpha,1-\alpha\}\log(\mathsf{SNR})\\
&= \alpha\log(\mathsf{SNR}). \qquad (34)
\end{aligned}$$

Comparing (33) with (34) we see that the difference $R_{UB_{new}}-R\to 0$ as $\mathsf{SNR},\mathsf{INR}\to\infty$ and therefore in regime 2 the symmetric capacity is given by:

$$C_{\text{sym}}=\log(\mathsf{INR}). \qquad (35)$$

We note that both in regimes 1 and 2 we have some flexibility in setting the private message power to asymptotically achieve the symmetric capacity. In regime 1 we can choose any private message power as long as $\mathsf{INR}_p\to\infty$ when $\mathsf{SNR},\mathsf{INR}\to\infty$. In a similar way, in regime 2 we can use any private message power that satisfies $\mathsf{INR}_p\to 0$ as $\mathsf{SNR},\mathsf{INR}\to\infty$. In both cases setting $\mathsf{INR}_p=1$ does not asymptotically achieve the symmetric capacity, but results in a symmetric rate no smaller than 1 bit/s/Hz from it. Unfortunately, in regime 3 the only choice of private message power that achieves a symmetric rate with bounded difference from the upper bound (12) is $\mathsf{INR}_p=$ constant, and this choice of private message power does not result in a gap that vanishes as $\mathsf{SNR},\mathsf{INR}\to\infty$.

In the strong interference regime the symmetric capacity is given by:

$$C_{\text{sym}}=\frac{1}{2}\log(1+\mathsf{INR}+\mathsf{SNR}) \qquad (36)$$

which asymptotically approaches $\log(\sqrt{\mathsf{INR}})$ for $1<\alpha\leq 2$ as $\mathsf{SNR},\mathsf{INR}\to\infty$.

Finally in the very strong interference regime the symmetric capacity is given by:

$$C_{\text{sym}} = \log\left(1 + \mathsf{SNR}\right) \tag{37}$$

which asymptotically approaches $\log(\mathsf{SNR})$ for $\mathsf{SNR} \to \infty$.

We summarize the results of this subsection in the following theorem.

***Theorem* 2.** *Let* $\alpha = (\log \mathsf{INR}/\log \mathsf{SNR})$. *For* $0 < \alpha < 1/2$, $1/2 < \alpha < 2/3$ *and* $\alpha > 1$, *the approximation*

$$C_{\text{sym}} \approx \begin{cases} \log(\frac{\mathsf{SNR}}{\mathsf{INR}}) & \log \mathsf{INR} < \frac{1}{2}\log \mathsf{SNR} \\ \log \mathsf{INR} & \frac{1}{2}\log \mathsf{SNR} < \log \mathsf{INR} < \frac{2}{3}\log \mathsf{SNR} \\ \log \sqrt{\mathsf{INR}} & \log \mathsf{SNR} < \log \mathsf{INR} < 2\log \mathsf{SNR} \\ \log \mathsf{SNR} & \log \mathsf{INR} \geq 2\log \mathsf{SNR} \end{cases} \tag{38}$$

*is asymptotically tight in the sense that the difference between* $C_{\text{sym}}$ *and the approximation goes to zero as* $\mathsf{SNR}, \mathsf{INR}$ *go to infinity with* $\alpha$ *fixed.*

# 4 Within One Bit of the General Capacity Region

In the previous section, we showed that a simple Han-Kobayashi scheme can achieve to within one bit of the symmetric rate of the symmetric Gaussian interference channel. We will show that this is also true for the whole capacity region of the general two-user Gaussian interference channel (not necessarily symmetric). Depending on the parameters of the Gaussian interference channel ($\mathsf{SNR}_1$, $\mathsf{SNR}_2$, $\mathsf{INR}_1$ and $\mathsf{INR}_2$), we can divide the analysis of the Gaussian interference channel into the following three cases.

1. Weak interference channel

   In this case, the parameters of the Gaussian interference channel satisfy $\mathsf{INR}_1 < \mathsf{SNR}_2$ and $\mathsf{INR}_2 < \mathsf{SNR}_1$.

2. Mixed interference channel

   In this case, the parameters of the Gaussian interference channel satisfy $\mathsf{INR}_1 \geq \mathsf{SNR}_2$ and $\mathsf{INR}_2 < \mathsf{SNR}_1$, or $\mathsf{INR}_1 < \mathsf{SNR}_2$ and $\mathsf{INR}_2 \geq \mathsf{SNR}_1$.

3. Strong interference channel

   In this case, the parameters of the Gaussian interference channel satisfy $\mathsf{INR}_1 \geq \mathsf{SNR}_2$ and $\mathsf{INR}_2 \geq \mathsf{SNR}_1$.

The capacity region of the strong interference channel is already known [7]. In the following, we will show that we can get to within one bit of the capacity region of the Gaussian interference channel for both the weak interference channel and the mixed interference channel.

## 4.1 Outer bound on the capacity region of the Gaussian interference channel

Since the existing bounds on the capacity region of the Gaussian interference channel can be arbitrarily loose, we need a new outer bound. In the following subsections we provide a new outer bound on capacity region of the weak and mixed interference channels.

### 4.1.1 Outer bound for weak interference channel

For the weak interference channel, i.e., $\mathsf{INR}_1 < \mathsf{SNR}_2$ and $\mathsf{INR}_2 < \mathsf{SNR}_1$, we have the following outer bound on the capacity region.

***Theorem* 3.** *The capacity region of the weak interference channel is contained within the set of rate pairs* $(R_1, R_2)$ *satisfying:*

$$\begin{aligned}
R_1 &\le \log\left(1+\mathsf{SNR}_1\right) \\
R_2 &\le \log\left(1+\mathsf{SNR}_2\right) \\
R_1+R_2 &\le \log\left(1+\mathsf{SNR}_1\right)+\log\left(1+\frac{\mathsf{SNR}_2}{1+\mathsf{INR}_2}\right) \\
R_1+R_2 &\le \log\left(1+\mathsf{SNR}_2\right)+\log\left(1+\frac{\mathsf{SNR}_1}{1+\mathsf{INR}_1}\right) \\
R_1+R_2 &\le \log\left(1+\mathsf{INR}_1+\frac{\mathsf{SNR}_1}{1+\mathsf{INR}_2}\right)+\log\left(1+\mathsf{INR}_2+\frac{\mathsf{SNR}_2}{1+\mathsf{INR}_1}\right) \\
2R_1+R_2 &\le \log\left(1+\mathsf{SNR}_1+\mathsf{INR}_1\right)+\log\left(1+\mathsf{INR}_2+\frac{\mathsf{SNR}_2}{1+\mathsf{INR}_1}\right)+\log\left(\frac{1+\mathsf{SNR}_1}{1+\mathsf{INR}_2}\right) \\
R_1+2R_2 &\le \log\left(1+\mathsf{SNR}_2+\mathsf{INR}_2\right)+\log\left(1+\mathsf{INR}_1+\frac{\mathsf{SNR}_1}{1+\mathsf{INR}_2}\right)+\log\left(\frac{1+\mathsf{SNR}_2}{1+\mathsf{INR}_1}\right).
\end{aligned} \tag{39}$$

*Proof.* We prove the bounds in (39) one by one.

1. The bounds on $R_1$ and $R_2$ are just the point to point capacity of the AWGN channel obtained by removing the interference from the other user.
2. The first bound on $R_1 + R_2$ is just the capacity of the one-sided interference channel resulting from the genie-aided channel in which a genie gives $x_2$ to receiver 1 (see Figure 6). We already used this bound in the symmetric case. This bound is known from previous results [8] but we provide here an alternative derivation since the same bounding techniques will be useful for obtaining the bounds for

$2R_1 + R_2$ and $R_1 + 2R_2$. Using Fano's inequality we can write for any codebook of block length $n$:

$$\begin{aligned} n(R_1+R_2) &\le I(\mathbf{x}_1^n;\mathbf{y}_1^n) + I(\mathbf{x}_2^n;\mathbf{y}_2^n) + n\epsilon \\ &\le I(\mathbf{x}_1^n;\mathbf{y}_1^n,\mathbf{x}_2^n) + I(\mathbf{x}_2^n;\mathbf{y}_2^n) + n\epsilon \\ &= I(\mathbf{x}_1^n;\mathbf{y}_1^n|\mathbf{x}_2^n) + I(\mathbf{x}_2^n;\mathbf{y}_2^n) + n\epsilon \\ &= h(\mathbf{y}_1^n|\mathbf{x}_2^n) - h(\mathbf{y}_1^n|\mathbf{x}_1^n,\mathbf{x}_2^n) + h(\mathbf{y}_2^n) - h(\mathbf{y}_2^n|\mathbf{x}_2^n) + n\epsilon \\ &= h(h_{11}\mathbf{x}_1^n + \mathbf{z}_1^n) - h(\mathbf{z}_1^n) + h(\mathbf{y}_2^n) - h(h_{12}\mathbf{x}_1^n + \mathbf{z}_2^n) + n\epsilon \\ &= h(\mathbf{y}_2^n) - h(\mathbf{z}_1^n) + h(h_{11}h_{12}\mathbf{x}_1^n + h_{12}\mathbf{z}_1^n) - h(h_{11}h_{12}\mathbf{x}_1^n + h_{11}\mathbf{z}_2^n) \\ &\quad -n\log|h_{12}|^2 + n\log|h_{11}|^2 + n\epsilon \\ &= h(\mathbf{y}_2^n) - h(\mathbf{z}_1^n) + h(\tilde{\mathbf{x}}_1^n + \tilde{\mathbf{z}}_1^n) - h(\tilde{\mathbf{x}}_1^n + \tilde{\mathbf{z}}_2^n) \\ &\quad -n\log|h_{12}|^2 + n\log|h_{11}|^2 + n\epsilon, \end{aligned} \tag{40}$$

where we defined

$$\begin{aligned} \tilde{x}_{1i} &= h_{11}h_{12}x_{2i} \\ \tilde{z}_{1i} &= h_{12}z_{1i} \\ \tilde{z}_{2i} &= h_{11}z_{2i}. \end{aligned} \tag{41}$$

In the weak interference case we have $\mathsf{INR}_2 < \mathsf{SNR}_1$, which implies

$$\mathbb{E}[\tilde{z}_{1i}^2] < \mathbb{E}[\tilde{z}_{2i}^2]. \tag{42}$$

Since the capacity region of Gaussian interference channel only depends on the marginal distribution of $z_1$ and $z_2$, we can assume that there exists an i.i.d Gaussian random vector $\mathbf{z}^n$, with $z_i \sim \mathcal{CN}(0, \mathbb{E}[\tilde{z}_{2i}^2] - \mathbb{E}[\tilde{z}_{1i}^2])$, such that

$$\tilde{z}_{2i} = \tilde{z}_{1i} + z_i. \tag{43}$$

Thus we have

$$\begin{aligned} n(R_1+R_2) &\le h(\mathbf{y}_2^n) - h(\mathbf{z}_1^n) + h(\tilde{\mathbf{x}}_1^n + \tilde{\mathbf{z}}_1^n) - h(\tilde{\mathbf{x}}_1^n + \tilde{\mathbf{z}}_1^n + \mathbf{z}^n) \\ &\quad -n\log|h_{12}|^2 + n\log|h_{11}|^2 + n\epsilon \\ &= h(\mathbf{y}_2^n) - h(\mathbf{z}_1^n) - I(\mathbf{z}^n, \mathbf{z}^n + \tilde{\mathbf{x}}_1^n + \tilde{\mathbf{z}}_1^n) \\ &\quad -n\log|h_{12}|^2 + n\log|h_{11}|^2 + n\epsilon. \end{aligned} \tag{44}$$

Using the worst case noise result [9], we can see that $-I(\mathbf{z}^n, \mathbf{z}^n + \tilde{\mathbf{x}}_1^n + \tilde{\mathbf{z}}_1^n)$ is maximized when $\mathbf{x}_1^n$ is i.i.d Gaussian random vector with $x_{1i} \sim \mathcal{CN}(0, P_1)$. Note that $h(\mathbf{y}_2^n)$ is maximized when $\mathbf{x}_1^n$ is a Gaussian random vector with i.i.d. components $x_{1i} \sim \mathcal{CN}(0, P_1)$ and $\mathbf{x}_2^n$ is a Gaussian random vector with i.i.d. components $x_{2i} \sim \mathcal{CN}(0, P_2)$. A simple calculation leads to

$$\begin{aligned} R_1 + R_2 &\le \log\left(1 + \frac{|h_{11}|^2 P_1}{N_0}\right) + \log\left(1 + \frac{|h_{22}|^2 P_2}{N_0 + |h_{12}|^2 P_1}\right) + \epsilon \\ &= \log\left(1 + \mathsf{SNR}_1\right) + \log\left(1 + \frac{\mathsf{SNR}_2}{1 + \mathsf{INR}_2}\right) + \epsilon. \end{aligned} \tag{45}$$

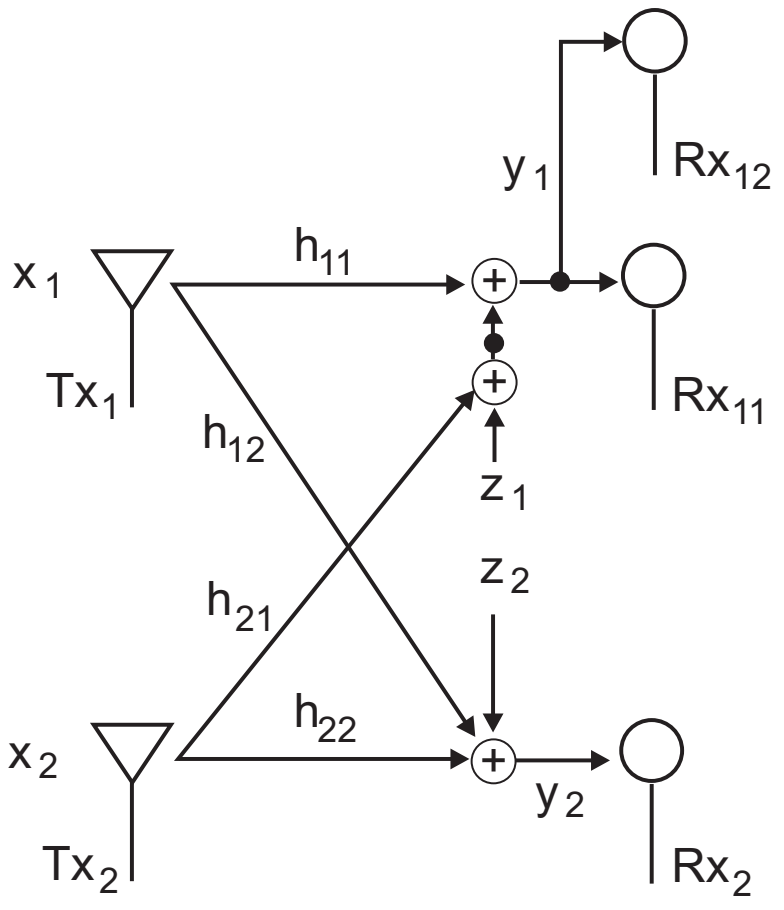

Figure 13: Gaussian interference channel with three receivers.

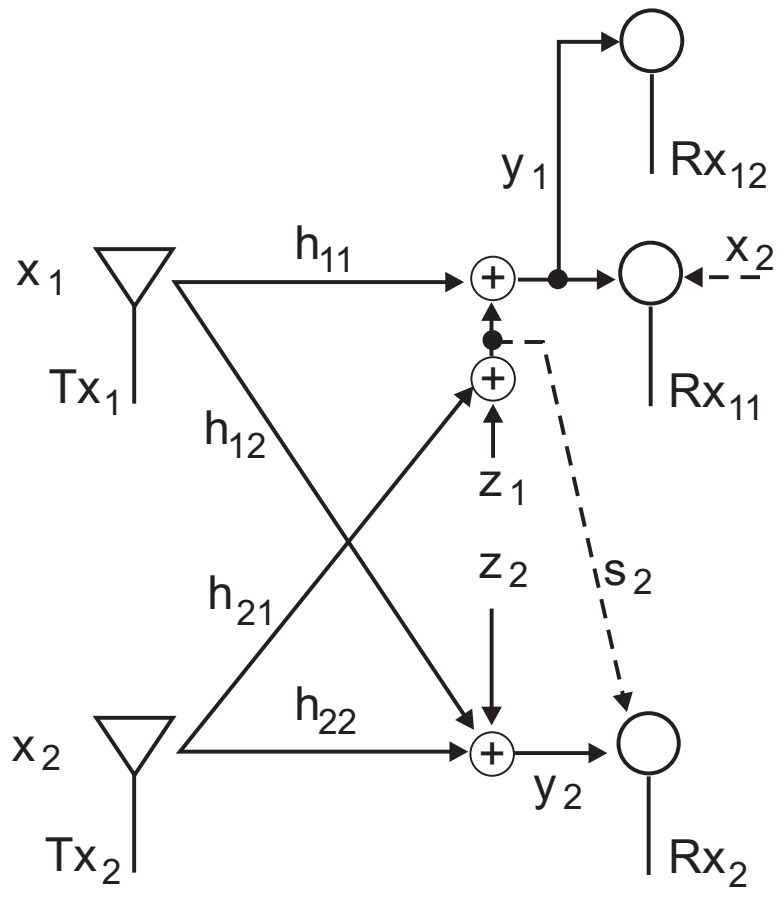

Figure 14: Genie-aided three-receiver Gaussian interference channel. A genie provides $x_2$ to receiver $Rx_{11}$ and $s_2$ to receiver $Rx_2$.

3. The second bound on $R_1 + R_2$ can be derived similarly by using the genie-aided channel in which a genie gives $x_1$ to receiver 2.

4. The third bound on $R_1 + R_2$ has been proved in Section 3.4.

5. Next we bound $2R_1 + R_2$. We consider the interference channel drawn in Figure 13, in which there are two identical receivers ($Rx_{11}$ and $Rx_{12}$) for user 1's message, and one receiver for user 2's message. We can think of $2R_1 + R_2$ as the sum of the rates at the three receivers. To derive an upper bound, we consider the genie-aided channel where a genie provides $s_1$ to receiver $Rx_{11}$ and $s_2$ to receiver $Rx_2$ (see Figure 14). For any codebook of block-length $n$ we can write:

$$
\begin{aligned}
n(2R_1 + R_2) &\leq I(\mathbf{x}_1^n; \mathbf{y}_1^n) + I(\mathbf{x}_2^n; \mathbf{y}_2^n) + I(\mathbf{x}_1^n; \mathbf{y}_1^n) + n\epsilon \\
&\leq I(\mathbf{x}_1^n; \mathbf{y}_1^n) + I(\mathbf{x}_2^n; \mathbf{y}_2^n \mathbf{s}_2^n) + I(\mathbf{x}_1^n; \mathbf{y}_1^n | \mathbf{x}_2^n) + n\epsilon \\
&= h(\mathbf{y}_1^n) - h(\mathbf{y}_1^n | \mathbf{x}_1^n) + h(\mathbf{y}_2^n \mathbf{s}_2^n) - h(\mathbf{y}_2^n \mathbf{s}_2^n | \mathbf{x}_2^n) + h(\mathbf{y}_1^n | \mathbf{x}_2^n) - h(\mathbf{y}_1^n | \mathbf{x}_1^n \mathbf{x}_2^n) + n\epsilon \\
&= h(\mathbf{y}_1^n) - h(\mathbf{s}_2^n) + h(\mathbf{s}_2^n) + h(\mathbf{y}_2^n | \mathbf{s}_2^n) - h(\mathbf{y}_2^n | \mathbf{x}_2^n) - h(\mathbf{s}_2^n | \mathbf{y}_2^n \mathbf{x}_2^n) \\
&\quad + h(\mathbf{y}_1^n | \mathbf{x}_2^n) - h(\mathbf{y}_1^n | \mathbf{x}_1^n \mathbf{x}_2^n) + n\epsilon \\
&= h(\mathbf{y}_1^n) + h(\mathbf{y}_2^n | \mathbf{s}_2^n) - h(\mathbf{y}_2^n | \mathbf{x}_2^n) - h(\mathbf{z}_1^n) + h(\mathbf{y}_1^n | \mathbf{x}_2^n) - h(\mathbf{z}_1^n) + n\epsilon \\
&\leq \sum_{i=1}^{n} [h(y_{1i}) + h(y_{2i} | s_{2i}) - 2h(z_{1i}))] - h(\mathbf{y}_2^n | \mathbf{x}_2^n) + h(\mathbf{y}_1^n | \mathbf{x}_2^n) + n\epsilon \\
&= \sum_{i=1}^{n} [h(y_{1i}) + h(y_{2i} | s_{2i}) - 2h(z_{1i}))] - h(h_{12}\mathbf{x}_1^n + \mathbf{z}_2^n) + h(h_{11}\mathbf{x}_1^n + \mathbf{z}_1^n) + n\epsilon.
\end{aligned}
\tag{46}
$$

Using similar reasons as those in the proof of the third bound on $R_1 + R_2$, we have

$$\begin{aligned}\frac{1}{n}\sum_{i=1}^{n} h(y_{1i}) &\leq \frac{1}{n}\log \pi e\left(|h_{11}|^2 P_{1i} + |h_{21}|^2 P_{2i} + N_0\right) \\ &\leq \log \pi e\left(|h_{11}|^2 \frac{1}{n}\sum_{i=1}^{n} P_{1i} + |h_{21}|^2 \frac{1}{n}\sum_{i=1}^{n} P_{2i} + N_0\right) \\ &\leq \log \pi e\left(|h_{11}|^2 P_1 + |h_{21}|^2 P_2 + N_0\right), \qquad (47)\end{aligned}$$

and

$$\frac{1}{n}\sum_{i=1}^{n} h(y_{2i}|s_{2i}) \leq \log \pi e\left(N_0 + |h_{12}|^2 P_1 + \frac{|h_{22}|^2 P_2 N_0}{N_0 + |h_{21}|^2 P_2}\right). \qquad (48)$$

Since $\mathsf{INR}_2 < \mathsf{SNR}_1$, we can use the worst case noise result to bound $-h(h_{12}\mathbf{x}_1^n + \mathbf{z}_2^n) + h(h_{11}\mathbf{x}_1^n + \mathbf{z}_1^n)$ as

$$-h(h_{12}\mathbf{x}_1^n + \mathbf{z}_2^n) + h(h_{11}\mathbf{x}_1^n + \mathbf{z}_1^n) \leq \log\left(\frac{1 + \mathsf{SNR}_1}{1 + \mathsf{INR}_2}\right). \qquad (49)$$

Combining all the above, we have

$$\begin{aligned}2R_1 + R_2 &\leq \log\left(1 + \mathsf{SNR}_1 + \mathsf{INR}_1\right) + \log\left(1 + \mathsf{INR}_2 + \frac{\mathsf{SNR}_2}{1 + \mathsf{INR}_1}\right) \\ &\quad + \log\left(\frac{1 + \mathsf{SNR}_1}{1 + \mathsf{INR}_2}\right) + \epsilon. \qquad (50)\end{aligned}$$

6. Similarly we can derive the bound for $R_1 + 2R_2$.

□

***Remark* 1.** *As mentioned, our first bound on $R_1 + R_2$ is also an outer bound on the sum rate of the one-sided interference channel (also known as Z-channel) [4, 5, 8] generated by removing the link from transmitter 2 to receiver 1. In the one-sided channel, this outer bound can actually be achieved by both users when they communicate using codebooks generated from i.i.d. samples of a Gaussian distribution at full power and receiver 2 treats the signal from transmitter 1 as noise. Hence we have a simple derivation of the sum capacity of the one-sided interference channel. Note that the proof of the sum capacity of the one-sided interference channel of [4, 5, 8] is quite indirect. In [5], the degraded Gaussian interference channel is introduced, and the capacity region of the degraded Gaussian interference channel is shown to be included in the capacity region of a degraded Gaussian broadcast channel. Moreover, the boundaries of the two regions are shown to touch at one point $A$. Later it was shown in [4] that the one-sided interference channel is equivalent to the degraded Gaussian interference channel. Recently, the author of [8] points out that through a slope calculation in [5], the sum capacity of the degraded Gaussian interference channel is achieved at point $A$, thus establishing the sum capacity of the Gaussian one-sided interference channel.*

Note that in our derivation of the first and second outer bounds on the sum rate, the outer bound on $2R_1 + R_2$, and the outer bound on $R_1 + 2R_2$, we used the conditions $\mathsf{INR}_2 < \mathsf{SNR}_1$ and $\mathsf{INR}_1 < \mathsf{SNR}_2$. For this reason this outer bound only holds for the weak interference channel. Next we present an outer bound on the capacity region of the mixed interference channel.

### 4.1.2 Outer bound for mixed interference channel

For the mixed interference channel, i.e., $\mathsf{INR}_1 \geq \mathsf{SNR}_2$ and $\mathsf{INR}_2 < \mathsf{SNR}_1$, we have the following outer bound on the capacity region.

***Theorem 4.*** *For the Gaussian mixed interference channel, the capacity region is contained within the set of rate pairs $(R_1, R_2)$ satisfying:*

$$\begin{aligned} R_1 &\leq \log(1+\mathsf{SNR}_1) \\ R_2 &\leq \log(1+\mathsf{SNR}_2) \\ R_1+R_2 &\leq \log(1+\mathsf{SNR}_1) + \log\left(1+\frac{\mathsf{SNR}_2}{1+\mathsf{INR}_2}\right) \\ R_1+R_2 &\leq \log(1+\mathsf{SNR}_1+\mathsf{INR}_1) \\ R_1+2R_2 &\leq \log(1+\mathsf{SNR}_2+\mathsf{INR}_2) + \log\left(1+\mathsf{INR}_1+\frac{\mathsf{SNR}_1}{1+\mathsf{INR}_2}\right) + \log\left(1+\frac{\mathsf{SNR}_2}{1+\mathsf{INR}_1}\right). \end{aligned} \tag{51}$$

*Proof.* We prove this outer bound by examining the proof of the bounds in (39).

1. The bounds on $R_1$ and $R_2$ still hold.
2. The proof of the first upper bound on the sum rate in (39) needs the condition $\mathsf{INR}_2 < \mathsf{SNR}_1$, which still holds in this mixed interference channel. So we have the same bound.
3. The proof of the second upper bound on the sum rate in (39) needs the condition $\mathsf{INR}_1 < \mathsf{SNR}_2$, which does not hold. Note that this bound is actually the sum capacity of the one-sided interference channel in which the link from transmitter 1 to receiver 2 is removed. When $\mathsf{INR}_1 \geq \mathsf{SNR}_2$, we are dealing with a one-sided interference channel with strong interference. It is shown in [4] that a sum rate outer bound for this channel is the sum rate of the MAC at receiver 1, i.e.

$$R_1 + R_2 \leq \log(1+\mathsf{SNR}_1+\mathsf{INR}_1), \tag{52}$$

   and it is obviously an upper bound on sum rate of the mixed interference channel.
4. The third upper bound on the sum rate in (39) still holds; however, it is straightforward to show that the third bound on $R_1+R_2$ is larger than the second bound on $R_1+R_2$ in (51), so we do not need to include this bound.
5. The upper bound on $2R_1+R_2$ in (39) still holds since $\mathsf{INR}_2 < \mathsf{SNR}_1$ still holds. However, it is straightforward to show that the bound on $2R_1+R_2$ is larger than the sum of the bound on $R_1$ and the second bound on $R_1+R_2$ in (51), so we do not need to include this bound.
6. The proof of the bound on $R_1+2R_2$ in (39) needs the condition of $\mathsf{INR}_1 < \mathsf{SNR}_2$, which is no longer true. To obtain a bound that does not require the condition $\mathsf{INR}_1 < \mathsf{SNR}_2$, we consider the interference channel drawn in Figure 15, in which there are two identical receivers ($Rx_{21}$ and $Rx_{22}$) for user 2's message, and one receiver for user 1's message. We can think of $R_1+2R_2$ as the sum rate at the three receivers of this new interference channel. To derive an upper bound, we consider the genie-aided channel where a genie provides both $s_2$ and $x_1$ to receiver $Rx_{21}$ and $s_1$ to receiver $Rx_1$ (see Figure 16).

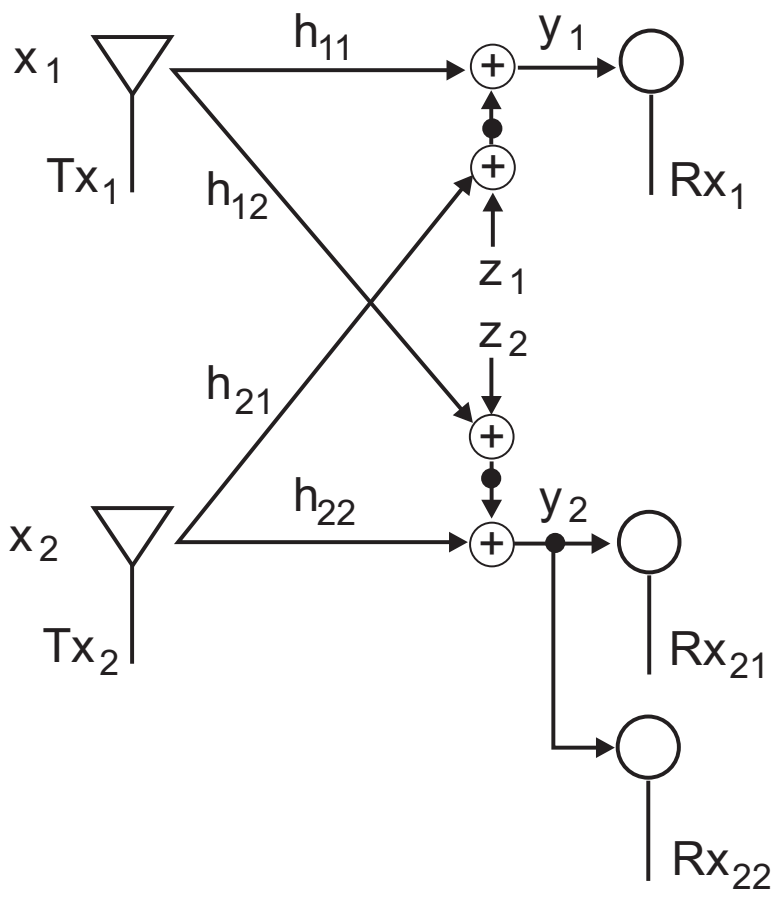

Figure 15: Gaussian interference channel with three receivers.

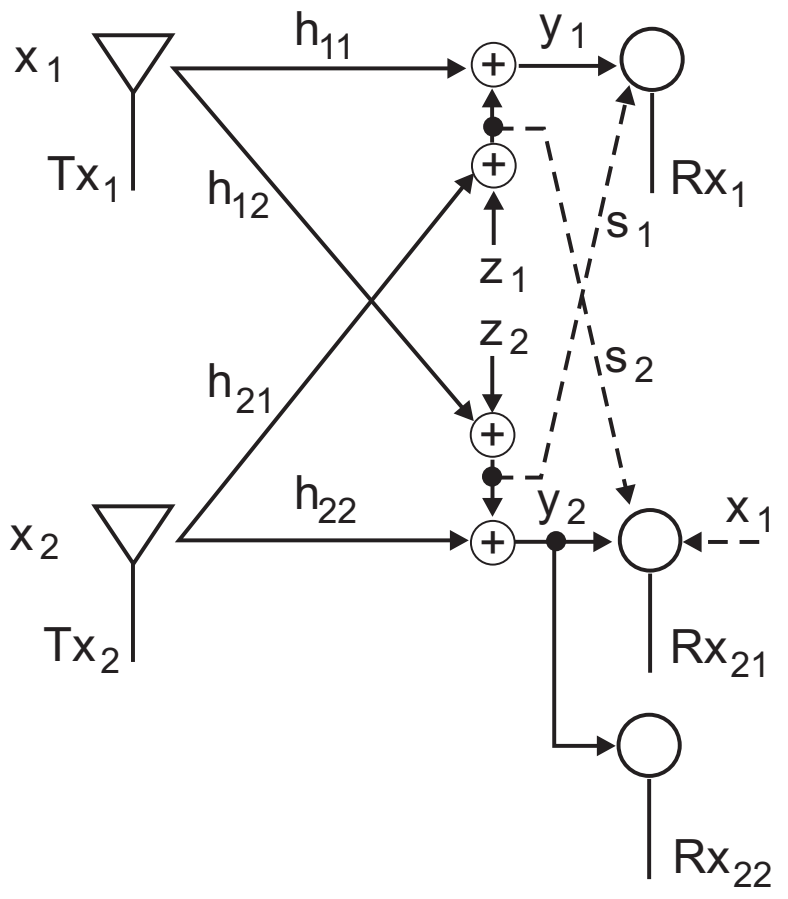

Figure 16: Genie-aided three-receiver Gaussian interference channel. A genie provides signals $s_2$ and $x_1$ to receiver $Rx_{21}$, and $s_1$ to $Rx_1$.

$$
\begin{aligned}
n(R_1+2R_2) &\leq I(\mathbf{x}_1^n;\mathbf{y}_1^n)+I(\mathbf{x}_2^n;\mathbf{y}_2^n)+I(\mathbf{x}_2^n;\mathbf{y}_2^n)+n\epsilon \\
&\leq I(\mathbf{x}_1^n;\mathbf{y}_1^n,\mathbf{s}_1^n)+I(\mathbf{x}_2^n;\mathbf{y}_2^n)+I(\mathbf{x}_2^n;\mathbf{y}_2^n\mathbf{s}_2^n|\mathbf{x}_1^n)+n\epsilon \\
&= I(\mathbf{x}_1^n;\mathbf{s}_1^n)+I(\mathbf{x}_1^n;\mathbf{y}_1^n|\mathbf{s}_1^n)+I(\mathbf{x}_2^n;\mathbf{y}_2^n)+I(\mathbf{x}_2^n;\mathbf{s}_2^n|\mathbf{x}_1^n)+I(\mathbf{x}_2^n;\mathbf{y}_2^n|\mathbf{s}_2^n\mathbf{x}_1^n)+n\epsilon \\
&= h(\mathbf{s}_1^n)-h(\mathbf{s}_1^n|\mathbf{x}_1^n)+h(\mathbf{y}_1^n|\mathbf{s}_1^n)-h(\mathbf{y}_1^n|\mathbf{x}_1^n\mathbf{s}_1^n)+h(\mathbf{y}_2^n)-h(\mathbf{y}_2^n|\mathbf{x}_2^n) \\
&\quad +h(\mathbf{s}_2^n|\mathbf{x}_1^n)-h(\mathbf{s}_2^n|\mathbf{x}_1^n\mathbf{x}_2^n)+h(\mathbf{y}_2^n|\mathbf{s}_2^n\mathbf{x}_1^n)-h(\mathbf{y}_2^n|\mathbf{s}_2^n\mathbf{x}_1^n\mathbf{x}_2^n)+n\epsilon \\
&= h(\mathbf{s}_1^n)-h(\mathbf{z}_2^n)+h(\mathbf{y}_1^n|\mathbf{s}_1^n)-h(\mathbf{s}_2^n)+h(\mathbf{y}_2^n)-h(\mathbf{s}_1^n) \\
&\quad +h(\mathbf{s}_2^n)-h(\mathbf{z}_1^n)+h(\mathbf{y}_2^n|\mathbf{s}_2^n\mathbf{x}_1^n)-h(\mathbf{z}_2^n)+n\epsilon \\
&\leq h(\mathbf{y}_1^n|\mathbf{s}_1^n)+h(\mathbf{y}_2^n)+h(\mathbf{y}_2^n|\mathbf{s}_2^n\mathbf{x}_1^n)-h(\mathbf{z}_1^n)-2h(\mathbf{z}_2^n)+n\epsilon \\
&\leq \sum_{i=1}^{n}\left[h(y_{1i}|s_{1i})+h(y_{2i})+h(y_{2i}|s_{2i}x_{1i})-h(z_{1i})-2h(z_{2i})\right]+n\epsilon.
\end{aligned}
$$

Using similar reasons as those in the proof of Theorem 1 in Section 3.4 we can write:

$$
\begin{aligned}
\frac{1}{n}\sum_{i=1}^{n}h(y_{1i}|s_{1i}) &\leq \frac{1}{n}\log\left[\pi e\left(N_0+|h_{21}|^2P_{2i}+\frac{|h_{11}|^2P_{1i}N_0}{N_0+|h_{12}|^2P_{1i}}\right)\right] \\
&\leq \log\left[\pi e\left(N_0+|h_{21}|^2\left(\frac{1}{n}\sum_{i=1}^{n}P_{2i}\right)+\frac{|h_{11}|^2\left(\frac{1}{n}\sum_{i=1}^{n}P_{1i}\right)N_0}{N_0+|h_{12}|^2\left(\frac{1}{n}\sum_{i=1}^{n}P_{1i}\right)}\right)\right] \\
&\leq \log\left[\pi e\left(N_0+|h_{21}|^2P_2+\frac{|h_{11}|^2P_1N_0}{N_0+|h_{12}|^2P_1}\right)\right], \qquad (53)
\end{aligned}
$$

together with

$$
\begin{aligned}
\frac{1}{n}\sum_{i=1}^{n} h(y_{2i}) &\leq \frac{1}{n}\log \pi e \left(|h_{22}|^2 P_{2i} + |h_{12}|^2 P_{1i} + N_0\right) \\
&\leq \log \pi e \left(|h_{22}|^2 \frac{1}{n}\sum_{i=1}^{n} P_{2i} + |h_{12}|^2 \frac{1}{n}\sum_{i=1}^{n} P_{1i} + N_0\right) \\
&\leq \log \pi e \left(|h_{22}|^2 P_2 + |h_{12}|^2 P_1 + N_0\right), \qquad (54)
\end{aligned}
$$

and

$$
\begin{aligned}
\frac{1}{n}\sum_{i=1}^{n} h(y_{2i}|s_{2i}x_{1i}) &= \frac{1}{n}\sum_{i=1}^{n} h(|h_{22}|^2 x_{2i} + z_{2i}|s_{2i}) \\
&\leq \frac{1}{n}\sum_{i=1}^{n} \log \pi e \left(N_0 + \frac{|h_{22}|^2 N_0 P_{2i}}{N_0 + |h_{21}|^2 P_{2i}}\right) \\
&\leq \log \pi e \left(N_0 + \frac{|h_{22}|^2 N_0 \left(\frac{1}{n}\sum_{i=1}^{n} P_{2i}\right)}{N_0 + |h_{21}|^2 \left(\frac{1}{n}\sum_{i=1}^{n} P_{2i}\right)}\right) \\
&\leq \log \pi e \left(N_0 + \frac{|h_{22}|^2 N_0 P_2}{N_0 + |h_{21}|^2 P_2}\right). \qquad (55)
\end{aligned}
$$

Thus we have

$$
\begin{aligned}
R_1 + 2R_2 &\leq \log \pi e \left(|h_{22}|^2 P_2 + |h_{12}|^2 P_1 + N_0\right) - \log \pi e N_0 \\
&\quad + \log \pi e \left(N_0 + |h_{21}|^2 P_2 + \frac{|h_{11}|^2 P_1 N_0}{N_0 + |h_{12}|^2 P_1}\right) - \log \pi e N_0 \\
&\quad + \log \pi e \left(N_0 + \frac{|h_{22}|^2 N_0 P_2}{N_0 + |h_{21}|^2 P_2}\right) - \log \pi e N_0 + \epsilon \\
&= \log \left(1 + \frac{|h_{22}|^2 P_2}{N_0} + \frac{|h_{12}|^2 P_1}{N_0}\right) + \log \left(1 + \frac{|h_{21}|^2 P_2}{N_0} + \frac{|h_{11}|^2 P_1/N_0}{1 + |h_{12}|^2 P_1/N_0}\right) \\
&\quad + \log \left(1 + \frac{|h_{22}|^2 P_2/N_0}{1 + |h_{21}|^2 P_2/N_0}\right) + \epsilon \\
&= \log\left(1 + \mathsf{SNR}_2 + \mathsf{INR}_2\right) + \log \left(1 + \mathsf{INR}_1 + \frac{\mathsf{SNR}_1}{1 + \mathsf{INR}_2}\right) \\
&\quad + \log \left(1 + \frac{\mathsf{SNR}_2}{1 + \mathsf{INR}_1}\right) + \epsilon. \qquad (56)
\end{aligned}
$$

□

## 4.2 Achievable scheme

The Han-Kobayashi scheme [1] is the best known achievable scheme for the interference channel. Recently a simplified yet equivalent Han-Kobayashi achievable region was given in [10], which we state in the following lemma.

***Lemma* 1.** *Let $\mathcal{P}^*$ be the set of joint probability distributions $P^*(\cdot)$ that factor as*

$$P^*(q, w_1, w_2, x_1, x_2, y_1, y_2) = P(q) \cdot P(w_1, x_1|q) \cdot P(w_2, x_2|q) \cdot P(y_1, y_2|x_1, x_2). \tag{57}$$

*For a fixed $P^* \in \mathcal{P}^*$, let $\mathcal{R}(P^*)$ be the set of $(R_1, R_2)$ satisfying:*

$$\begin{aligned}
R_1 &\leq I(x_1; y_1|w_2 q) \\
R_2 &\leq I(x_2; y_2|w_1 q) \\
R_1 + R_2 &\leq I(x_2 w_1; y_2|q) + I(x_1; y_1|w_1 w_2 q) \\
R_1 + R_2 &\leq I(x_1 w_2; y_1|q) + I(x_2; y_2|w_1 w_2 q) \\
R_1 + R_2 &\leq I(x_1 w_2; y_1|w_1 q) + I(x_2 w_1; y_2|w_2 q) \\
2R_1 + R_2 &\leq I(x_1 w_2; y_1|q) + I(x_1; y_1|w_1 w_2 q) + I(x_2 w_1; y_2|w_2 q) \\
R_1 + 2R_2 &\leq I(x_2 w_1; y_2|q) + I(x_2; y_2|w_1 w_2 q) + I(x_1 w_2; y_1|w_1 q).
\end{aligned} \tag{58}$$

*Then the Han-Kobayashi achievable region is given by $\mathcal{R} = \bigcup_{P^* \in \mathcal{P}^*} \mathcal{R}(P^*)$.*

In (58), $w_1$ ($w_2$) is the common information of user 1 (user 2) that can be decoded at both receivers, and $q$ is the time sharing parameter. For the Gaussian interference channel, if we use Gaussian codebooks, and use $u_1$ and $u_2$ to denote the private information of user 1 and user 2 respectively, we can write

$$\begin{aligned}
x_1 &= u_1 + w_1 \\
x_2 &= u_2 + w_2,
\end{aligned} \tag{59}$$

where $u_1$, $u_2$, $w_1$ and $w_2$ are independent complex Gaussian random variables. Different $P^* \in \mathcal{P}^*$ correspond to different power splits between common and private messages, and different time-sharing strategies between the power splits.

Consider a fixed power splitting (i.e. we don't do time sharing) between private information and common information of the two users. Let $P_{u1}$ and $P_{u2}$ be the power of user 1 and user 2's private messages respectively. We define $\mathsf{INR}_{p_2}$ to be the interference to noise ratio of user 1's private message at receiver 2 and $\mathsf{INR}_{p_1}$ to be the interference to noise ratio of user 2's private message at receiver 1, i.e.

$$\begin{aligned}
\mathsf{INR}_{p_2} &= \frac{|h_{12}|^2 P_{u1}}{N_0} \\
\mathsf{INR}_{p_1} &= \frac{|h_{21}|^2 P_{u2}}{N_0}.
\end{aligned} \tag{60}$$

It is clear that $0 \leq \mathsf{INR}_{p_2} \leq \mathsf{INR}_2$ and $0 \leq \mathsf{INR}_{p_1} \leq \mathsf{INR}_1$. With this definition, the signal to noise ratio of user 1's private message at receiver 1 is $\mathsf{SNR}_{p_1} = \mathsf{INR}_{p_2} \frac{\mathsf{SNR}_1}{\mathsf{INR}_2}$ and the signal to noise ratio of user 2's private message at receiver 2 is $\mathsf{SNR}_{p_2} = \mathsf{INR}_{p_1} \frac{\mathsf{SNR}_2}{\mathsf{INR}_1}$. We can parameterize a Han-Kobayashi achievable scheme with a fixed power splitting by using $\mathsf{INR}_{p_2}$ and $\mathsf{INR}_{p_1}$. We denote the Han-Kobayashi scheme with parameters $\mathsf{INR}_{p_2}$, $\mathsf{INR}_{p_1}$ as HK($\mathsf{INR}_{p_2}$, $\mathsf{INR}_{p_1}$), and the corresponding achievable region as $\mathscr{R}(\mathsf{INR}_{p_2}, \mathsf{INR}_{p_1})$.

Note that HK($\mathsf{INR}_{p_2}$, $\mathsf{INR}_{p_1}$) and $\mathscr{R}(\mathsf{INR}_{p_2}, \mathsf{INR}_{p_1})$ correspond to a Han-Kobayashi scheme where there is a fixed private and common message power split and there is no time sharing (i.e. the time sharing random variable $q$ is a constant). Therefore, $\mathscr{R}(\mathsf{INR}_{p_2}, \mathsf{INR}_{p_1}) \subset \mathcal{R}$, where $\mathcal{R}$ is the general Han-Kobayashi achievable region given in Lemma 1. In general, the inclusion is strict, that is, varying the power allocations and time sharing between multiple private and common message power splits allows to achieve a larger rate region. However, we will see that the region achievable with a clever choice of a fixed private and common message power split and without time sharing is close to the capacity region of the channel.

To evaluate the Han-Kobayashi region (58) for the Gaussian interference channel, even if we restrict ourselves to use only Gaussian codebooks, we need to consider all possible power splits and different time sharing strategies among them. This is in general very complicated and a calculation of a subset of the Han-Kobayashi achievable region using some special choices of power splitting and time sharing strategies can be found in [8]. However, from the intuition we built in Section 3, we know that a good power splitting should have the property that $\mathsf{INR}_{p_2} = 1$ and $\mathsf{INR}_{p_1} = 1$, i.e., the interference to noise ratio of each user's private message at the other user's receiver is one. We also showed that this power splitting can achieve to within one bit the symmetric rate capacity of the symmetric Gaussian interference channel. In the next section, we will show that this is also a good splitting for the entire capacity region. More specifically, we will show that by choosing $\mathsf{INR}_{p_2}$, $\mathsf{INR}_{p_1}$ as close to 1 as possible, we can achieve rates within one bit of the whole capacity region.

## 4.3 Within one bit of the capacity region

Equipped with the new outer bound derived in Section 4.1.1 and the intuition of a good power splitting in Section 3, we are now ready to prove our main result: a simple Han-Kobayashi scheme can achieve to within one bit of the capacity region of the Gaussian interference channel. First we provide a formal definition of the *within one bit* notion.

***Definition* 1.** *An achievable region is said to be* within one bit of the capacity region *if for any rate pair $(R_1, R_2)$ on the boundary of the achievable region, the rate pair $(R_1 + 1, R_2 + 1)$ is not achievable. Equivalently, $(R_1 - 1, R_2 - 1)$ is in the achievable region for any rate pair $(R_1, R_2)$ in the capacity region.*

Since the outer bound of the weak interference channel and the outer bound of the mixed interference channel are different, we treat these two channels separately in the following two subsections.

### 4.3.1 Weak interference channel

Our main result is stated in the following theorem.

***Theorem* 5.** *The achievable region*

$$\mathscr{R}\big(\min(1, \mathsf{INR}_2), \min(1, \mathsf{INR}_1)\big) \tag{61}$$

*is within one bit of the capacity region of the Gaussian weak interference channel.*

***Remark* 2.** *The reason to consider the region $\mathscr{R}\big(\min(1, \mathsf{INR}_2), \min(1, \mathsf{INR}_1)\big)$ is because when $\mathsf{INR}_1 < 1$ or $\mathsf{INR}_2 < 1$, we can not use HK(1,1). However, say in the case of $\mathsf{INR}_1 < 1$, the interference caused by user*

*2 at receiver 1 is even weaker than the additive Gaussian noise. Thus we won't lose much of the optimality by simply treating all of user 2's signal as noise at receiver 1, i.e., letting* $\mathsf{INR}_{p_1} = \mathsf{INR}_1$.

*Proof.* It can be seen from (39) and (58) that both the outer bound to the capacity region and the achievable region $\mathscr{R}\big(\min(1, \mathsf{INR}_2), \min(1, \mathsf{INR}_1)\big)$ are piecewise linear, and only consist of straight lines with slopes $0$, $-1/2$, $-1$, $-2$, and $\infty$. We define $\delta_{R_1}$ to be the difference between the outer bound on $R_1$ in (39) (first constraint) and achievable $R_1$ in $\mathscr{R}\big(\min(1, \mathsf{INR}_2), \min(1, \mathsf{INR}_1)\big)$, and similarly define $\delta_{R_2}$, $\delta_{R_1+R_2}$, $\delta_{2R_1+R_2}$ and $\delta_{R_1+2R_2}$. Note that if the rate pair $(R_1, R_2)$ is on the boundary of the achievable region $\mathscr{R}\big(\min(1, \mathsf{INR}_2), \min(1, \mathsf{INR}_1)\big)$, it must be on one of the bounding straight lines. Thus if the following holds,

$$\begin{aligned} \delta_{R_1} &< 1 \\ \delta_{R_2} &< 1 \\ \delta_{R_1+R_2} &< 2 \\ \delta_{2R_1+R_2} &< 3 \\ \delta_{R_1+2R_2} &< 3, \end{aligned} \tag{62}$$

then the rate pair $(R_1{+}1, R_2{+}1)$ would be outside the outer bound (39), and hence $\mathscr{R}\big(min(1, \mathsf{INR}_2), min(1, \mathsf{INR}_1)\big)$ is within one bit of the capacity region of the Gaussian weak interference channel. We now show that (62) holds for the different parameter ranges.

1. <u>$\mathsf{INR}_1 \geq 1$ and $\mathsf{INR}_2 \geq 1$</u>

   In this case, we have

$$\mathscr{R}\big(\min(1, \mathsf{INR}_2), \min(1, \mathsf{INR}_1)\big) = \mathscr{R}(1, 1). \tag{63}$$

   It is straightforward to evaluate $\mathscr{R}(1, 1)$ from Lemma 1. The result is provided in the following corollary.

   ***Corollary 1.*** *The achievable rate region* $\mathscr{R}(1, 1)$ *contains all the rate pairs* $(R_1, R_2)$ *satisfying:*

$$\begin{aligned} R_1 &\leq \log(2 + \mathsf{SNR}_1) - 1 \\ R_2 &\leq \log(2 + \mathsf{SNR}_2) - 1 \\ R_1 + R_2 &\leq \log(2\mathsf{INR}_2 + \mathsf{SNR}_1) + \log\left(1 + \frac{1 + \mathsf{SNR}_2}{\mathsf{INR}_2}\right) - 2 \\ R_1 + R_2 &\leq \log(2\mathsf{INR}_1 + \mathsf{SNR}_2) + \log\left(1 + \frac{1 + \mathsf{SNR}_1}{\mathsf{INR}_1}\right) - 2 \\ R_1 + R_2 &\leq \log\left(1 + \mathsf{INR}_1 + \frac{\mathsf{SNR}_1}{\mathsf{INR}_2}\right) + \log\left(1 + \mathsf{INR}_2 + \frac{\mathsf{SNR}_2}{\mathsf{INR}_1}\right) - 2 \\ 2R_1 + R_2 &\leq \log(1 + \mathsf{SNR}_1 + \mathsf{INR}_1) + \log\left(1 + \mathsf{INR}_2 + \frac{\mathsf{SNR}_2}{\mathsf{INR}_1}\right) + \log\left(2 + \frac{\mathsf{SNR}_1}{\mathsf{INR}_2}\right) - 3 \\ R_1 + 2R_2 &\leq \log(1 + \mathsf{SNR}_2 + \mathsf{INR}_2) + \log\left(1 + \mathsf{INR}_1 + \frac{\mathsf{SNR}_1}{\mathsf{INR}_2}\right) + \log\left(2 + \frac{\mathsf{SNR}_2}{\mathsf{INR}_1}\right) - 3. \end{aligned} \tag{64}$$

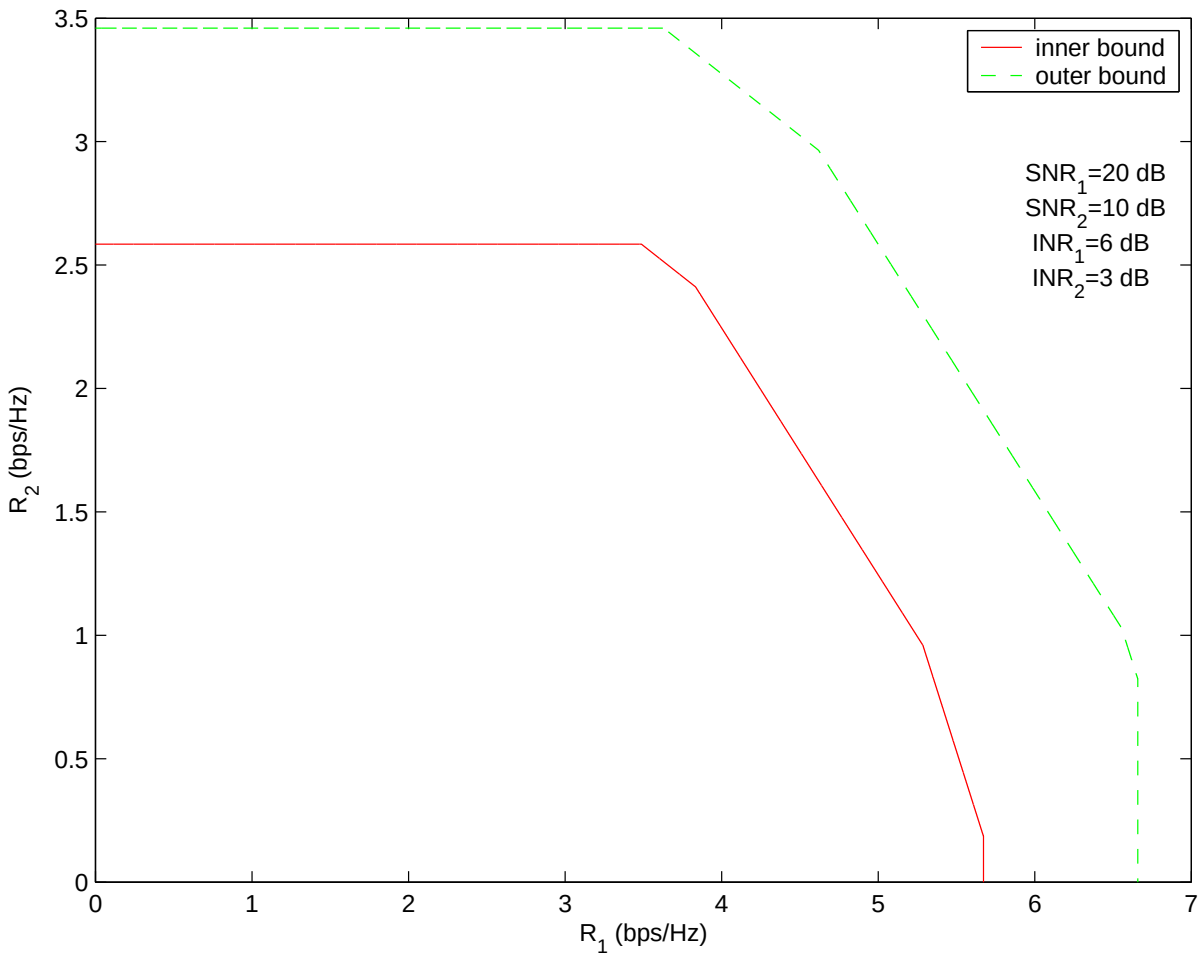

Figure 17: Comparison of Han-Kobayashi achievable region $\mathscr{R}(1,1)$ and outer bound (39).

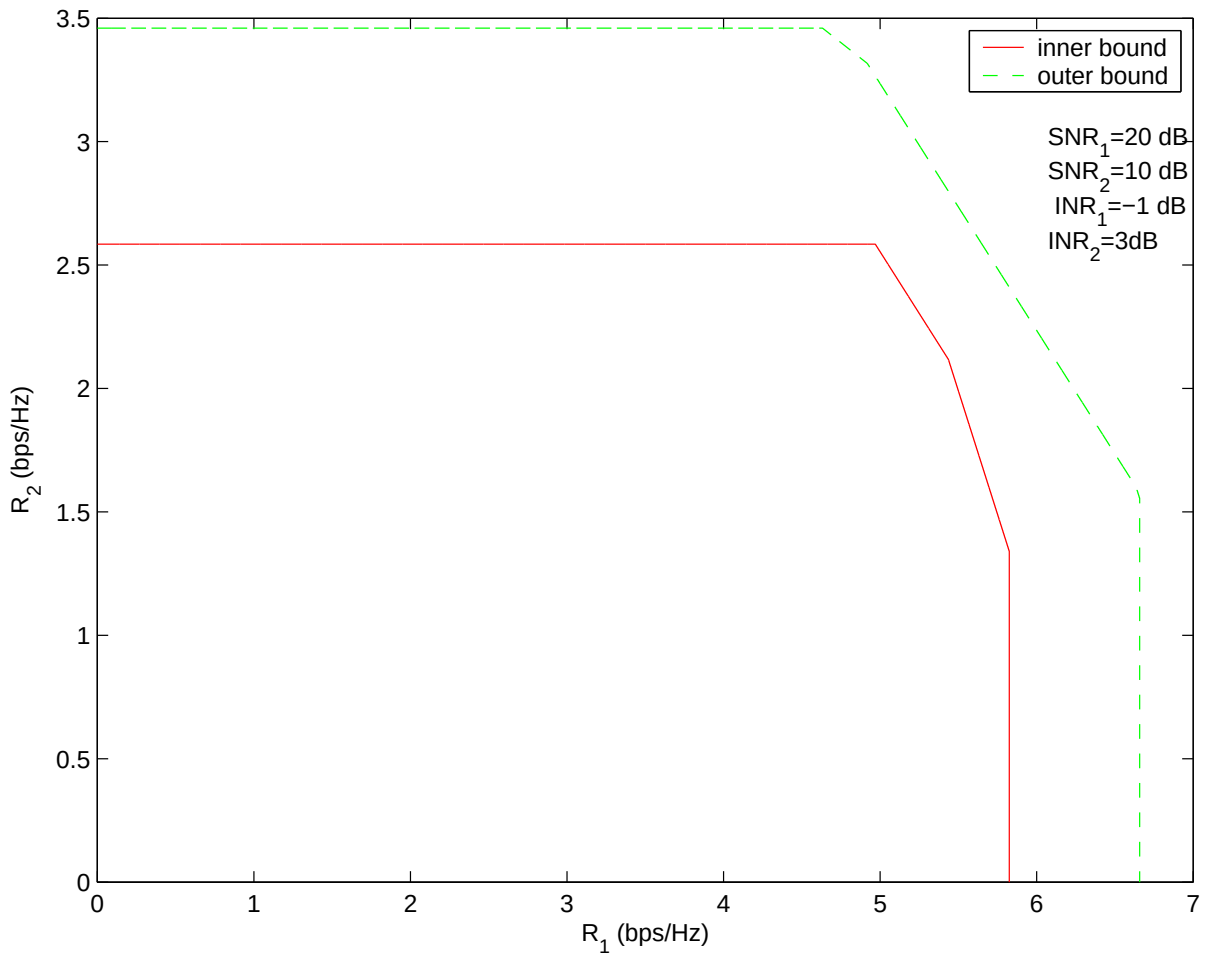

Figure 18: Comparison of Han-Kobayashi achievable region $\mathscr{R}(1,\mathsf{INR}_1)$ and outer bound (39).

If we denote the three outer bounds on the sum rate in (39) by $a_1, a_2, a_3$ respectively, and the three inner bounds on the sum rate in (64) by $b_1, b_2, b_3$ respectively, we have

$$\delta_{R_1+R_2} = \min\{a_1,a_2,a_3\} - \min\{b_1,b_2,b_3\} \le \max\{a_1-b_1, a_2-b_2, a_3-b_3\}, \qquad (65)$$

and hence we can simply upper bound $\delta_{R_1+R_2}$ by the maximum of the differences between the $i$-th bound on $R_1+R_2$ in the outer bound (39) and the corresponding $i$-th bound on $R_1+R_2$ in the inner bound (64) for $i=1,2,3$. We can readily compare (39) and (64) term by term, and see that (62) is true. (See Figure 17). For example,

$$\begin{aligned}\delta_{2R_1+R_2} &= \log(1+\mathsf{SNR}_1+\mathsf{INR}_1) + \log\left(1+\mathsf{INR}_2+\frac{\mathsf{SNR}_2}{1+\mathsf{INR}_1}\right) + \log\left(\frac{1+\mathsf{SNR}_1}{1+\mathsf{INR}_2}\right)\\ &\quad - \left[\log(1+\mathsf{SNR}_1+\mathsf{INR}_1) + \log\left(1+\mathsf{INR}_2+\frac{\mathsf{SNR}_2}{\mathsf{INR}_1}\right) + \log\left(2+\frac{\mathsf{SNR}_1}{\mathsf{INR}_2}\right) - 3\right]\\ &= \left[\log\left(1+\mathsf{INR}_2+\frac{\mathsf{SNR}_2}{1+\mathsf{INR}_1}\right) - \log\left(1+\mathsf{INR}_2+\frac{\mathsf{SNR}_2}{\mathsf{INR}_1}\right)\right]\\ &\quad + \left[\log\left(\frac{1+\mathsf{SNR}_1}{1+\mathsf{INR}_2}\right) - \log\left(2+\frac{\mathsf{SNR}_1}{\mathsf{INR}_2}\right)\right] + 3\\ &< 3.\end{aligned} \qquad (66)$$

2. $\underline{\mathsf{INR}_1 < 1 \text{ and } \mathsf{INR}_2 \ge 1}$

In this case, we have

$$\mathscr{R}\big(\min(1,\mathsf{INR}_2), \min(1,\mathsf{INR}_1)\big) = \mathscr{R}(1,\mathsf{INR}_1). \qquad (67)$$

Evaluating the achievable region (58) with $\mathsf{INR}_{p_2}=1$ and $\mathsf{INR}_{p_1}=\mathsf{INR}_1$, we have the following achievable region $\mathscr{R}(1, \mathsf{INR}_1)$.

$$\begin{aligned}
R_1 &\le \log\left(1 + \frac{\mathsf{SNR}_1}{1+\mathsf{INR}_1}\right) \\
R_2 &\le \log\left(2 + \mathsf{SNR}_2\right) - 1 \\
R_1 + R_2 &\le \log\left(\mathsf{INR}_2 + \frac{\mathsf{SNR}_1}{1+\mathsf{INR}_1}\right) + \log\left(1 + \frac{1+\mathsf{SNR}_2}{\mathsf{INR}_2}\right) - 1 \\
R_1 + R_2 &\le \left(1 + \frac{\mathsf{SNR}_1}{1+\mathsf{INR}_1}\right) + \log\left(2 + \mathsf{SNR}_2\right) - 1 \\
R_1 + R_2 &\le \log\left(\mathsf{INR}_2 + \frac{\mathsf{SNR}_1}{1+\mathsf{INR}_1}\right) + \log\left(1 + \frac{1+\mathsf{SNR}_2}{\mathsf{INR}_2}\right) - 1 \\
2R_1 + R_2 &\le \log\left(1 + \mathsf{SNR}_1 + \mathsf{INR}_1\right) + \log\left(1 + \mathsf{INR}_2 + \mathsf{SNR}_2\right) \\
&\quad + \log\left(1 + \mathsf{INR}_1 + \frac{\mathsf{SNR}_1}{\mathsf{INR}_2}\right) - \log 2(1+\mathsf{INR}_1)^2 \\
R_1 + 2R_2 &\le \log(2 + \mathsf{SNR}_2) + \log\left(\mathsf{INR}_2 + \frac{\mathsf{SNR}_1}{1+\mathsf{INR}_1}\right) + \log\left(1 + \frac{1+\mathsf{SNR}_2}{\mathsf{INR}_2}\right) - 2.
\end{aligned} \tag{68}$$

Since the second bound on $R_1 + R_2$ is the sum of the bound on $R_1$ and the bound on $R_2$, the first and the third bounds on $R_1 + R_2$ are the same, and the bound on $R_1 + 2R_2$ is the sum of of the bound on $R_2$ and the first bound on $R_1 + R_2$, these three bounds are redundant and we have the following simplified achievable region $\mathscr{R}(1, \mathsf{INR}_1)$.

$$\begin{aligned}
R_1 &\le \log\left(1 + \frac{\mathsf{SNR}_1}{1+\mathsf{INR}_1}\right) \\
R_2 &\le \log\left(2 + \mathsf{SNR}_2\right) - 1 \\
R_1 + R_2 &\le \log\left(\mathsf{INR}_2 + \frac{\mathsf{SNR}_1}{1+\mathsf{INR}_1}\right) + \log\left(1 + \frac{1+\mathsf{SNR}_2}{\mathsf{INR}_2}\right) - 1 \\
2R_1 + R_2 &\le \log\left(1 + \mathsf{SNR}_1 + \mathsf{INR}_1\right) + \log\left(1 + \mathsf{INR}_2 + \mathsf{SNR}_2\right) \\
&\quad + \log\left(1 + \mathsf{INR}_1 + \frac{\mathsf{SNR}_1}{\mathsf{INR}_2}\right) - \log 2(1+\mathsf{INR}_1)^2.
\end{aligned} \tag{69}$$

Comparing $\mathscr{R}(1, \mathsf{INR}_1)$ with the corresponding bounds in (39) (use the first bound on $R_1 + R_2$ in (39)), and using the fact that $\mathsf{INR}_1 < 1$, we can see that (62) is true. (See Figure 18).

3. $\mathsf{INR}_1 \ge 1$ and $\mathsf{INR}_2 < 1$

   In this case, we have

$$\mathscr{R}\big(\min(1, \mathsf{INR}_2), \min(1, \mathsf{INR}_1)\big) = \mathscr{R}(\mathsf{INR}_2, 1). \tag{70}$$

   This case is similar to the previous case and we can show that (62) is true. (See Figure 19).

4. $\mathsf{INR}_1 < 1$ and $\mathsf{INR}_2 < 1$

   In this case, we have

$$\mathscr{R}\big(\min(1, \mathsf{INR}_2), \min(1, \mathsf{INR}_1)\big) = \mathscr{R}(\mathsf{INR}_2, \mathsf{INR}_1). \tag{71}$$

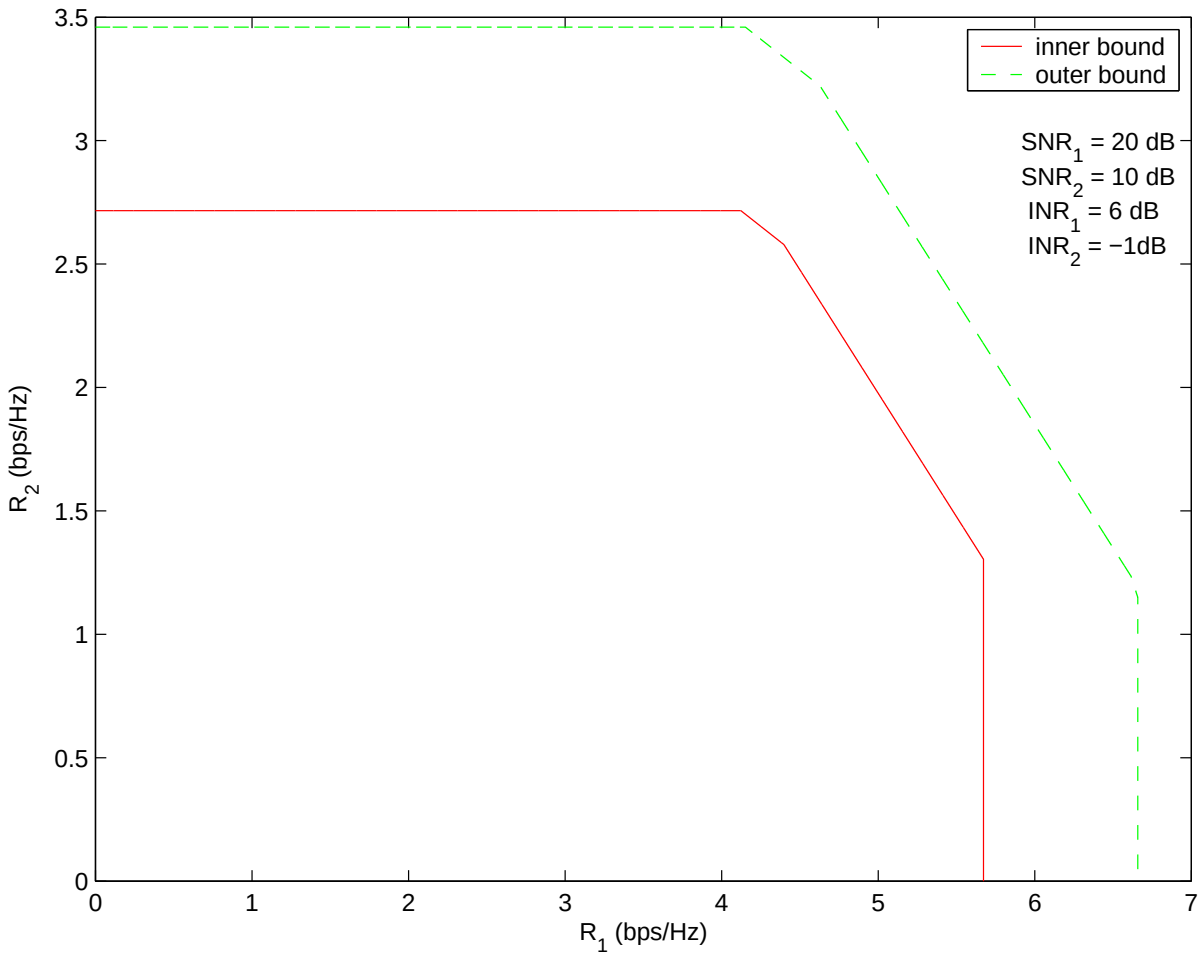

Figure 19: Comparison of Han-Kobayashi achievable region $\mathscr{R}(\mathsf{INR}_2, 1)$ and outer bound (39).

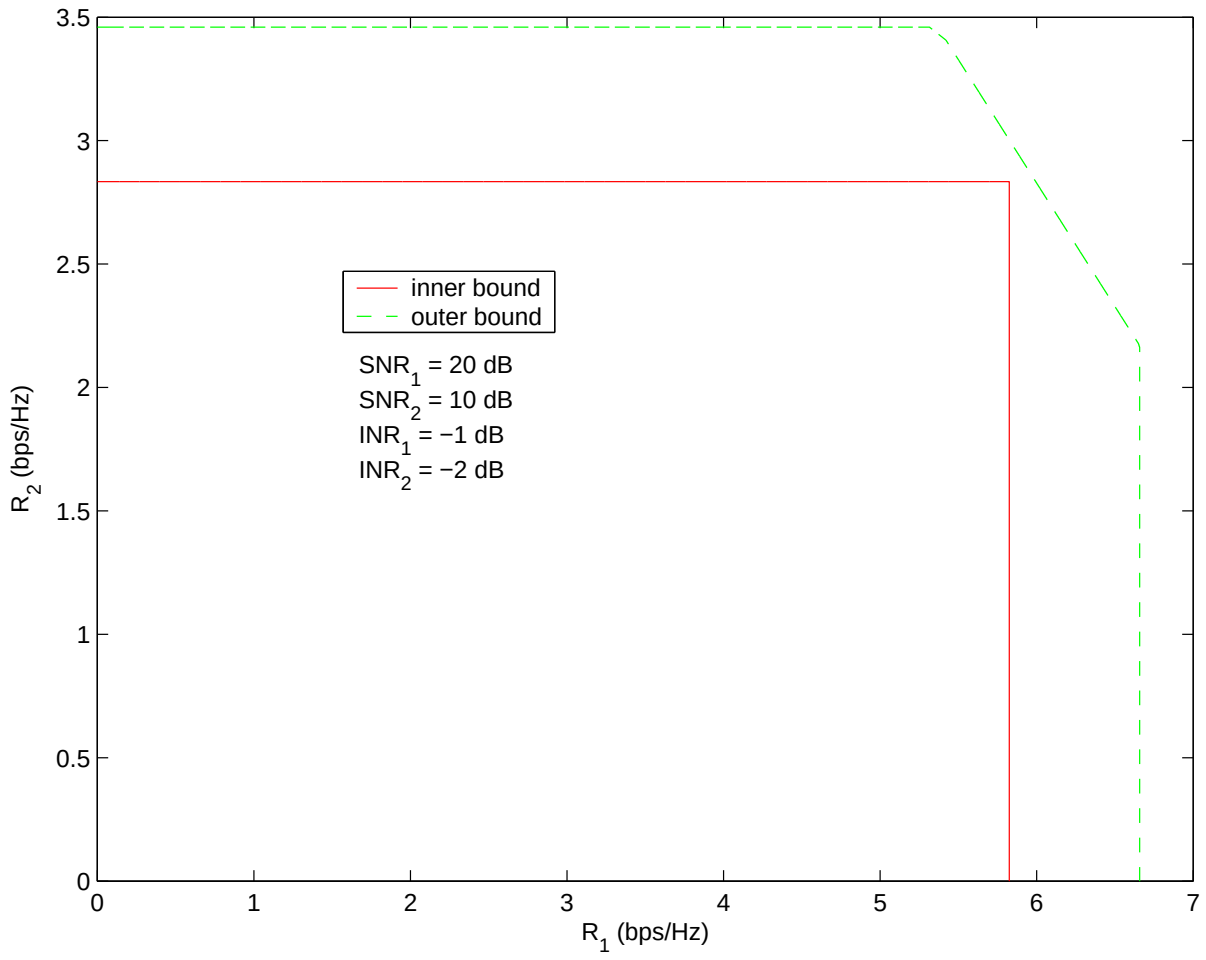

Figure 20: Comparison of Han-Kobayashi achievable region $\mathscr{R}(\mathsf{INR}_2, \mathsf{INR}_1)$ and outer bound (39).

Evaluating the achievable region (58) with $\mathsf{INR}_{p2} = \mathsf{INR}_2$ and $\mathsf{INR}_{p1} = \mathsf{INR}_1$ and getting rid of redundant bounds, we have the following region $\mathscr{R}(\mathsf{INR}_2, \mathsf{INR}_1)$.

$$\begin{aligned} R_1 &\leq \log\left(1 + \frac{\mathsf{SNR}_1}{1 + \mathsf{INR}_1}\right) \\ R_2 &\leq \log\left(1 + \frac{\mathsf{SNR}_2}{1 + \mathsf{INR}_2}\right). \end{aligned} \tag{72}$$

Note that $\mathscr{R}(\mathsf{INR}_2, \mathsf{INR}_1)$ is the achievable region obtained by each user treating the other user's signal as noise. Comparing this region with the outer bound (39), we can see that (62) is true. (See Figure 20)

Combining the above four cases, we have shown that (62) is true for all values of $\mathsf{SNR}_1$, $\mathsf{SNR}_2$, $\mathsf{INR}_1$ and $\mathsf{INR}_2$, given that $\mathsf{INR}_1 < \mathsf{SNR}_2$, $\mathsf{INR}_2 < \mathsf{SNR}_1$. Thus we have proved that the achievable region

$$\mathscr{R}\big(\min(1, \mathsf{INR}_2), \min(1, \mathsf{INR}_1)\big)$$

is within one bit of the capacity region of the Gaussian weak interference channel. □

### 4.3.2 Mixed interference channel

We assume that $\mathsf{INR}_1 \geq \mathsf{SNR}_2$ and $\mathsf{INR}_2 < \mathsf{SNR}_1$ in the mixed interference channel. A remarkable feature of this channel is that user 2's message can be fully decoded at receiver 1. Using this fact, a natural scheme for user 2 is to use all of his power on the common message, i.e. set $\mathsf{INR}_{p1} = 0$. We also let $\mathsf{INR}_{p2}$ to be as close to 1 as possible using the intuition derived from the weak interference channel. We have the following result.

***Theorem* 6.** *The achievable region*

$$\mathscr{R}\big(\min(1, \mathsf{INR}_2), 0\big)$$

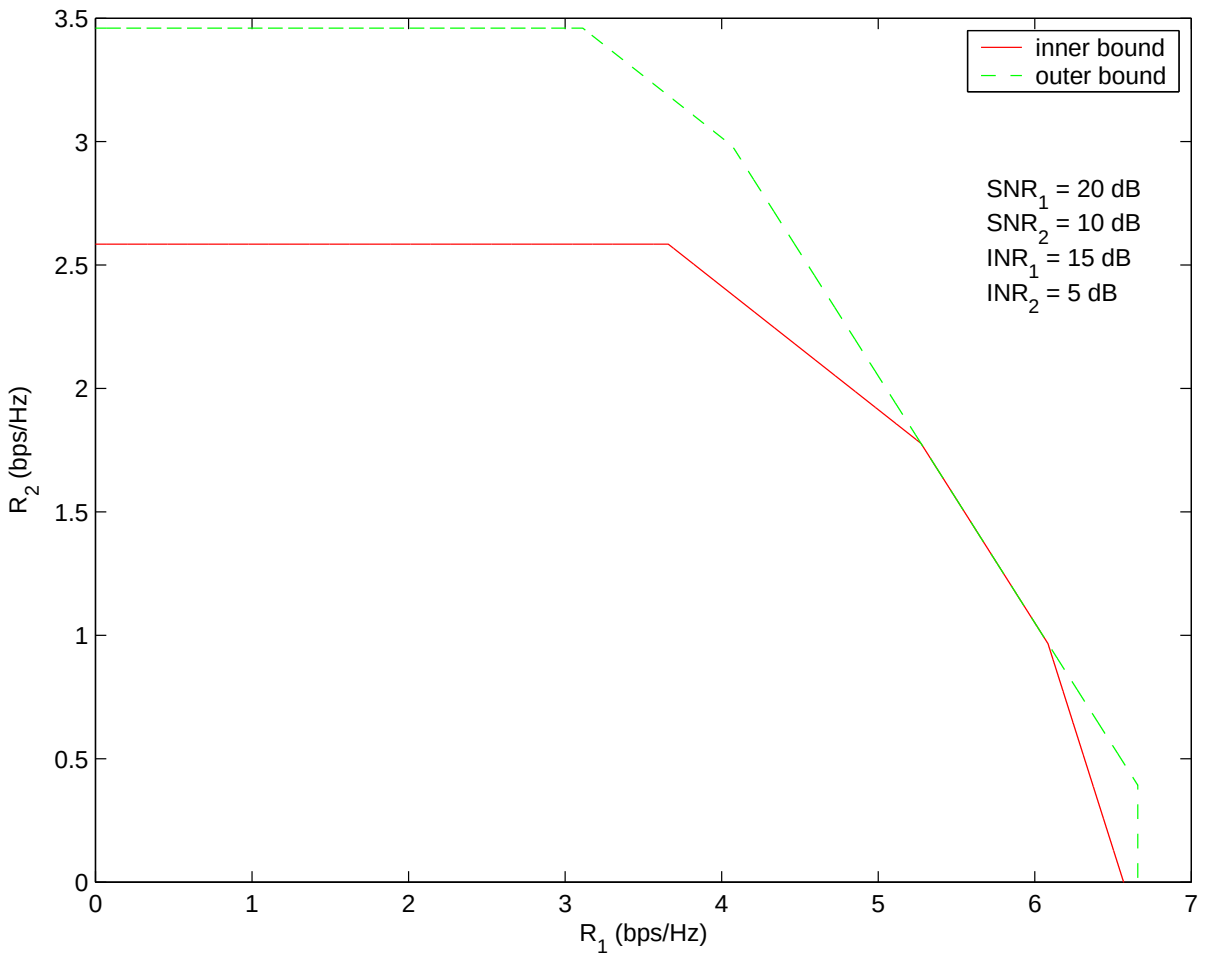

Figure 21: Comparison of Han-Kobayashi achievable region $\mathscr{R}(1,0)$ and outer bound (39).

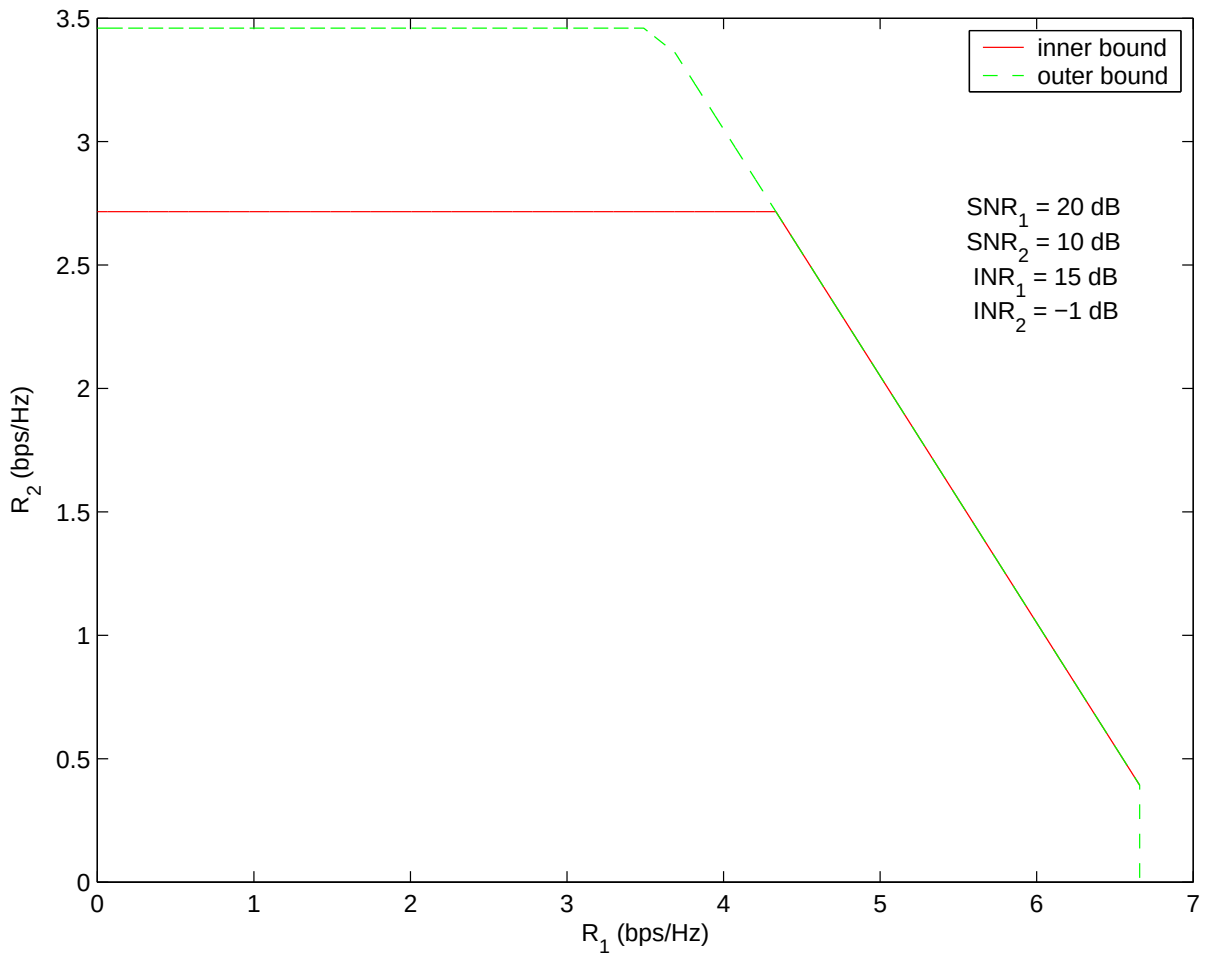

Figure 22: Comparison of Han-Kobayashi achievable region $\mathscr{R}(\mathsf{INR}_2, 0)$ and outer bound (39).

*is within one bit of the capacity region of the Gaussian interference channel when* $\mathsf{INR}_1 \geq \mathsf{SNR}_2$, $\mathsf{INR}_2 < \mathsf{SNR}_1$.

*Proof.* We only need to prove that (62) is true. There are two cases to consider:

1. $\underline{\mathsf{INR}_2 > 1}$

   In this case we use the Han-Kobayashi scheme HK(1,0). By evaluating (58), we have the following result.

   ***Corollary* 2.** *The achievable rate region* $\mathscr{R}(1,0)$ *contains all the rate pairs* $(R_1, R_2)$ *satisfying:*

$$\begin{aligned}
R_1 &\leq \log\left(1 + \mathsf{SNR}_1\right) \\
R_2 &\leq \log\left(2 + \mathsf{SNR}_2\right) - 1 \\
R_1 + R_2 &\leq \log\left(\mathsf{INR}_2 + \mathsf{SNR}_1\right) + \log\left(1 + \frac{1+\mathsf{SNR}_2}{\mathsf{INR}_2}\right) - 1 \\
R_1 + R_2 &\leq \log\left(1 + \mathsf{INR}_1 + \mathsf{SNR}_1\right) \\
R_1 + R_2 &\leq \log\left(1 + \mathsf{INR}_1 + \frac{\mathsf{SNR}_1}{\mathsf{INR}_2}\right) + \log(1+\mathsf{INR}_2) - 1 \\
2R_1 + R_2 &\leq \log(1+\mathsf{INR}_2) + \log(1+\mathsf{SNR}_1+\mathsf{INR}_1) + \log\left(1 + \frac{\mathsf{SNR}_1}{\mathsf{INR}_2}\right) - 1 \\
R_1 + 2R_2 &\leq \log\left(1 + \mathsf{SNR}_2 + \mathsf{INR}_2\right) + \log\left(1 + \mathsf{INR}_1 + \frac{\mathsf{SNR}_1}{\mathsf{INR}_2}\right) - 1.
\end{aligned} \tag{73}$$

   Comparing $\mathscr{R}(1,0)$ in (73) with the outer bound (51), we can see that (62) is true. (See Figure 21).

2. $\underline{\mathsf{INR}_2 < 1}$

We use the Han-Kobayashi scheme HK$(\mathsf{INR}_2, 0)$, and get the following Han-Kobayashi achievable region $\mathscr{R}(\mathsf{INR}_2, 0)$:

$$\begin{aligned} R_1 &\leq \log\left(1+\mathsf{SNR}_1\right) \\ R_2 &\leq \log\left(1+\frac{\mathsf{SNR}_2}{1+\mathsf{INR}_2}\right) \\ R_1+R_2 &\leq \log(1+\mathsf{SNR}_1)+\log\left(1+\frac{\mathsf{SNR}_2}{1+\mathsf{INR}_2}\right) \\ R_1+R_2 &\leq \log\left(1+\mathsf{SNR}_1+\mathsf{INR}_1\right) \\ R_1+R_2 &\leq \log\left(1+\mathsf{SNR}_1+\mathsf{INR}_1\right) \\ 2R_1+R_2 &\leq \left(1+\mathsf{SNR}_1\right)+\log\left(1+\mathsf{SNR}_1+\mathsf{INR}_1\right) \\ R_1+2R_2 &\leq \log\left(1+\frac{\mathsf{SNR}_2}{1+\mathsf{INR}_2}\right)+\log\left(1+\mathsf{SNR}_1+\mathsf{INR}_1\right). \end{aligned} \tag{74}$$

We can see that the bounds on $2R_1 + R_2$ and $R_1 + 2R_2$ are redundant. Comparing this region with the outer bound (51), we can see that (62) is true. (See figure 22).

Thus we have shown that (62) is true and we have proved theorem 6. □

### 4.4 Discussion on one-bit result

The achievable region discussed in the previous section is not the largest possible. In fact, we can easily improve the achievable region by using other private-common message power splits. For example, Costa [4] pointed out that if we require receiver 2 to fully decode user 1's message before decoding his own message, and we require receiver 1 to treat user 2's signal as noise, then the rate pair

$$\begin{aligned} R_1 &= \log\left(1+\frac{\mathsf{INR}_2}{1+\mathsf{SNR}_2}\right) \\ R_2 &= \log\left(1+\mathsf{SNR}_2\right) \end{aligned} \tag{75}$$

is achievable. This rate pair is not inside $\mathscr{R}(1,1)$. However, we can achieve this rate pair by using the scheme HK$(\mathsf{INR}_2, 0)$, i.e., user 1 has only common message, and user 2 has only private message. We do not intend to optimize over all possible Han-Kobayashi strategies to get the largest achievable region, which can be a very complicated task. In fact, the most important point that we want to make with our one-bit result is that we do not lose much by using a simple Han-Kobayashi strategy.

Our one-bit result shows that $\mathscr{R}(1,1)$ is a good approximation to the capacity region in the high SNR, INR regime, since one bit is relatively a small number compared to the rates of the users. The high SNR, INR regime corresponds to the interference-limited situation where interference plays a major role in communication. The low SNR, INR regime ($\mathsf{SNR} << 1$ and $\mathsf{INR} << 1$) is not very interesting since the effect of interference is smaller than that of the additive Gaussian noise. Nevertheless, a loss of one bit in this regime may be large compared to the rates of the users. However, in the low SNR, INR regime we can achieve the following region by simply treating interference as noise:

$$\begin{aligned} R_1 &\le \log\left(1+\frac{\mathsf{SNR}_1}{1+\mathsf{INR}_1}\right) \\ R_2 &\le \log\left(1+\frac{\mathsf{SNR}_2}{1+\mathsf{INR}_2}\right). \end{aligned} \tag{76}$$

Comparing this region with the simple point-to-point outer bound:

$$\begin{aligned} R_1 &\le \log\left(1+\mathsf{SNR}_1\right) \\ R_2 &\le \log\left(1+\mathsf{SNR}_2\right), \end{aligned} \tag{77}$$

we can see that if $(R_1, R_2)$ is on the boundary of the achievable region (76), then $(2R_1, 2R_2)$ is outside of the capacity region. We say that region (76) is *within half of the capacity region*. This is a complementary result to our one-bit result. Note that treating interference as noise is one special case of our Han-Kobayashi scheme. In fact, we have similar results for all parameter values for our scheme.

***Theorem* 7.** *The achievable region*

$$\mathscr{R}\big(\min(1, \mathsf{INR}_2), \min(1, \mathsf{INR}_1)\big) \tag{78}$$

*is within half of the capacity region of the Gaussian weak interference channel.*

***Theorem* 8.** *The achievable region*

$$\mathscr{R}\big(\min(1, \mathsf{INR}_2), 0\big)$$

*is within half of the capacity region of the Gaussian interference channel when* $\mathsf{INR}_1 \ge \mathsf{SNR}_2$, $\mathsf{INR}_2 < \mathsf{SNR}_1$.

*Proof.* By comparing the achievable region with the corresponding outer bound, we can prove these results after some algebraic manipulation. The proofs of these two theorems are similar to the proofs of theorems 5 and 6, and hence are omitted. □

We conclude this section two additional remarks:

1. We can achieve fairly good performance by using a simple choice of the Han-Kobayashi scheme where $\mathsf{INR}_p$ is chosen as close to 1 as possible. We will provide additional insights about why this choice is a natural one in Section 6.

2. We derived a new outer bound on the capacity region of the Gaussian interference channel. The one-bit result shows that this outer bound is quite good in the high SNR, INR regime. The new bound is motivated by the results for a certain class of deterministic interference channels of [3]. We will investigate the connection between our results and those of [3] in Section 7.

# 5 Generalized Degrees of Freedom Region

At high SNR and INR, we can generalize the notion of degrees of freedom for the symmetric capacity of the symmetric Gaussian interference channel to the entire region for all values of parameters by focusing only on the first order terms in $\log \mathsf{SNR}_1$,$\log \mathsf{SNR}_2$,$\log \mathsf{INR}_1$ and $\log \mathsf{INR}_2$. More precisely, we use approximations such as:

$$\begin{aligned} \log(1+\mathsf{SNR}_1+\mathsf{INR}_1) &\approx \max(\log \mathsf{SNR}_1, \log \mathsf{INR}_1) \\ \log\left(1+\frac{\mathsf{SNR}_1}{1+\mathsf{INR}_1}\right) &\approx \left(\log \frac{\mathsf{SNR}_1}{\mathsf{INR}_1}\right)^+ \end{aligned} \tag{79}$$

to provide an expansion of the capacity region of the Gaussian interference channel which is accurate to first order. These first order approximations satisfy the property that the higher order terms are $O(1)$. Therefore the approximation error relative to $\log \mathsf{SNR}_1$ etc. vanishes as $\mathsf{SNR}_1 \to \infty$. This property will be useful in the derivation of the generalized degrees of freedom region to be considered next.

Let $\mathcal{C}(\mathsf{SNR}_1, \mathsf{SNR}_2, \mathsf{INR}_1, \mathsf{INR}_2)$ denote the capacity region of the interference channel with parameters $\mathsf{SNR}_1, \mathsf{SNR}_2, \mathsf{INR}_1, \mathsf{INR}_2$. Let $\tilde{\mathcal{D}}$ be a scaled version of $\mathcal{C}(\mathsf{SNR}_1, \mathsf{SNR}_2, \mathsf{INR}_1, \mathsf{INR}_2)$ given by:

$$\tilde{\mathcal{D}}(\mathsf{SNR}_1, \mathsf{SNR}_2, \mathsf{INR}_1, \mathsf{INR}_2) = \left\{\left(\frac{R_1}{\log \mathsf{SNR}_1}, \frac{R_2}{\log \mathsf{SNR}_2}\right) : (R_1, R_2) \in \mathcal{C}(\mathsf{SNR}_1, \mathsf{SNR}_2, \mathsf{INR}_1, \mathsf{INR}_2)\right\}$$

and let

$$\begin{aligned} \alpha_1 &= \frac{\log \mathsf{SNR}_2}{\log \mathsf{SNR}_1} \\ \alpha_2 &= \frac{\log \mathsf{INR}_1}{\log \mathsf{SNR}_1} \\ \alpha_3 &= \frac{\log \mathsf{INR}_2}{\log \mathsf{SNR}_1}. \end{aligned}$$

We define the generalized degrees of freedom region as:

$$\mathcal{D}(\alpha_1, \alpha_2, \alpha_3) = \lim_{\substack{\mathsf{SNR}_1, \mathsf{SNR}_2, \mathsf{INR}_1, \mathsf{INR}_2 \to \infty \\ \alpha_1, \alpha_2, \alpha_3 \text{ fixed}}} \tilde{\mathcal{D}}(\mathsf{SNR}_1, \mathsf{SNR}_2, \mathsf{INR}_1, \mathsf{INR}_2).$$

With this definition, the capacity region can be approximately expressed as the set of rate pairs $(R_1, R_2)$ such that:

$$\begin{aligned} R_1 &= d_1 \log \mathsf{SNR}_1 \\ R_2 &= d_2 \log \mathsf{SNR}_2 \end{aligned}$$

for $(d_1, d_2) \in \mathcal{D}$.

The generalized degrees of freedom $d_1, d_2$ give a sense of how interference affects communication. In the absence of interference, each user can achieve a rate $R_i \approx \log \mathsf{SNR}_i$. Due to interference, the single user capacity is scaled by a factor $d_i$.

In the following subsections we first compute the generalized degrees of freedom region $\mathcal{D}$ for the interference channels, then present the generalized degrees of freedom region for two examples: the symmetric channel and the one-sided interference channel.

## 5.1 Generalized degrees of freedom region of interference channel

### 5.1.1 Weak interference channel

For the weak interference channel, by applying the approximations like the ones of (79) to the outer bound (39) and the achievable region $\mathscr{R}(1,1)$ (equation (64)), we can easily see that the first order expansions of the corresponding bounds are equal, and the resulting first order expansion of the capacity region has the following form:

$$\begin{aligned}
R_1 &\leq \log \mathsf{SNR}_1 \\
R_2 &\leq \log \mathsf{SNR}_2 \\
R_1 + R_2 &\leq \log \mathsf{SNR}_1 + \left(\log\left(\frac{\mathsf{SNR}_2}{\mathsf{INR}_2}\right)\right)^+ \\
R_1 + R_2 &\leq \log \mathsf{SNR}_2 + \left(\log\left(\frac{\mathsf{SNR}_1}{\mathsf{INR}_1}\right)\right)^+ \\
R_1 + R_2 &\leq \max\left(\log \mathsf{INR}_1, \log\left(\frac{\mathsf{SNR}_1}{\mathsf{INR}_2}\right)\right) + \max\left(\log \mathsf{INR}_2, \log\left(\frac{\mathsf{SNR}_2}{\mathsf{INR}_1}\right)\right) \\
2R_1 + R_2 &\leq \max(\log \mathsf{SNR}_1, \log \mathsf{INR}_1) + \max\left(\log \mathsf{INR}_2, \log\frac{\mathsf{SNR}_2}{\mathsf{INR}_1}\right) + \log\frac{\mathsf{SNR}_1}{\mathsf{INR}_2} \\
R_1 + 2R_2 &\leq \max(\log \mathsf{SNR}_2, \log \mathsf{INR}_2) + \max\left(\log \mathsf{INR}_1, \log\frac{\mathsf{SNR}_1}{\mathsf{INR}_2}\right) + \log\frac{\mathsf{SNR}_2}{\mathsf{INR}_1}.
\end{aligned} \tag{80}$$

From (80), we have that the generalized degrees of freedom region is given by:

$$\begin{aligned}
d_1 &\leq 1 \\
d_2 &\leq 1 \\
d_1 + \alpha_1 d_2 &\leq 1 + (\alpha_1 - \alpha_3)^+ \\
d_1 + \alpha_1 d_2 &\leq \alpha_1 + (1 - \alpha_2)^+ \\
d_1 + \alpha_1 d_2 &\leq \max(\alpha_2, 1 - \alpha_3) + \max(\alpha_3, \alpha_1 - \alpha_2) \\
2d_1 + \alpha_1 d_2 &\leq \max(1, \alpha_2) + \max(\alpha_3, \alpha_1 - \alpha_2) + 1 - \alpha_3 \\
d_1 + 2\alpha_1 d_2 &\leq \max(\alpha_1, \alpha_3) + \max(\alpha_2, 1 - \alpha_3) + \alpha_1 - \alpha_2.
\end{aligned} \tag{81}$$

### 5.1.2 Mixed interference channel

For mixed interference channel, by applying the approximations like the ones of (79) on the outer bound (51) and the inner bound (73), we can easily see that the first order expansions of the corresponding bounds are equal, and the resulting first order expansion of the capacity region has the following form:

$$\begin{aligned}
R_1 &\le \log \mathsf{SNR}_1 \\
R_2 &\le \log \mathsf{SNR}_2 \\
R_1 + R_2 &\le \ \log \mathsf{SNR}_1 + \left(\log\left(\frac{\mathsf{SNR}_2}{\mathsf{INR}_2}\right)\right)^+ \\
R_1 + R_2 &\le \max(\log \mathsf{SNR}_1, \log \mathsf{INR}_1) \\
R_1 + R_2 &\le \max\left(\log \mathsf{INR}_1, \log\left(\frac{\mathsf{SNR}_1}{\mathsf{INR}_2}\right)\right) + \log \mathsf{INR}_2 \\
2R_1 + R_2 &\le \max(\log \mathsf{SNR}_1 + \log \mathsf{INR}_1, 2\log \mathsf{SNR}_1) \\
R_1 + 2R_2 &\le \max(\log \mathsf{SNR}_2, \log \mathsf{INR}_2) + \max\left(\log \mathsf{INR}_1, \log \frac{\mathsf{SNR}_1}{\mathsf{INR}_2}\right).
\end{aligned} \tag{82}$$

Note that the third bound on $R_1 + R_2$ can be written as:

$$R_1 + R_2 \le \max\left(\log \mathsf{INR}_1, \log\left(\frac{\mathsf{SNR}_1}{\mathsf{INR}_2}\right)\right) + \log \mathsf{INR}_2 = \max\left(\log \mathsf{INR}_1 + \log \mathsf{INR}_2, \log \mathsf{SNR}_1\right), \tag{83}$$

which is larger than the second bound on $R_1 + R_2$. Hence the third bound on $R_1 + R_2$ is redundant. The bound on $2R_1 + R_2$ is the same as the sum of the bound on $R_1$ and the second bound on $R_1 + R_2$, so it is also redundant. So we end up with a first order expansion of the capacity region of the form:

$$\begin{aligned}
R_1 &\le \log \mathsf{SNR}_1 \\
R_2 &\le \log \mathsf{SNR}_2 \\
R_1 + R_2 &\le \log \mathsf{SNR}_1 + \left(\log\left(\frac{\mathsf{SNR}_2}{\mathsf{INR}_2}\right)\right)^+ \\
R_1 + R_2 &\le \max(\log \mathsf{SNR}_1, \log \mathsf{INR}_1) \\
R_1 + 2R_2 &\le \max\left(\log \mathsf{SNR}_2, \log \mathsf{INR}_2\right) + \max\left(\log \mathsf{INR}_1, \log\left(\frac{\mathsf{SNR}_1}{\mathsf{INR}_2}\right)\right).
\end{aligned} \tag{84}$$

Using (84) we obtain the generalized degrees of freedom region for the mixed interference channel:

$$\begin{aligned}
d_1 &\le 1 \\
d_2 &\le 1 \\
d_1 + \alpha_1 d_2 &\le \ 1 + (\alpha_1 - \alpha_3)^+ \\
d_1 + \alpha_1 d_2 &\le \max(1, \alpha_2) \\
d_1 + 2\alpha_1 d_2 &\le \max(\alpha_1, \alpha_3) + \max\left(\alpha_2, 1 - \alpha_3\right).
\end{aligned} \tag{85}$$

### 5.1.3 Strong interference channel

The capacity region of the strong interference channel is shown to be the intersection of that of two multiple access channels, and is given by

$$\begin{aligned} R_1 &\le \log(1+\mathsf{SNR}_1) \\ R_2 &\le \log(1+\mathsf{SNR}_2) \\ R_1+R_2 &\le \log(1+\mathsf{SNR}_1+\mathsf{INR}_1) \\ R_1+R_2 &\le \log(1+\mathsf{SNR}_2+\mathsf{INR}_2). \end{aligned} \tag{86}$$

By applying the approximations like the ones of (79), we obtain the generalized degrees of freedom region for the strong interference channel.

$$\begin{aligned} d_1 &\le 1 \\ d_2 &\le 1 \\ d_1+\alpha_1 d_2 &\le \max(1,\alpha_2) \\ d_1+\alpha_1 d_2 &\le \max(\alpha_1,\alpha_3). \end{aligned} \tag{87}$$

## 5.2 Example 1: the symmetric channel

For the symmetric Gaussian interference channel with $\mathsf{SNR}_1=\mathsf{SNR}_2=\mathsf{SNR}$, $\mathsf{INR}_1=\mathsf{INR}_2=\mathsf{INR}$ and $\mathsf{SNR}>>\mathsf{INR}$, we have $\alpha_1=1$ and $\alpha_2=\alpha_3=\alpha$. In this case, for weak interference channel in which $0<\alpha<1$, we have the following generalized degrees of freedom region:

$$\begin{aligned} d_1 &\le 1 \\ d_2 &\le 1 \\ d_1+d_2 &\le \min\left\{2-\alpha, 2\max\left(\alpha,1-\alpha\right)\right\} \\ 2d_1+d_2 &\le 2-\alpha+\max\left(\alpha,1-\alpha\right) \\ d_1+2d_2 &\le 2-\alpha+\max\left(\alpha,1-\alpha\right). \end{aligned} \tag{88}$$

For strong interference channel in which $\alpha\ge 1$, we have the following generalized degrees of freedom region:

$$\begin{aligned} d_1 &\le 1 \\ d_2 &\le 1 \\ d_1+d_2 &\le \alpha. \end{aligned} \tag{89}$$

The generalized degrees of freedom region of the symmetric channel is plotted in Figure 23. Note that the diagram for $\alpha\ge 2$ corresponds to the very strong interference case, in which interference does not reduce the available degrees of freedom of the channel. We see that the degrees of freedom region is *not* monotonically decreasing with $\mathsf{INR}$ in the weak interference regime. As in the case of the symmetric rate discussed in Section 3, there are three regimes in which the degrees of freedom region shows different qualitative behaviors, namely $0\le\alpha<1/2$, $1/2\le\alpha<2/3$, and $2/3\le\alpha<1$.

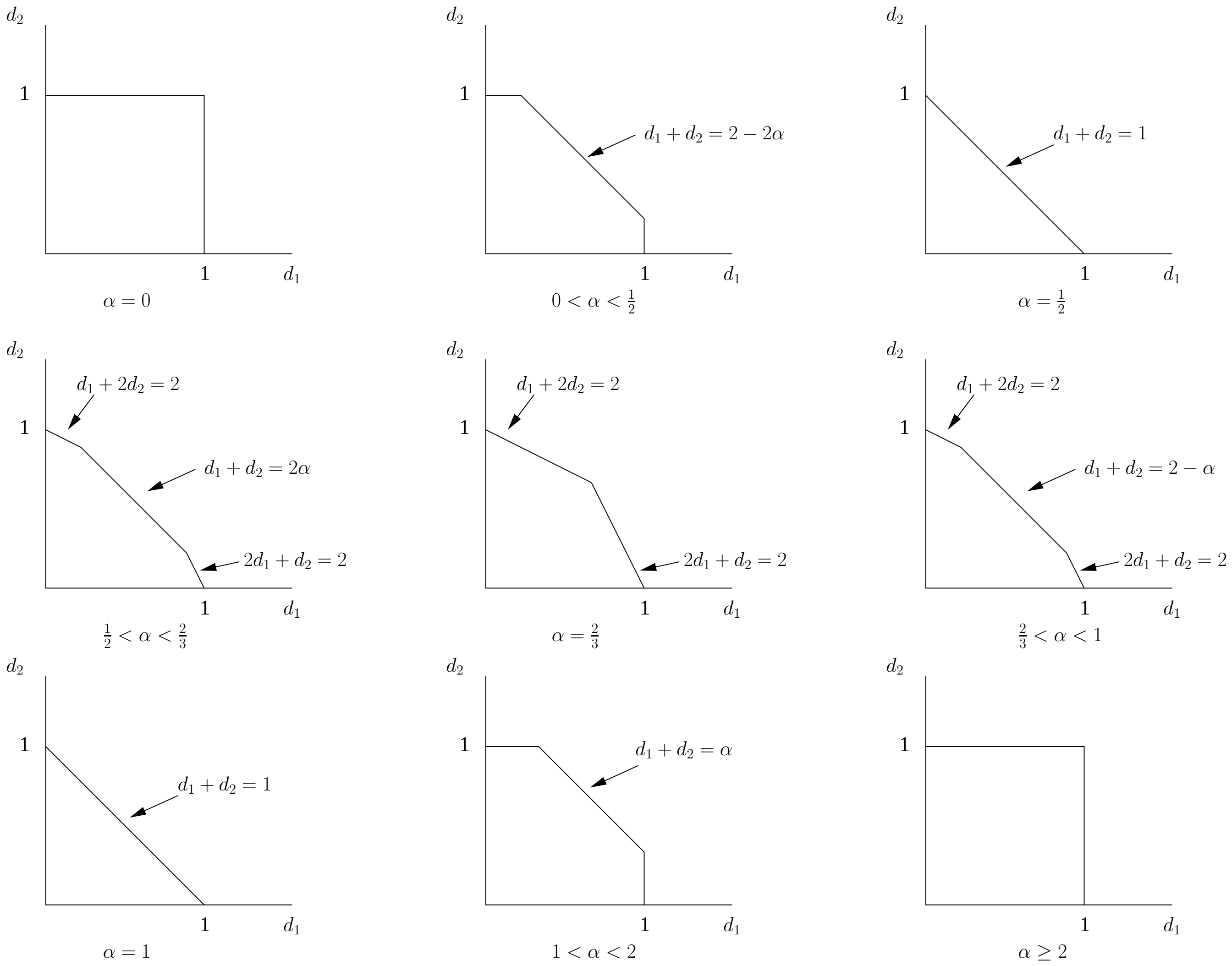

Figure 23: Generalized degrees of freedom region for the symmetric Gaussian interference channel.

If we use an orthogonalizing strategy, the generalized degrees of freedom region that we can achieve is shown in Figure 24. If we treat interference as noise, the generalized degrees of freedom region that we can achieve is shown in Figure 25. So an orthogonalizing strategy is strictly sub-optimal except when $\alpha = \frac{1}{2}$ and $\alpha = 1$, and treating interference as noise is strictly sub-optimal except for $\alpha \leq \frac{1}{2}$, as was already shown in Section 3.6 for the symmetric rate.

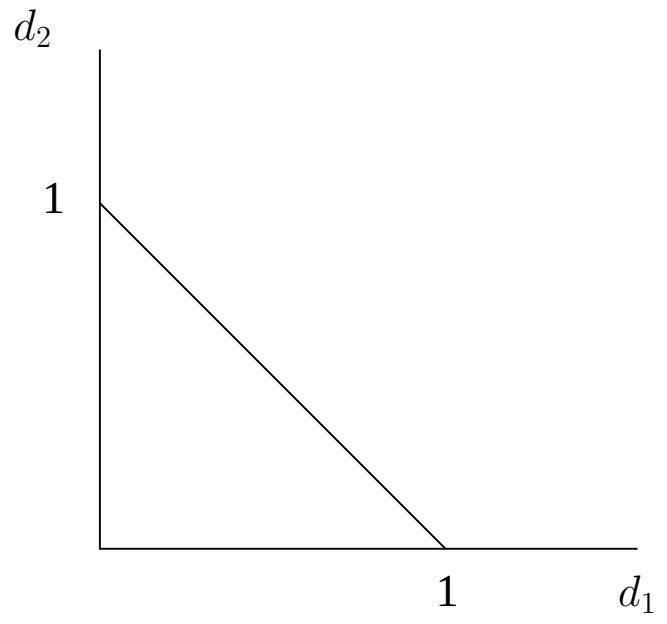

Figure 24: Generalized degrees of freedom region for the symmetric Gaussian weak interference channel using orthogonalizing scheme.

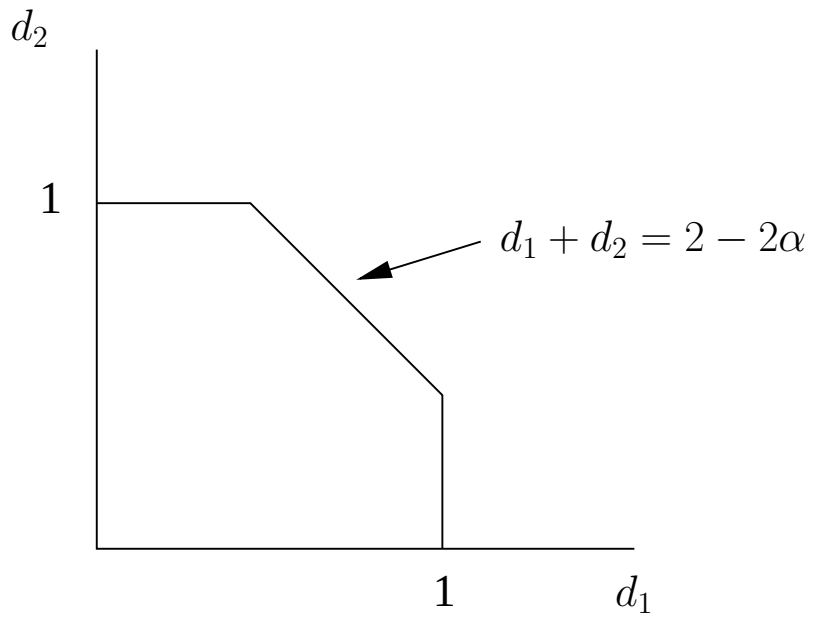

Figure 25: Generalized degrees of freedom region for the symmetric Gaussian weak interference channel when treating interference as noise.

## 5.3 Example 2: the one-sided interference channel

For the one-sided interference channel shown in Figure 7, we have $\mathsf{INR}_1 = 0$. In weak interference case $(\mathsf{SNR}_1 > \mathsf{INR}_2)$, by applying the approximations like the ones of (79) to the achievable region $\mathscr{R}(1, \mathsf{INR}_1)$ given in (69), we get the following first order expansions of the achievable region

$$\begin{aligned} R_1 &\leq \log(\mathsf{SNR}_1) \\ R_2 &\leq \log(\mathsf{SNR}_2) \\ R_1 + R_2 &\leq \log(\mathsf{SNR}_1) + \left(\log\left(\frac{\mathsf{SNR}_2}{\mathsf{INR}_2}\right)\right)^+ . \end{aligned} \tag{90}$$

By examining the proof of the outer bound (39) we can easily see that the previous first order expansions of the achievable region is actually tight. Thus we get the following generalized degrees of freedom region for the one-sided interference channel.

$$\begin{aligned} d_1 &\leq 1 \\ d_2 &\leq 1 \\ d_1 + \alpha_1 d_2 &\leq \max(1, 1 + \alpha_1 - \alpha_3) . \end{aligned} \tag{91}$$

In Figure 26 we plot the generalized degrees of freedom of the weak one-sided interference channel. There are two different cases.

1. $\alpha_1 > \alpha_3$

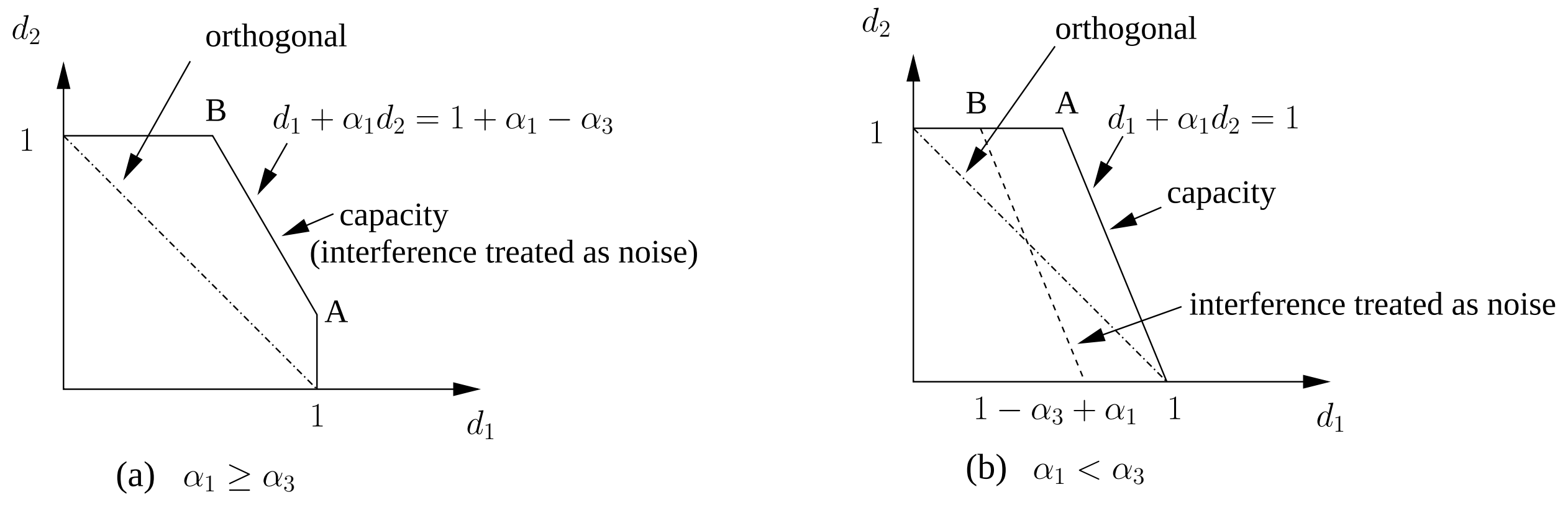

Figure 26: Generalized degrees of freedom region of the weak one-sided interference channel for treating interference as noise, orthogonalizing, and optimal power splitting.

In this case, the generalized degrees of freedom region is:

$$\begin{aligned} d_1 &\leq 1 \\ d_2 &\leq 1 \\ d_1 + \alpha_1 d_2 &\leq 1 + \alpha_1 - \alpha_3. \end{aligned} \tag{92}$$

The entire region can be achieved by adjusting the power of the long-range link and treating its interference as noise. In particular, to achieve the corner point $A = \left(1, 1 - \frac{\alpha_3}{\alpha_1}\right)$, user 1 transmits at full power and user two treats user 1's interference as noise. To achieve the corner point $B = (1 - \alpha_3, 1)$, user 1's transmitting power needs to be reduced so that the received signal to noise ratio at receiver 1 is $\frac{\mathsf{SNR}_1}{\mathsf{INR}_2}$, and user two treats user 1's interference as noise.

2. $\alpha_1 < \alpha_3$

In this case, the generalized degrees of freedom region is:

$$\begin{aligned} d_1 &\leq 1 \\ d_2 &\leq 1 \\ d_1 + \alpha_1 d_2 &\leq 1. \end{aligned} \tag{93}$$

The corner point $A = (1 - \alpha_1, 1)$ has to be achieved by a private common split in user 1's message. Treating interference as noise will only get to the point $B = (1 - \alpha_3, 1)$ which is strictly smaller.

We also draw the performance of the orthogonalizing scheme in Figure 26, which is suboptimal in both cases.

In the strong interference case ($\mathsf{SNR}_1 \leq \mathsf{INR}_2$), the capacity of the one-sided interference channel is known. For completeness, we present the corresponding generalized degrees of freedom region in the following.

$$\begin{aligned} d_1 &\leq 1 \\ d_2 &\leq 1 \\ d_1 + \alpha_1 d_2 &\leq \max(\alpha_1, \alpha_3). \end{aligned} \tag{94}$$

## 6 Private versus common information

In Section 4 we have shown that the simple Han-Kobayashi scheme that sets the private message power so that $\mathsf{INR}_{p1} = 1$ and $\mathsf{INR}_{p2} = 1$ achieves to within 1 bit/s/Hz of the capacity region. We also argued that setting $\mathsf{INR}_{p1} = 1$ and $\mathsf{INR}_{p2} = 1$ achieves a good tradeoff between obtaining a good direct link rate and not causing excessive interference to the other link. In this section we will provide an alternative analysis that justifies the choice $\mathsf{INR}_{p1} = 1$ and $\mathsf{INR}_{p2} = 1$. In brief, we will argue that the information received at the non-intending receiver that is above the noise level should essentially be decodable, and hence can be thought of as common information.

Consider a communication scheme that splits the message to be sent into many sub-messages of small rate and power. The transmitted message is the superposition of these sub-messages, and has total power $P$. Receiver 2 is able to decode the message transmitted from its own transmitter, and subtract it from the received signal $y_2$, obtaining the signal $\tilde{y}_2$. We further make the optimistic assumption that receiver 1 can also decode and subtract the interference received from transmitter 2, obtaining a signal $\tilde{y}_1$. This interference cancellation may not always be possible, and therefore we will obtain an upper bound on the rate of user 1. The resulting channels are:

$$\begin{aligned} \tilde{y}_1 &= \sqrt{\mathsf{SNR}_1}\tilde{x}_1 + z_1 \\ \tilde{y}_2 &= \sqrt{\mathsf{INR}_2}\tilde{x}_1 + z_2 \end{aligned} \tag{95}$$

where $z_1, z_2 \sim \mathcal{CN}(0,1)$ and we normalized $\tilde{x}_1$ so that $E[|\tilde{x}_1|^2] \leq 1$.

We define the differential rates:

$$\begin{aligned} r_1(z) &= \frac{\mathsf{SNR}_1}{1 + \mathsf{SNR}_1 z} \\ r_2(z) &= \frac{\mathsf{INR}_2}{1 + \mathsf{INR}_2 z} \end{aligned} \tag{96}$$

which can be interpreted in the following way: $r_1(z) \cdot dz$ ($r_2(z) \cdot dz$) is the rate that can be achieved in a sub-message of power $dz$ facing an interference of power $z \cdot \mathsf{SNR}_1$ ($z \cdot \mathsf{INR}_2$) in channel 1 (2). (See [11] p. 2802, where this concept is introduced). These functions can also be interpreted as the marginal increase in rate at interference level $z \cdot \mathsf{SNR}_1$ (or $z \cdot \mathsf{INR}_2$).

Imagine we plot these two functions and let us see what happens as $z$ goes from 0 to 1. For $\mathsf{INR}_2 \cdot z \ll 1$, $r_1(z) \gg r_2(z)$. The marginal increase in rate in the direct link is much larger than in the indirect link, and therefore the other receiver has no hope of decoding this sub-message if information is sent at this rate. Thus, at this signal level, information should be private, only decodable by receiver 1. When $\mathsf{INR}_2 \cdot z \gg 1$, $r_2(z) \approx r_1(z) \approx 1/z$ and any information sent in the direct link at this rate can also be decoded by the other receiver. At this signal level, we should therefore be sending common information. Figure 27 shows how the differential rates $r_1(z)$ and $r_2(z)$ vary as a function of $z$ for a symmetric channel where $\mathsf{SNR}_1 = \mathsf{SNR}_2 = 20$ dB and $\mathsf{INR}_1 = \mathsf{INR}_2 = 10$ dB. We see that when $z \approx -10$ dB, which corresponds to $\mathsf{INR}_p = 0$ dB, the differential rates are approximately equal.

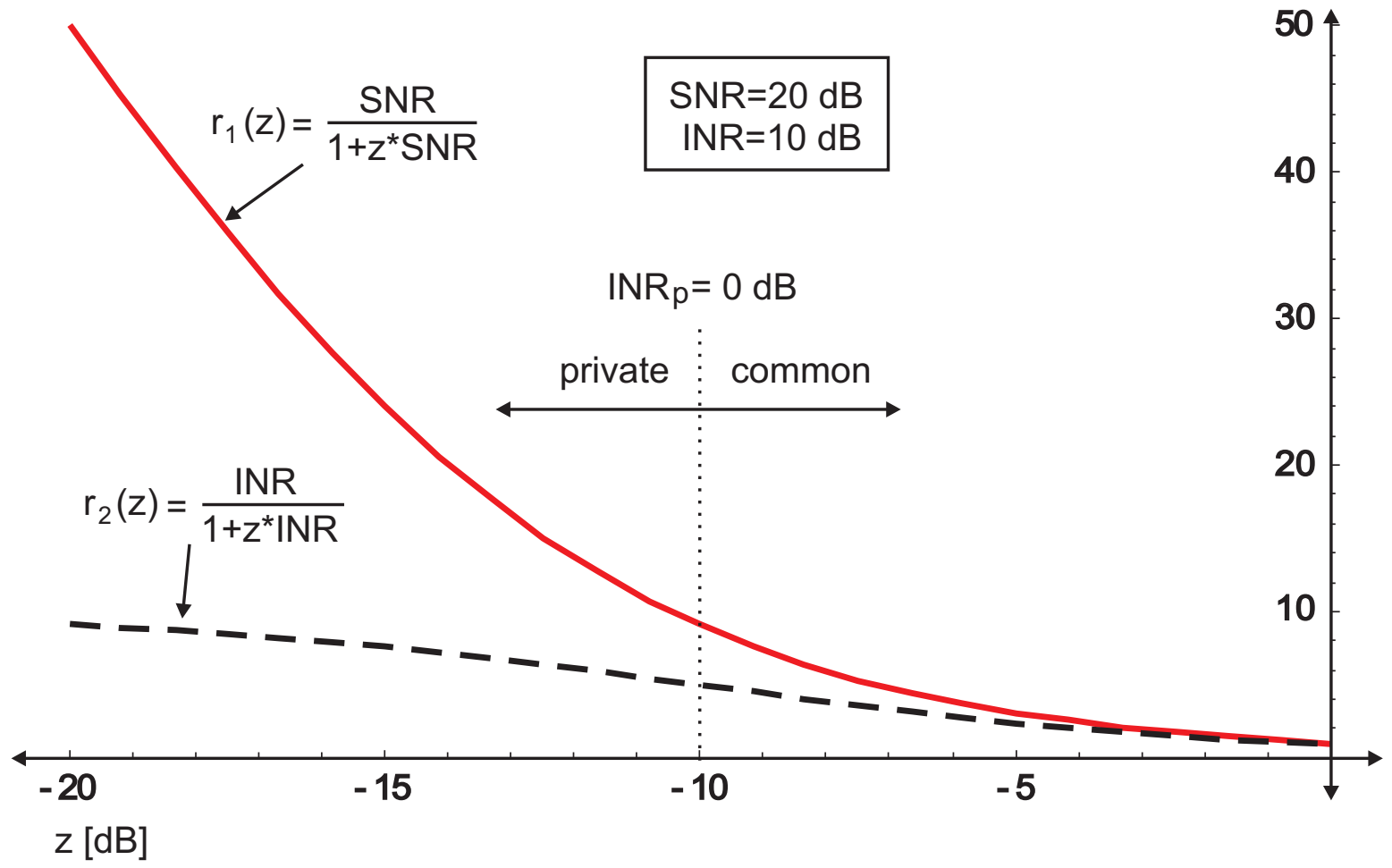

Figure 27: Differential rates $r_1(z)$ and $r_2(z)$ for a symmetric channel where $\mathsf{SNR} = 20$ dB and $\mathsf{INR} = 10$ dB.

The above argument shows that the sub-messages that are decoded first and that face an interference level $z \cdot \mathsf{INR}_2 > 1$ at receiver 2 can be decoded by both receivers, and are therefore common information.

We will now analyze in what situations there is a gain in sending common information. Suppose we start with a nominal strategy of sending all private information at full power on both links. Each receiver treats the interference as noise.

How can we improve this strategy? From the discussion above, without loss of optimality we can convert the part of the signals in both links above the other receiver's noise level into common information. We focus on the part of user 1's signal above receiver 2's noise level, which we call $c_1$. Assume that this signal is getting a rate $R$ in the nominal strategy of treating interference as noise. This can be viewed as common information, decoded in the following ways by the two receivers: receiver 1 decodes $c_1$ first, treating the component sent by transmitter 1 received at receiver 2 below noise level plus the interference from transmitter 2 as noise. From receiver 2 point of view, $c_1$ acts as interference to its own signal and therefore we can view receiver 2 as decoding $c_1$ last, after decoding it's own information. Note that since the decoding order is different at the two receivers, only receiver 1 is limiting the rate of $c_1$. In receiver 2, there is still slack: even if we increased the rate of $c_1$ beyond $R$, receiver 2 would still be able to decode. By the same logic, the part of the signal above noise level sent from transmitter 2 (call it $c_2$) also has slack at receiver 1.

This suggests that we can improve the performance by changing the decoding order of $c_1$ in receiver 1. If we decode an $\epsilon$ part of $c_1$ after decoding an $\epsilon$ part of $c_2$ ($\epsilon$ swap of ordering in receiver 1), then the rate assigned to $c_1$ can be improved from $R$ to $R + \delta$. Note that $c_1$ can still be decoded by receiver 2, since there was slack in the first place. Also, the mutual information achieved for $c_2$ at receiver 1 has decreased because its ordering is a slightly less favorable, but because there was slack for $c_2$ at receiver 1, $c_2$ can still be decoded.

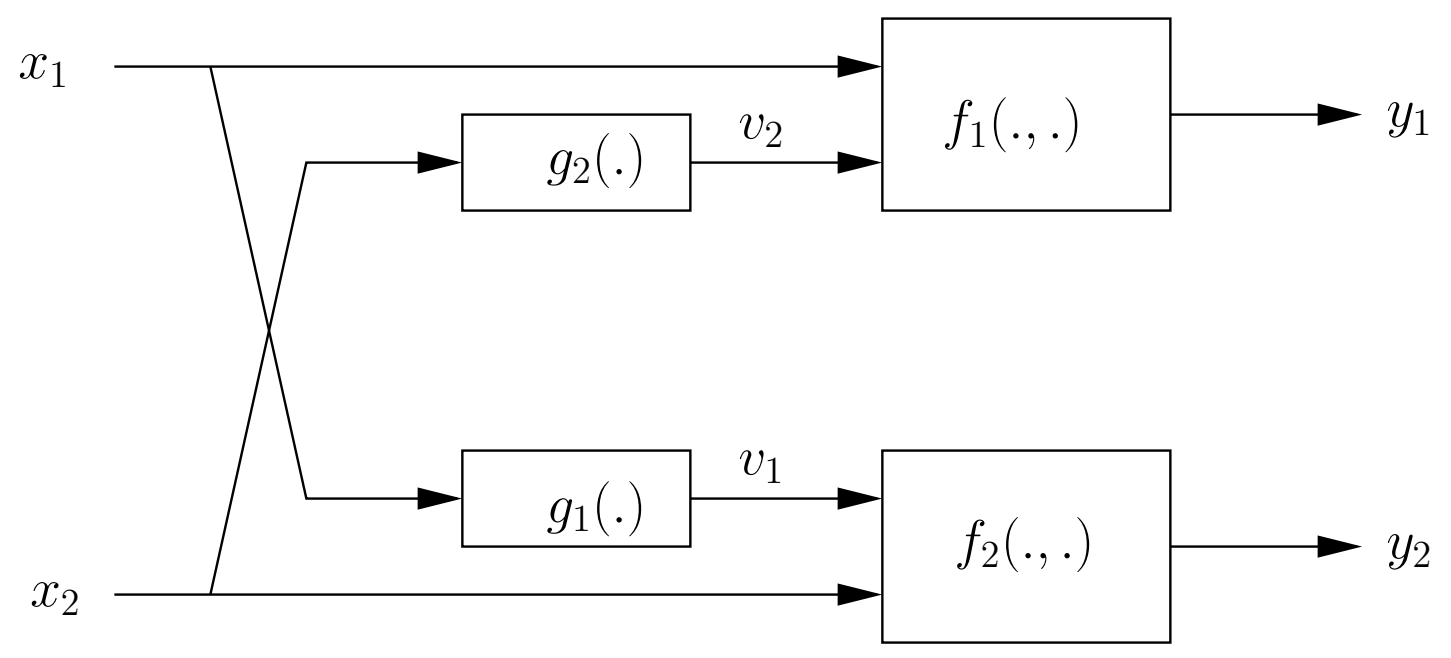

Figure 28: Deterministic interference channel.

Thus, we have improved the rate of user 1 while keeping the rate of user 2 invariant. Therein lies the power of viewing the signal above noise level as common information: there is flexibility in changing the decoding order. When viewed as private information, the decoding order is fixed and there is slack in one of the two receivers that cannot be exploited. That is in essence the "structure" in the interfering signals that is not exploited in treating interference as noise. By changing the cancellation ordering, we are reducing the slack in one of the receivers of the common information.

# 7 Connection to a Deterministic Interference Channel

In Section 6 we argued that the portion of the received interfering signal above the noise level should be common information and that hidden below the noise level should be private. In other words, the part of the received interfering signal that is most visible to the other link is made common while the rest is made private. This argument is only approximate, as the part of the interfering signal below the noise level still has some visibility to the other link. Therefore the proposed strategy still has up to one-bit gap to capacity.

There is in fact a channel in which part of the interfering signal is *completely* invisible to the other link. This channel, introduced by El Gamal and Costa in [3], is a special type of a deterministic interference channel. Because of the complete invisibility of part of the signal, they can show that a Han-Kobayashi strategy of assigning common information to the visible part and private information to the invisible part is *exactly* optimal. Our approach to the Gaussian interference channel is in fact based on drawing analogies to this deterministic channel.

The deterministic interference channel of [3] is shown in Figure 28. In this channel, $x_1$ and $x_2$ are the inputs, $v_1 = g_1(x_1)$ is the interference caused by $x_1$ at receiver 2 and $v_2 = g_2(x_2)$ is the interference caused by $x_2$ at receiver 1. $y_1$ and $y_2$ are the outputs, and they are deterministic functions of $(x_1, v_2)$ and $(x_2, v_1)$,

respectively:

$$y_1 = f_1(x_1, v_2)$$
$$y_2 = f_2(x_2, v_1). \tag{97}$$

In addition, there is an important assumption about the interference signals $v_1$ and $v_2$ given by the following conditions:

$$H(y_1|x_1) = H(v_2)$$
$$H(y_2|x_2) = H(v_1). \tag{98}$$

The previous conditions are equivalent to the existence of functions $h_1(.,.)$ and $h_2(.,.)$ such that

$$v_1 = h_2(x_2, y_2)$$
$$v_2 = h_1(x_1, y_1). \tag{99}$$

These conditions guarantee that each receiver can observe a clean version of the interfering signal after decoding the own message. This assumption is key for deriving the capacity region of the channel. One can argue that regardless of the communication strategy, the signals $v_1$ and $v_2$ are common information, since they can be cleanly observed after decoding the own message. In addition, due to the functions $g_1(\cdot)$ and $g_2(\cdot)$ part of the transmitted message is completely invisible to the non-intending receiver. This part of the message becomes private information.

In the deterministic channel, the conditions (98) seem artificial, but in the Gaussian channel analogous conditions arise more naturally. For the Gaussian interference channel, receiver 2 can observe $s_1 = h_{12}x_1 + z_2$ after decoding $x_2$. Similarly, we can define $s_2 = h_{21}x_2 + z_1$, which is the signal that receiver 1 can observe after decoding the message $x_1$. As we argued in the case of the deterministic channel, the signals $s_1$ and $s_2$ can be thought of as the common information that can be observed after decoding the own message. The role of $z_1$ ($z_2$) in $s_2$ ($s_1$) can be compared to the role of the function $g_2(\cdot)$ ($g_1(\cdot)$) in the deterministic channel, that is, hiding the private information to the non-intending receiver.

The outer bounds derived in [3] to establish the capacity region of the deterministic channel can be interpreted in terms of genie aided channels, with various combinations of $x_1, x_2, v_1, v_2$ given to the receivers. Analogous genie aided channels, with appropriate modifications of the side information, were used in Section 4 to derive the outer bounds for the Gaussian interference channel with weak and mixed interference.

## Appendix A: Analysis of upper bound of [6] Theorem 2

In this appendix we will show that the upper bound of [6] Theorem 2 achieves a worst case gap of 1 bits/s/Hz in the parameter range $\mathcal{B}_1$ with respect to the symmetric rate of our simple Han-Kobayashi scheme of Section 3.2. In addition, the gap between this bound and the Han-Kobayashi scheme can be unbounded in the parameter range $\mathcal{B}_2$. Therefore, the bound of [6] Theorem 2 does not give better performance than the bound (12) in terms of characterizing the symmetric capacity of the symmetric channel to within 1 bit/s/Hz.

The results of [6] are derived for the normalized interference channel, i.e. $|h_d|^2 = 1$, $N_0 = 1$, so in order to use the results, we need to replace: $|h_c|^2 \to |h_c|^2/|h_d|^2$, $P \to |h_d|^2 P/N_0$. Specializing [6] Theorem 2 to the complex symmetric interference channel with $0 < |h_c|^2/|h_d|^2 < 1$ ($0 < \mathsf{INR} < \mathsf{SNR}$) one obtains the symmetric rate upper bound:

$$\begin{aligned} R_{UB_K} &= \log\left[1+\frac{-(|h_c|^2+|h_d|^2)+\sqrt{(|h_c|^2+|h_d|^2)^2+4|h_c|^2|h_d|^2(P/N_0)(|h_c|^2+|h_d|^2)}}{2|h_c|^2}\right] \\ &= \log\left[2-\left(1+\frac{\mathsf{SNR}}{\mathsf{INR}}\right)+\sqrt{\left(1+\frac{\mathsf{SNR}}{\mathsf{INR}}\right)^2+4\cdot\mathsf{SNR}\left(1+\frac{\mathsf{SNR}}{\mathsf{INR}}\right)}\right]-1 \end{aligned} \tag{100}$$

Consider the case in which the first term of the $\min\{\cdot,\cdot\}$ of (7) is active, that is, the parameter range $\mathcal{B}_1$. Then we can write:

$$\begin{aligned} R_{UB_K}-R_{HK_1} &= \log\left[2-\left(1+\frac{\mathsf{SNR}}{\mathsf{INR}}\right)+\sqrt{\left(1+\frac{\mathsf{SNR}}{\mathsf{INR}}\right)^2+4\cdot\mathsf{SNR}\left(1+\frac{\mathsf{SNR}}{\mathsf{INR}}\right)}\right]-1 \\ &\quad -\frac{1}{2}\log\left(1+\mathsf{SNR}+\mathsf{INR}\right)-\frac{1}{2}\log\left(2+\frac{\mathsf{SNR}}{\mathsf{INR}}\right)+1 \\ &= \frac{1}{2}\log\left\{\left[1-\frac{\mathsf{SNR}}{\mathsf{INR}}+\sqrt{\left(1+\frac{\mathsf{SNR}}{\mathsf{INR}}\right)^2+4\cdot\mathsf{SNR}\left(1+\frac{\mathsf{SNR}}{\mathsf{INR}}\right)}\right]^2\right\} \\ &\quad -\frac{1}{2}\log\left(1+\mathsf{SNR}+\mathsf{INR}\right)-\frac{1}{2}\log\left(2+\frac{\mathsf{SNR}}{\mathsf{INR}}\right) \\ &\le \frac{1}{2}\log\left\{2\left[1+\left(\frac{\mathsf{SNR}}{\mathsf{INR}}\right)^2\right]+4\cdot\mathsf{SNR}\left(1+\frac{\mathsf{SNR}}{\mathsf{INR}}\right)+2\left[1-\left(\frac{\mathsf{SNR}}{\mathsf{INR}}\right)^2\right]\right\} \\ &\quad -\frac{1}{2}\log\left[\left(1+\mathsf{SNR}+\mathsf{INR}\right)\cdot\left(2+\frac{\mathsf{SNR}}{\mathsf{INR}}\right)\right] \\ &= \frac{1}{2}\log\left\{\frac{4}{2+(\mathsf{SNR}/\mathsf{INR})}\cdot\frac{1+\mathsf{SNR}\cdot(1+\mathsf{SNR}/\mathsf{INR})}{1+\mathsf{SNR}+\mathsf{INR}}\right\} \\ &\le \frac{1}{2}\log\left[\frac{4}{2+(\mathsf{SNR}/\mathsf{INR})}\cdot\frac{\mathsf{SNR}}{\mathsf{INR}}\right] \\ &= \frac{1}{2}\log\left[\frac{4}{2(\mathsf{INR}/\mathsf{SNR})+1}\right] \\ &< 1 \end{aligned} \tag{101}$$

where we used $\sqrt{1+x} \geq 1$ for $x \geq 0$ in the first inequality and the assumption $\mathsf{INR} \leq \mathsf{SNR}$ in the second inequality. It is easy to check that for example, if $\mathsf{INR} = \mathsf{SNR}^{3/4}$, $R_{UB_1} - R_{HK_1} \to 1$ as $\mathsf{SNR} \to \infty$. Therefore the worst case difference of 1 bit/s/Hz in (101) can actually occur.

We see that in the parameter range $\mathcal{B}_1$, the achievable strategy and the upper bound differ for at most 1 bit/s/Hz. Therefore, in this parameter range our simple scheme gives a bounded (and small) gap with respect to the upper bound.

We will now show that in the parameter range $\mathcal{B}_2$, the gap between our scheme and the upper bound (100)

can be arbitrarily large. In this parameter range we can write:

$$\begin{aligned} R_{UB_K} - R_{HK_2} &= \log\left[1 - \frac{\mathsf{SNR}}{\mathsf{INR}} + \sqrt{\left(1+\frac{\mathsf{SNR}}{\mathsf{INR}}\right)^2 + 4\cdot\mathsf{SNR}\left(1+\frac{\mathsf{SNR}}{\mathsf{INR}}\right)}\right] \\ &\quad - \log\left(1+\mathsf{INR}+\frac{\mathsf{SNR}}{\mathsf{INR}}\right) \\ &\geq \log\left[1 - \frac{\mathsf{SNR}}{\mathsf{INR}} + 2\sqrt{\mathsf{SNR}\left(1+\frac{\mathsf{SNR}}{\mathsf{INR}}\right)}\right] \\ &\quad - \log\left(1+\mathsf{INR}+\frac{\mathsf{SNR}}{\mathsf{INR}}\right) \end{aligned}$$

where the inequality follows from discarding $(1+\mathsf{SNR}/\mathsf{INR})^2$ in the square root. To show that this difference can be unbounded, take $\mathsf{INR} = \sqrt{\mathsf{SNR}}$. With this choice of parameters we get:

$$R_{UB_K} - R_{HK_2} \geq \log\left[\frac{1-\sqrt{\mathsf{SNR}}+2\sqrt{\mathsf{SNR}(1+\sqrt{\mathsf{SNR}})}}{1+2\sqrt{\mathsf{SNR}}}\right]$$

where the right hand side goes to infinity for $\mathsf{SNR} \to \infty$.